\documentclass[twocolumn,apj,numberedappendix]{openjournal}

\usepackage[dvipsnames]{xcolor}
\usepackage[T1]{fontenc}
\usepackage[utf8]{inputenc}
\usepackage[english]{babel}
\usepackage{natbib}
\usepackage{amsmath}
\usepackage{listings}
\definecolor{lbcolor}{rgb}{0.9,0.9,0.9}
\definecolor{cosmiclatte}{rgb}{1.0, 0.97, 0.91}
\usepackage{hyperref}
\hypersetup{
    unicode, 
    colorlinks=true,
    linkcolor=linkcolor,
    citecolor=linkcolor,
    filecolor=linkcolor,
    urlcolor=linkcolor,
}
\usepackage{color,colortbl}
\definecolor{linkcolor}{rgb}{0.0,0.3,0.5}
\DeclareGraphicsExtensions{.bmp,.png,.jpg,.pdf}
\usepackage{orcidlink}

\usepackage{savesym}
\savesymbol{tablenum}
\usepackage{siunitx}
\restoresymbol{SIX}{tablenum}
\usepackage{float}
\usepackage{marginnote}
\usepackage{tabularray}
\usepackage{booktabs}
\usepackage{bm}
\usepackage{tikz}
\usepackage[normalem]{ulem}
\usepackage{xcolor}

\usepackage{microtype}

\UseTblrLibrary{booktabs}
\usetikzlibrary{calc}

\DeclareSIUnit{\erg}{erg}

\usepackage{xspace}

\newcommand{\isotm}[2]{{}^{#2}\mathrm{#1}}

\newcommand{\gcc}{\mathrm{g~cm^{-3} }}

\newcommand{\mesa}{{\sffamily MESA}\xspace}
\newcommand{\maestroex}{{\sffamily MAESTROeX}\xspace}
\newcommand{\castro}{{\sffamily Castro}\xspace}
\newcommand{\amrex}{{\sffamily AMReX}\xspace}
\newcommand{\microphysics}{{\sffamily Microphysics}\xspace}
\newcommand{\pynucastro}{{\sffamily pynucastro}\xspace}
\newcommand{\flash}{{\sffamily Flash}\xspace}
\newcommand{\yt}{{\sffamily yt}\xspace}

\newcommand{\unyt}{{\sffamily unyt}\xspace}
\newcommand{\athenapp}{{\sffamily Athena}{\footnotesize ++}\xspace}
\newcommand{\phoebus}{{\sffamily Phoebus}\xspace}

\newcommand{\nucleus}{{\tt Nucleus}\xspace}

\newcommand{\rate}{{\tt Rate}\xspace}
\newcommand{\reaclibrate}{{\tt ReacLibRate}\xspace}
\newcommand{\derivedrate}{{\tt DerivedRate}\xspace}
\newcommand{\modifiedrate}{{\tt ModifiedRate}\xspace}
\newcommand{\branchedrate}{{\tt BranchedRate}\xspace}
\newcommand{\approximaterate}{{\tt ApproximateRate}\xspace}
\newcommand{\tabularrate}{{\tt TabularWeakRate}\xspace}
\newcommand{\temperaturetabularrate}{{\tt TemperatureTabularRate}\xspace}
\newcommand{\starlibrate}{{\tt StarLibRate}\xspace}

\newcommand{\library}{{\tt Library}\xspace}
\newcommand{\reacliblibrary}{{\tt ReacLibLibrary}\xspace}
\newcommand{\tabularlibrary}{{\tt TabularWeakLibrary}\xspace}
\newcommand{\starliblibrary}{{\tt StarLibLibrary}\xspace}

\newcommand{\networksolution}{{\tt NetworkSolution}\xspace}

\newcommand{\fermiintegral}{{\tt FermiIntegral}\xspace}
\newcommand{\electroneos}{{\tt ElectronEOS}\xspace}
\newcommand{\stellareos}{{\tt StellarEOS}\xspace}

\newcommand{\ratecollection}{{\tt RateCollection}\xspace}
\newcommand{\pythonnetwork}{{\tt PythonNetwork}\xspace}
\newcommand{\simplecxxnetwork}{{\tt SimpleCxxNetwork}\xspace}
\newcommand{\amrexastrocxxnetwork}{{\tt AmrexAstroCxxNetwork}\xspace}
\newcommand{\fortrannetwork}{{\tt FortranNetwork}\xspace}

\newcommand{\nsenetwork}{{\tt NSENetwork}\xspace}

\newcommand{\composition}{{\tt Composition}\xspace}
\newcommand{\solarcomposition}{{\tt SolarComposition}\xspace}

\newcommand{\supnote}[1]{{(supplemental notebook: {\tt #1})}}

\newcommand{\cxx}{C\nolinebreak\hspace{-.05em}\raisebox{.4ex}{\tiny\bf +}\nolinebreak\hspace{-.10em}\raisebox{.4ex}{\tiny\bf +}\xspace}

\makeatletter
\newcommand{\MarginPar}[1]{%
  \marginnote{%
    \tiny\sffamily\color{red}%
    \if@firstcolumn
      \raggedleft
      \parfillskip=10pt
    \else
      \raggedright
    \fi
    #1%
  }%
}
\makeatother
\newcommand{\numpy}{{\sffamily NumPy}\xspace}
\newcommand{\matplotlib}{{\sffamily matplotlib}\xspace}
\newcommand{\scipy}{{\sffamily SciPy}\xspace}
\newcommand{\sympy}{{\sffamily SymPy}\xspace}
\newcommand{\numba}{{\sffamily Numba}\xspace}
\newcommand{\networkx}{{\sffamily NetworkX}\xspace}

\usepackage{xparse}

\ExplSyntaxOn

\seq_new:N \l__nuclei_output_seq
\seq_new:N \l__nuclei_match_seq

\NewDocumentCommand{\nucleilist}{m}
 {
  \seq_clear:N \l__nuclei_output_seq

  \clist_map_inline:nn { #1 }
   {
    \__nuclei_parse:n { ##1 }
   }

  \seq_set_map:NNn \l_tmpa_seq \l__nuclei_output_seq { $##1$ }
  \seq_use:Nn \l_tmpa_seq { ,~ }
 }

\cs_new_protected:Nn \__nuclei_parse:n
 {
  \regex_extract_once:nnNTF
   { \A \s* ([A-Za-z]+) ([0-9]+) \s* \Z }
   { #1 }
   \l__nuclei_match_seq
   {
    \seq_put_right:Nx \l__nuclei_output_seq
     {
      \exp_not:N \isotm
       { \seq_item:Nn \l__nuclei_match_seq { 2 } }
       { \seq_item:Nn \l__nuclei_match_seq { 3 } }
     }
   }
   {
    \seq_put_right:Nn \l__nuclei_output_seq { #1 }
   }
 }

\ExplSyntaxOff

\journalinfo{The Open Journal of Astrophysics}

\begin{document}

\title{\pynucastro 3: A community library for nuclear astrophysics}

\author{Zhi Chen$^1$\orcidlink{0000-0002-2839-107X},
        Khanak Bhargava$^1$\orcidlink{0000-0003-0385-7918},
        Eric T. Johnson$^1$\orcidlink{0000-0003-3603-6868},
        Reth Lopes$^1$\orcidlink{0009-0008-1874-4389},
        Melissa Rasmussen$^1$\orcidlink{0000-0002-0297-0313},
        Alexander Smith Clark$^2$\orcidlink{0000-0001-5961-1680},
        Michael Zingale$^1$\orcidlink{0000-0001-8401-030X},
        Brendan Boyd$^1$
        \orcidlink{0000-0002-5419-9751},
        Kiran Eiden$^3$
        \orcidlink{0000-0001-6191-4285},
        Sam Glosser
        \orcidlink{0009-0000-6648-3381}, and
        Don E. Willcox$^4$
        \orcidlink{0000-0003-2300-5165}}

\affiliation{$^1$Department of Physics and Astronomy,
             Stony Brook University,
             Stony Brook, NY 11794-3800, USA}

\affiliation{$^2$Institute for Gravitation and the Cosmos,
             Penn State University,
             University Park, PA 16802-6300, USA}

\affiliation{$^3$Department of Astronomy,
             University of California,
             Berkeley, CA 94720, USA}

\affiliation{$^4$Institute for Advanced Computational Science,
             Stony Brook University,
             Stony Brook, NY 11794-3800, USA}
             
\begin{abstract}
We describe the latest release of \pynucastro---a community python library
for nuclear astrophysics.  The goal of the \pynucastro project is to
build the tools needed to interactively explore nuclear properties,
reaction rates, and networks, and to export these networks to a
variety of simulation codes.  Major changes in \pynucastro since
the last major release include new rate
approximations, a stellar equation of state, support for the StarLib
library and rate uncertainties, and new tools for exploring networks.
\end{abstract}


\maketitle

\section{Introduction}


\pynucastro is a python library designed to bridge the nuclear
experiment and astrophysics communities by providing an interface to
nuclear data, methods to interactively explore nuclear properties and reaction rates,
build nuclear reaction networks, and to export these reaction networks to simulation codes. 
We described earlier releases of pynucastro in
\citet{pynucastro,pynucastro2,pynucastro2.1}.  Since then, a large
number of new features have been added, including support for
thermodynamics, new rate approximations, StarLib rates and rate uncertainties \citep{starlib}, 
and new capabilities for analyzing networks.  We've also extended the
capability for \pynucastro\ to export networks to simulation codes,
building new generic interfaces in addition to enhancing the support
for \amrex-Astrophysics simulation codes \citep{amrex-astro}.   Below we describe the
motivation for the library and the new features in the
\pynucastro\ 3.0 release.

Modeling stars means modeling astrophysical reacting flows---capturing
the interplay of hydrodynamics, gravity, and thermonuclear reactions
(and more) in a simulation code, often parallelized across 100s--1000s
of nodes on a supercomputer, and likely running on GPUs.  Today,
multi-dimensional simulations of stellar convection (e.g.,
\citealt{meakin:2007,gilet:2013,couch:2015,muller:2017,horst:2021,herwig:2023,rizzuti:2024}) and explosions
(e.g., \citealt{townsley:2007,mendoza-temis:2015,sandoval:2021,pakmor:2024,janka:2025}) are
filling in crucial gaps in our understanding of stellar evolution
and nucleosynthesis
\citep{horizons}.  Capturing the energy release and composition changes
from nuclear burning requires evolving a system of ODEs describing
the effects of reactions (the reaction network) coupled to
hydrodynamics and other physics in the simulation code.  At its core,
for a binary reaction $i(j,k)l$, the reaction network solves
(see, e.g., \citealt{hixmeyer}):
\begin{align}
\label{eq:reaction_ydot}
\frac{dY_i}{dt} = &- \sum_{j,k} \rho Y_i Y_j \phi_{ij} \lambda_{i(j,k)l} \nonumber \\ &+ \sum_{j,k} \rho Y_l Y_k \phi_{kl} \lambda_{l(k,j)i}  \equiv f_i({\bf Y}) \enskip,
\end{align}
where $\rho$ is the baryon mass density,
$Y_i$ is the molar fraction of species $i$ ($Y_i = X_i / A_i$,
with $X_i$ the mass fraction and $A_i$ the atomic weight), and the
$\lambda$'s are the temperature-dependent portion of the reaction rate,
\begin{equation}
\lambda_{i(j,k)l} = N_A \langle \sigma v \rangle_{i(j,k)l}
\end{equation}
with $N_A$ Avogadro's number and $\langle \sigma v\rangle$ the average
of the cross section, $\sigma$, and velocity, $v$, over the distribution
of relative velocities between the nuclei.  Finally, $\phi_{ij}$ is the Coulomb screening enhancement factor between nuclei $i$ and $j$.
Eq.~\ref{eq:reaction_ydot} says that the amount of species $i$ decreases (first term on the right)
due to the reaction $i(j,k)l$ and increases (second term on the right) due to
the reverse reaction, $l(k,j)i$.
There can be additional stoichiometric factors if multiple nuclei are consumed
or produced in a reaction and normalizations if nuclei $i$ and $j$ or $k$ and $l$
are identical (see \citealt{clayton:1968,hixmeyer}).  The form for decay reactions
or three-body reactions differ slightly but follow the same ideas.

There is a similar $dY_i/dt$ equation for each of the nuclei in the reaction
network.
We define ${\bf f}({\bf Y})$ as the righthand side
of the system of Eq.~\ref{eq:reaction_ydot}, which can depend on the vector of molar fractions, ${\bf Y}$.
The full ODE system for the network is then
\begin{equation}
\label{eq:net_Y}
\frac{d{\bf Y}}{dt} = {\bf f}({\bf Y}) \enskip.
\end{equation}
The Jacobian of this system, ${\bf J}$, is defined as
\begin{equation}
    J_{ij} = \frac{\partial f_i}{\partial Y_j}
    \label{eq:net_jacobian}
\end{equation}
for row $i$ and column $j$, and is often needed for numerical
integration.
The values of the $\lambda$'s are
determined from nuclear experiment or theoretical nuclear structure calculations, and provided in various
compilations.  The values of the screening factors, $\phi_{ij}$ come from detailed calculations of plasmas.  One of the earliest goals of \pynucastro was to make it easy for an astrophysicist to work with these different rate compilations by providing a simple, uniform interface.

Often Eq.~\ref{eq:net_Y} is
accompanied by a temperature or energy equation, especially when
coupled to a hydrodynamics code \citep{muller:1986,Cabezón_2004,strang_rnaas}.  
The
specific energy generation rate (erg / gram / second) from reactions,
$\epsilon_\mathrm{nuc}$, comes from the change in composition and is computed as:
\begin{equation}
\epsilon_\mathrm{nuc} = -N_A \sum_i \frac{dY_i}{dt} m_i c^2 \enskip,
\end{equation}
where $N_A$ is Avogadro's number and $m_i$ is the mass of nuclei $i$,
and $c$ is the speed of light \citep{hixmeyer}.  Additional energy terms
are usually accounted for in the energy equation, including thermal
neutrino losses, $\epsilon_{\nu,\mathrm{therm}}$ and neutrino losses from weak reaction rates, $\epsilon_{\nu,\mathrm{weak}}$.  

While there are many open-source, community multi-dimensional
simulation codes that are applied to reacting-flow problems in stellar astrophysics,
including \athenapp~\citep{athenapp}, \castro~\citep{castro,castro_joss},
\flash~\citep{flash}, \maestroex~\citep{maestroex}, and 
\phoebus~\citep{phoebus}, for the most part, only a few standard
reaction networks are used.  A challenge in multi-dimensional
simulations is that we want to keep the reaction network as small as
possible, to reduce the memory requirements and solution cost, while
still accurately capturing the energy release and nucleosynthesis.
For this reason, approximate networks, like the 13-isotope
$\alpha$-chain {\tt aprox13} \citep{iso7}, are popular---these group
rates together by assuming some equilibrium process in the flow
through the network, thereby cutting down the number of nuclei needed
to be carried while getting the benefit of a larger network.  These
approximate networks are often hard-coded, and some trace their
lineage back to the 1970s (like the venerable {\tt aprox21} network of
\citealt{Kepler} for massive star evolution).  As a result of their
design, it can be difficult to update the rates or add nuclei to these
hard-coded networks.  These one-size-fits-all networks can miss some critical nuclear pathways for some problems---examples
include the rate sequence
$\isotm{C}{12}(p,\gamma)\isotm{N}{13}(\alpha,p)\isotm{O}{16}$
discussed for its importance in thermonuclear supernova (SN Ia)
ignition in \citet{shenbildsten} and stable nickel isotopes, seen in
SN Ia observations \citep{Kumar:2026}.  Both of these processes are
absent in the {\tt aprox}-family of networks.

There are a number of libraries that allow one to build an arbitrary
nuclear reaction network by specifying nuclei at runtime (sometimes called
``soft-wired''), including {\sffamily Skynet}~\citep{skynet},
{\sffamily XNet}~\citep{hixthielemann:1999}, {\sffamily WinNet}~\citep{winnet}, and
{\sffamily Torch}~\citep{torch}. 
Dynamic reaction network libraries such as 
{\sffamily NuGrid}~\citep{nugrid} and
{\sffamily GridFire}~\citep{gridfire} have also been developed that 
dynamically select all the relevant nuclei and 
reaction rates under different thermodynamic conditions
using graph topology and reaction-rate based criteria.
However, for the most part, these are
not designed for coupling to hydrodynamics in multi-dimensional
simulations, and do not have mechanisms to do rate approximations that
can greatly accelerate multi-dimensional simulations.  Instead, these
networks are most often used for one-zone self-heating burns (e.g.,
\citealt{schatz:2001}) or nucleosynthetic postprocessing of simulation
results (e.g., \citealt{townsley:2016,Harris:2017,sieverding:2023}).

\pynucastro\ serves a different purpose than the above network
codes---\pynucastro is designed to allow for the interactive
exploration of rates and networks, and the easy creation and export
(to python or \cxx) of efficient networks for specific science
applications.  It supports up-to-date nuclear rates and data, together
with the commonly-used rate approximations. This allows one to build
custom networks for a simulation code that fit the problem, rather
than hoping that a stock network, like {\tt aprox13}, will do the job.
These exported networks can be used with operator-splitting or
spectral deferred corrections coupling to hydrodynamics
\citep{castro-simple-sdc}.  Furthermore, \pynucastro provides the tools
to assess the importance of rates and nuclei for a particular problem,
allowing for network reduction {\it prior} to exporting the network for use
in a simulation code. In contrast to the the dynamic reduction schemes,
this approach eliminates the runtime overhead associated with
reduction algorithm along with a smaller memory usage.  Outputting a fixed
network based on the science case also allows for excellent GPU optimization
of the reaction network code, which is critical
for today's exascale computers.

\subsection{Design}

The overall design of \pynucastro has not changed since the 2.0
release \citep{pynucastro2}.  Here we briefly review the major classes and
summarize the state of the library, noting new features along the way.

\pynucastro follows standard python conventions and provides an
interface familiar to the astrophysics community.  It leverages \numpy
\citep{numpy,numpy2020}, \matplotlib \citep{matplotlib}, \scipy \citep{scipy},
\sympy \citep{sympy}, \numba \citep{numba}, and \networkx \citep{networkx}
to provide its wide range of capabilities.
By convention, \pynucastro is imported as

\begin{lstlisting}
import pynucastro as pyna
\end{lstlisting}

The core classes of
\pynucastro\ are \nucleus, for describing a single nucleus, \rate, for
describing a single reaction rate, \library, for describing a
collection of rates (perhaps from a single source, like ReacLib, \citealt{reaclib}), and
\ratecollection\ as the base type for a reaction network that
understands the links between rates and how to construct and evaluate
the $d{\bf Y}/dt$ terms.  All of the capabilities of \pynucastro\ are
available for use in an interactive Jupyter environment, and where
needed, \numba is used for JIT-acceleration.  Evaluating the energy output and compositional
changes from a reaction network requires more than just the raw
reaction rates---electron screening, thermal neutrino losses, and
nuclear properties including masses, spins and partition functions must also be provided.
\pynucastro\ provides all of these pieces.  Furthermore, when working
with photodisintegration (reverse) rates, it is advised by ReacLib to
recompute these using detailed balance---this is handled easily by
\pynucastro.

\begin{figure}
    \centering
    \plotone{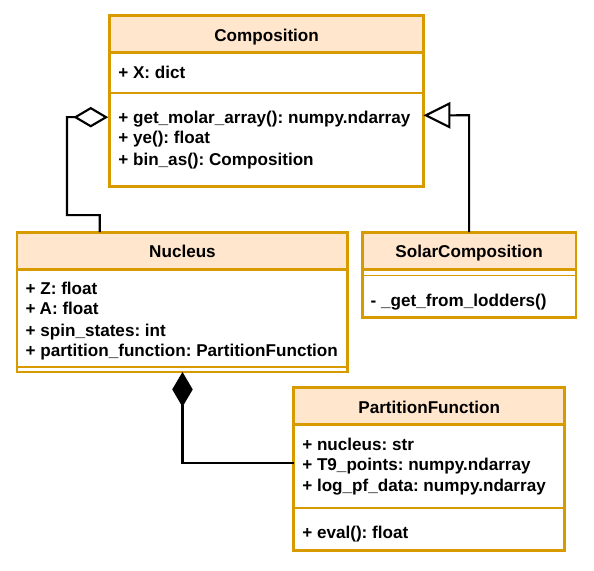}
    \caption{A UML class diagram showing relationships between different classes related to nuclear data. 
    Different arrowhead shapes represent different relationship types: hollow triangles for inheritance, 
    hollow diamond for aggregation, and filled diamond for composition.}
    \label{fig:uml-nuclear-data}
\end{figure}

Figure~\ref{fig:uml-nuclear-data} shows a Unified Modeling Language (UML) class diagram 
to illustrate the relationships among different classes 
used to represent nuclear data and composition.
The \nucleus class knows the properties of a single nucleus (number of protons
and neutrons, ground spin states, mass, half-life), relying on the experimental data
from the Nubase 2020 evaluation \citep{nubase:2020} and using the
CODATA physical constants \citep{codata:2018}, as obtained from \scipy
\citep{scipy}.  Additionally, temperature-dependent nuclear partition
functions from \citet{rauscher:1997,rauscher:2000,rauscher:2003} are
stored and interpolated as needed via the {\tt PartitionFunction} class.
The \composition class stores a collection of \nucleus objects and their associated
abundances. A new \solarcomposition class is introduced that provides the realistic
solar composition based on \citet{lodders2020}.
See Section \ref{sec:lodders} for more details.

Several functions, such as those used to evaluate the
network righthand side, $d\mathbf{Y}/dt$, and Jacobian, take $(\rho, T, \mathrm{comp})$ as input,
where $\mathrm{comp}$ is the \composition.
It is convenient to encapsulate these quantities in a single data class,
and for this purpose, \pynucastro introduced the {\tt ThermoState} class.
To avoid accidentally swapping the order of $\rho$ and $T$,
{\tt ThermoState} is implemented as a keyword-only data class.
Many \pynucastro functions that previously required
$(\rho, T, \mathrm{comp})$ have been updated to accept a {\tt ThermoState}
object instead. For backward compatibility, the previous $(\rho, T, \mathrm{comp})$
interface is still supported during the transition phase, 
but now generates a deprecation warning.
The following shows how to construct a {\tt ThermoState} object:
\begin{lstlisting}
nuclei = ["he4", "c12", "o16"]
comp = pyna.Composition(nuclei, init="solar")
state = pyna.ThermoState(rho=1e2, T=2e8,
                         comp=comp)
\end{lstlisting}

\begin{figure*}
    \centering
    \plotone{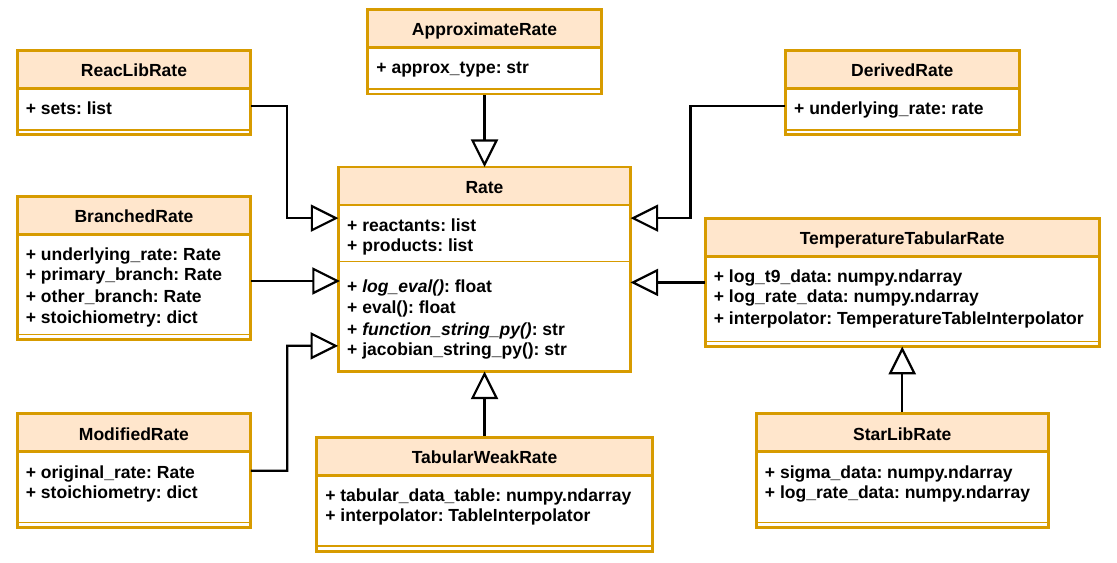}
    \caption{A UML class diagram showing subclasses inheriting from the base abstract \rate class. 
    Italicized function names represent abstract methods that must be implemented by the child class.}
    \label{fig:uml-rate}
\end{figure*}

Figure~\ref{fig:uml-rate} shows all different types of rates
available in \pynucastro, all of which subclass the abstract \rate class.
The \rate class provides the basic structure and metadata for a nuclear reaction rate,
while each subclass has its own implementation for evaluating
the temperature dependent $N_A \langle \sigma v\rangle$
portion of the rate (for strong-mediated reactions) or $\lambda$ directly from a density-temperature tabulation of weak rates.  Each subclass also has methods to output python or \cxx code to compute the rate.
Here we briefly discuss each individual rate subclasses: 
\begin{enumerate}
    \item \reaclibrate: a rate type that understands the ReacLib format
    \citep{reaclib} that evaluates $N_A\langle \sigma v\rangle$ using a
    7-parameter exponential with various fractional temperature powers, following
    \begin{equation}
        \label{eq:reaclib}
        N_A\langle \sigma v\rangle = \exp{\left( a_0 + \sum_{i=1}^{5} a_i T_9^{(2i-5)/3} + a_6 \log{T_9}\right)} \enskip,
    \end{equation}
    with $T_9 = T/(10^9 \ \mathrm{K})$.

    \item \tabularrate: a rate type that stores the two-dimensional
      table ($\rho Y_e$ and $T$) of values of electron-capture and
      $\beta$-decay rates, as well as energy loss rate due to neutrino emissions.
      Note since $\beta^-$ decay and $e^+$ capture, as well as
      $\beta^+$ decay and $e^-$ capture have the same parent/daughter nuclei,
      these rates are combined into a single 
      electron-capture/$\beta$-decay rate.
      It evaluates the rate using a bilinear interpolation scheme on the
      tabulated data. This was formerly called {\tt TabularRate}, but
      the changed to \tabularrate for clarity.  See Section~\ref{sec:weaktab}
      for an updated coverage of \tabularrate from various data
      sources.

    \item \temperaturetabularrate: a rate type that stores a table of 
    temperature and its corresponding value for $N_A\langle \sigma v\rangle$,
    and uses monotone cubic Hermite interpolation to evaluate the rate.
    See Section~\ref{sec:alternate-rates} for an example usage.

    \item \starlibrate: a rate type that inherits from \temperaturetabularrate
    whose temperature dependency and rate uncertainties are tabulated.
    \starlibrate assumes that rate uncertainties follow lognormal distribution
    and is intended to work with the data from StarLib Library \citep{starlib}. 
    See Section~\ref{sec:starlib} for more implementation details.

    \item \approximaterate: a rate type that combines rate sequences into
    a single effective rate by assuming equilibration of the intermediate nucleus.
    This class was first introduced in \citet{pynucastro2}, where the initial
    implementation supported the $(\alpha, p)(p,\gamma)$ approximation,
    combining the sequences $A(\alpha,\gamma)B$ and
    $A(\alpha,p)X(p,\gamma)B$ into a single effective rate for
    $A(\alpha,\gamma)B$ by assuming equilibrium through nucleus $X$
    (i.e., $dY_X/dt = 0$).  \approximaterate allows for the removal of $X$ from the network
    while keeping the enhanced flow that the $(\alpha, p)(p,\gamma)$ channel provides,
    and is a key approximation used in networks like {\tt aprox21}.
    Since the initial implementation, \pynucastro has introduced additional types of approximations
    that appear in {\tt aprox21}, including the
    double neutron-capture approximation (see Section~\ref{sec:double-n-capture})
    and approximations for carbon and oxygen burning (see Section~\ref{sec:co-burning}).

    \item \modifiedrate: a rate type that modifies reactants, products, as well
    as the stoichiometry of a given rate object, but still uses the underlying rate object
    to evaluate the $N_A\langle \sigma v\rangle$ term. This is a new class that 
    extends the {\tt modify\_products()} introduced in \citet{pynucastro2}.
    See Section~\ref{sec:modified_rates} for more details.

    \item \branchedrate: a rate type that computes a branching ratio
      for the endpoint of a reaction sequence.  Like \modifiedrate, it
      can also change the stoichiometry.  This is useful for
      approximating complex rate sequences like those that appear in
      the CNO cycle.  This is a new class and is described in Section
      \ref{sec:branched_rate}.

    \item \derivedrate: a rate type that takes any rate object representing a
    strong nuclear reaction and computes its inverse rate using detailed balance,
    utilizing the ground-state spin and partition functions of the nuclei involved.
    Given either the forward or reverse reaction, it constructs the corresponding reverse or forward rate, respectively. 
    See Appendix~\ref{appendix:nse} for details on the updated implementation.
\end{enumerate}

Electron-screening can greatly increase reaction rates, sometimes by
factors of 1000 or more \citep{woosley:2004}.  \pynucastro provides
screening formulations from several sources, including the collection
of \citet{graboske:1973,alastuey:1978,itoh:1979}, widely used in {\tt
  aprox}-networks (where it is usually called {\tt screen5}), and more
recent formulations from
\citet{chabrier_potekhin:1998,chugunov:2007,chugunov:2009}.  Any of
these screening formulations can be applied to a rate or throughout an
entire network.  This variety of methods allows for direct
comparison of screening implementations within a python environment.
Exported \cxx / Fortran networks currently use the implementation of \citet{chugunov:2007}.

\begin{figure}
    \centering
    \plotone{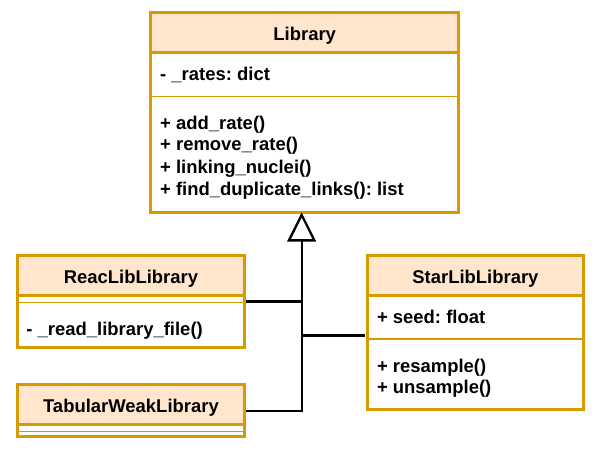}
    \caption{A UML class diagram showing data-source-specific \library subclasses
    of the base \library class, each with a unique constructor that reads
    rate data files and initializes the \library object.}
    \label{fig:uml-library}
\end{figure}

The \library class serves as a container for a collection of \rate objects.
In addition to storing reaction rates, it provides methods for
filtering rates based on certain properties, finding duplicate links,
and a basic validation method to check if any rates might be missing from a collection.
There are several \library subclasses with initialization methods
to read in different rate data source files and construct 
the \rate objects from it, including \reacliblibrary, \tabularlibrary, and \starliblibrary. 
A visual relationship among these classes is shown in Figure~\ref{fig:uml-library}.
To build a network, one typically starts by gathering all the desired reaction rates
from one or more data source specific \library objects and combine them into a \library object (\library objects can be added using the $+$ operator).
This \library object can then be used to create a network by
constructing a \ratecollection or one of its subclasses.


\begin{figure*}
    \centering
    \plotone{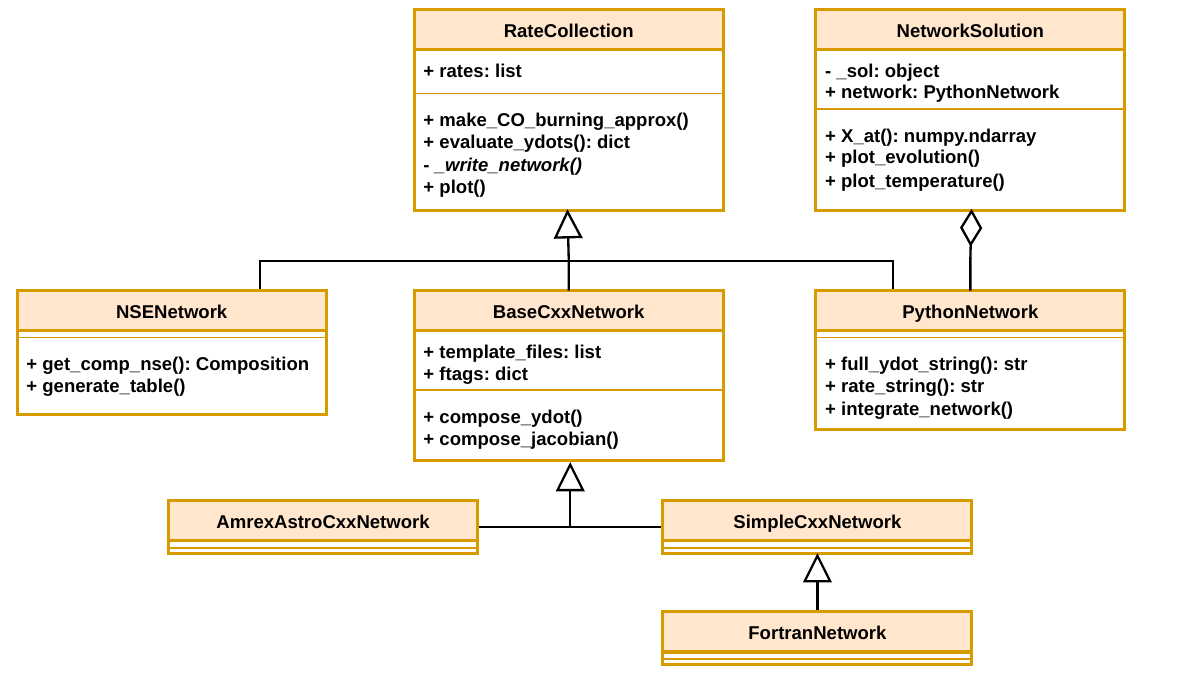}
    \caption{A UML class diagram showing subclasses inheriting from the base \ratecollection class.
    Different arrowhead shapes represent different relationship types: hollow triangles for inheritance 
    and hollow diamond for aggregation.
    Italicized function names represent abstract methods that must be implemented by the child class.}
    \label{fig:uml-rate-collection}
\end{figure*}

A \ratecollection is the base network type in \pynucastro that allows 
for evaluating the $dY_i/dt$ terms from all the rates, plotting the 
flow through a network (as a function of density, temperature, and composition),
and more. Unlike a \library, a \ratecollection is not allowed to have multiple rates (i.e.\ from different sources) for the same nuclear process.  Figure~\ref{fig:uml-rate-collection} shows a series of 
subclasses that inherits from \ratecollection.
A \pythonnetwork extends \ratecollection by
providing the ability to write a python module to disk that includes
all of the information needed to integrate a reaction network with
\scipy's {\tt solve\_ivp} set of integrators.
A \networksolution class was recently introduced to provide
a convenient interface for analyzing solutions produced by {\tt solve\_ivp}
when evolving a \pythonnetwork. It includes commonly used analysis routines,
such as plotting the evolution of abundances and temperature.
For \amrex-based codes,
the \amrexastrocxxnetwork is written to take advantage of GPU-offloading
and work with a large variety of integrators and coupling methods provided
by the \amrex-Astro \microphysics library \citep{Microphysics,microphysics-joss}.
An \nsenetwork\ can be used to solve for the nuclear
statistical equilibrium (NSE) abundances of a set of nuclei, and with just a few lines
of code, create a tabulation in terms of temperature, density, and
electron fraction, $Y_e$, including the time-dependent evolution
$dY_e/dt$ due to electron-/positron-captures and decays, following the
same ideas as \citet{seitenzahl:2009}. More details on
generating the NSE table can be found in Section~\ref{sec:nse-table}.
An NSE table created this way
was used for the core evolution in massive stars in \citet{Zingale:2024b}.
Finally, since the last major release, two additional network types have been
added: \simplecxxnetwork is a basic \cxx network without the dependency
on the \amrex library and \fortrannetwork is a set of Fortran wrappers
to \simplecxxnetwork.  These are described in Section \ref{sec:new_network_types}.

\subsection{Community Development}


\pynucastro follows a community development model: the project is
hosted on
github\footnote{\url{https://github.com/pynucastro/pynucastro}} and
new changes are done via pull requests, requiring a review from a
project member and passing a comprehensive test suite (described in
more detail in Section \ref{sec:testing}).  Since version 2.0.0
(described in \citealt{pynucastro2}), almost 800 pull requests have
been merged, from 14 different developers.  Feature requests, posted
via github issues, have led to the development of a simple \cxx
backend (without the GPU machinery of \amrexastrocxxnetwork) and
Fortran wrappers, to enable others to use \pynucastro networks in
their simulation codes.

\subsection{Applications}

Networks produced with \pynucastro\ have been used for a variety of
multidimensional simulations of stellar explosions.  Applications
using \castro\ \citep{castro} include flame propagation in X-ray
bursts \citep{Chen:2023,johnson:2024}, double-detonation models of
Type Ia supernovae \citep{Zingale:2024}, convection in massive stars
\citep{Zingale:2024b}, convection in classical novae
\citep{Smith2025}, and direct numerical simulations of oxygen flames
\citep{zhang:2025b}.  Using \maestroex \citep{maestroex}, applications
include convection in X-ray bursts \citep{Guichandut:2024} and
convective Urca in carbon/oxygen white dwarfs \citep{Boyd:2025}.
\pynucastro has also been used to find the nuclear statistical
equilibrium state for a study on neutron-star common-envelope systems
\citep{esteban:2025}.

The \simplecxxnetwork and \fortrannetwork classes were developed
based on community requests.  \simplecxxnetwork was used in
\citet{hasenour:2025}, demonstrating the ability to connect
\pynucastro-generated networks to multidimensional simulation codes that are independent of
the \amrex library.

As a python library, \pynucastro makes it easy to couple machine learning tools to nuclear data, reaction rates, or entire reaction networks. The exploratory study of \citet{Fan:2022} demonstrated how recurrent neural networks could be trained as surrogate models for carbon fusion in subsonic flame simulations. Later, \citet{grichener:2025} developed surrogate neural network models at a variety of time steps for larger reaction networks suitable for simulations of core collapse supernovae. Simultaneously, \citet{zhang:2025} trained neural networks as surrogate models for integration of \pynucastro reaction networks, introducing techniques for sampling the large parameter space of reaction network initial conditions. We expect training models on reaction
networks will continue to be a growing application in the future that \pynucastro can support.

\subsection{Code Examples and Supplemental Notebooks}

In the next section, we describe the new features added since \pynucastro 2.0.
We show code only in cases where we want to illustrate the API or demonstrate
a workflow.  For an example where we wish to show both input and output, the
standard python interpreted prompt, {\tt >}{\tt >}{\tt >}, is shown for the input.
Otherwise, we provide a Jupyter notebook to reproduce each figure in our supplemental
material \citep{pynucastro3_zenodo}.

\section{New features}


In this section, we describe the major new features since \pynucastro 2.0.

\subsection{New Helper Interfaces}
\label{sec:helper}
As typical workflows with \pynucastro have become clearer, new
interfaces have been implemented to automate some common tasks.
Perhaps the most significant is a general interface for creating a
network from a list of nuclei, {\tt network\_helper}.  For many
applications, this is the only function that is needed to create a
network.  It will (1) find all of the reactions linking the nuclei
from either ReacLib or StarLib (2) find all of the weak reactions from
tabular sources linking the same nuclei, (3) eliminate any duplicate
links (by default preferring the tabular weak rate sources over
ReacLib/StarLib), and (4) rederive the reverse rates using detailed
balance (as described in \citealt{pynucastro2}).  The network is then
returned as a \pythonnetwork, \simplecxxnetwork, or
\amrexastrocxxnetwork.  Rate approximations can then be done on the
network if desired, or it can be exported or integrated in an
interactive session.

\begin{figure}[t]
\centering
\plotone{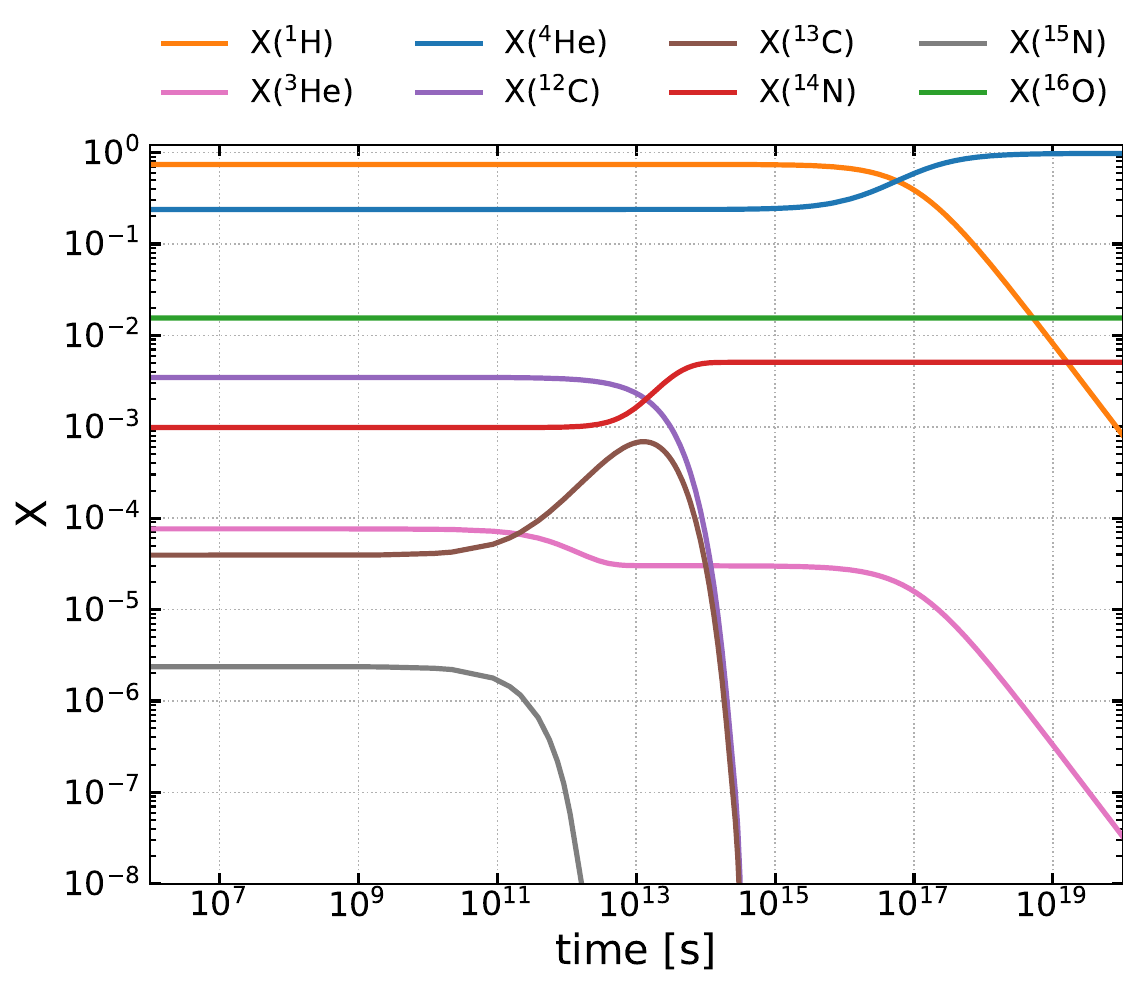}
\caption{\label{fig:integration} An example of integrating
a basic H-burning network.  The initial composition is solar
and we see that after about $10^{17}$~s, the hydrogen
mass fraction drops and the helium mass fraction rises.
The $\isotm{O}{16}$ is not consumed, and we see that the CNO elements
pile up in $\isotm{N}{14}$, as it represents the slowest link in the
CNO cycle.  \supnote{integration-example.ipynb}}
\end{figure}

New methods were also added to make integrating a network easy in a
Jupyter environment.  From a \pythonnetwork, {\tt integrate\_network}
will build a python module containing everything needed to solve the
system of ordinary differential equations (ODEs) for the network
(Eq.~\ref{eq:net_Y}), including the righthand side function, ${\bf f}({\bf Y})$, and
Jacobian.  This will then be loaded into memory, and the \scipy\ {\tt
  solve\_ivp} method will be used to evolve the system.  Finally, the
output can be visualized directly from the solution object returned
using {\tt plot\_evolution}.  An example of this workflow,
integrating a basic pp and CNO network, is shown below:
\begin{lstlisting}
nuc_list = ["h1", "h2", "he3", "he4", "c12",
            "c13", "n13", "n14", "n15", "o16"]
net = pyna.network_helper(nuc_list)

rho = 150
T = 1.5e7
comp = pyna.Composition(net.unique_nuclei,
                        init="solar")

tmax = 1.e20
sol = net.integrate_network(tmax, rho, T,
                            comp, atol=1.e-8)

fig = sol.plot_evolution(ymin=1.e-8, tmin=1.e6
      legend_outside=True, legend_framon=False,
                         ncol=4)
\end{lstlisting}
This creates a network connecting all the input nuclei,
defines a thermodynamic state, with the composition set to
solar, and then integrates for $10^{20}~\mathrm{s}$.  Notice
that when we integrate, we can specify the tolerance for the adaptive
stepping in the integrator.  Here, we set an absolute tolerance
of $10^{-8}$ (a relative tolerance can also be specified).
The result is shown in Figure~\ref{fig:integration}.
In Section \ref{sec:self-heating-burn}, we show how to include temperature evolution
using the equation of state.

\nucleus objects can now be added and subtracted, allowing for easy
interactive understanding of different processes.  For example,
we can see the outcome of adding an $\alpha$ to $\isotm{C}{12}$:
\begin{lstlisting}
>>> from pynucastro import Nucleus
>>> c12 = Nucleus("c12")
>>> a = Nucleus("a")
>>> c12 + a
O16
\end{lstlisting}
Several classes have gained {\tt .summary()} methods that
give a high-level overview of the current state of an
object.  For example, the \nucleus\ {\tt c12} created
above shows:
\begin{lstlisting}
>>> c12.summary()
C12 / carbon-12
---------------
  A: 12
  N: 6
  Z: 6

  mass: 11177.92923 MeV
  mass excess: 0.00000 MeV
  binding energy / nucleon: 7.68014 MeV

  half-life: stable

  partition function: not available
  spin states: 1

  dummy: False
  nse: False
  spin states are reliable: True
\end{lstlisting}

A key design feature of \pynucastro is the ability to
mix reaction rates from different sources into a single
network.  To facilitate this, \pynucastro now provides
the {\tt full\_library} method, which returns all of the
rates it knows about.  As an example, here we look
for electron-capture rates on ${}^{56}\mathrm{Ni}$:
%
%
\lstset{
  escapeinside={(*@}{@*)}
}
\begin{lstlisting}
>>> fl = pyna.full_library()
>>> rates = fl.get_rate_by_name("ni56(,)co56")
>>> for r in rates:
...    print(r, type(r).__name__,
...          r.source["Label"])
...
Ni56 (*@$\rightarrow$@*) Co56 + e(*@$^{+}$@*) + (*@$\nu$@*) ReacLibRate wc12
Ni56 (*@$\rightarrow$@*) Co56 + e(*@$^{+}$@*) + (*@$\nu$@*) StarLibRate au03
Ni56 + e(*@$^{-}$@*) (*@$\rightarrow$@*) Co56 + (*@$\nu$@*) TabularWeakRate langanke
Ni56 + e(*@$^{-}$@*) (*@$\rightarrow$@*) Co56 + (*@$\nu$@*) TabularWeakRate ffn
\end{lstlisting}
Here, {\tt rates} is a list that contains all of the reaction rates
that match our name.  We then loop over the rates and see that four
matches were found.  The first is from ReacLib, the second from
StarLib, and the last two are tabulated weak rates, from
\citet{langanke:2000} and \citet{ffn}.  For a reaction network, we can
only use one of these, since each physical process in a network can
have only a single implementation.  In this case, we expect the
tabulated weak rates to be more accurate, since those are functions of
density and temperature, instead of just temperature.  When we use
{\tt network\_helper}, any duplicate rates will be resolved by picking
the tabular weak rate by default, and the ordering of which source for
the tabular rates can also be specified.  Further, for these two
tabular rates, the electron-capture and positron-emission processes
are combined into a single effective rate in \pynucastro
\citep{pynucastro2.1}.

\subsection{Solar Composition}
\label{sec:lodders}
\pynucastro now provides a class, \solarcomposition, for
initializing compositions from the present-day solar isotopic
abundances reported by \citet{lodders2020}. It constructs a {\tt
  Composition} object from the isotope mass fractions, normalizes them, and gives a default solar
mixture with $X_\mathrm{H} \approx 0.746$, $X_\mathrm{He} \approx 0.239$,
and $Z \approx 0.0150$.

\begin{figure}
    \centering
    \plotone{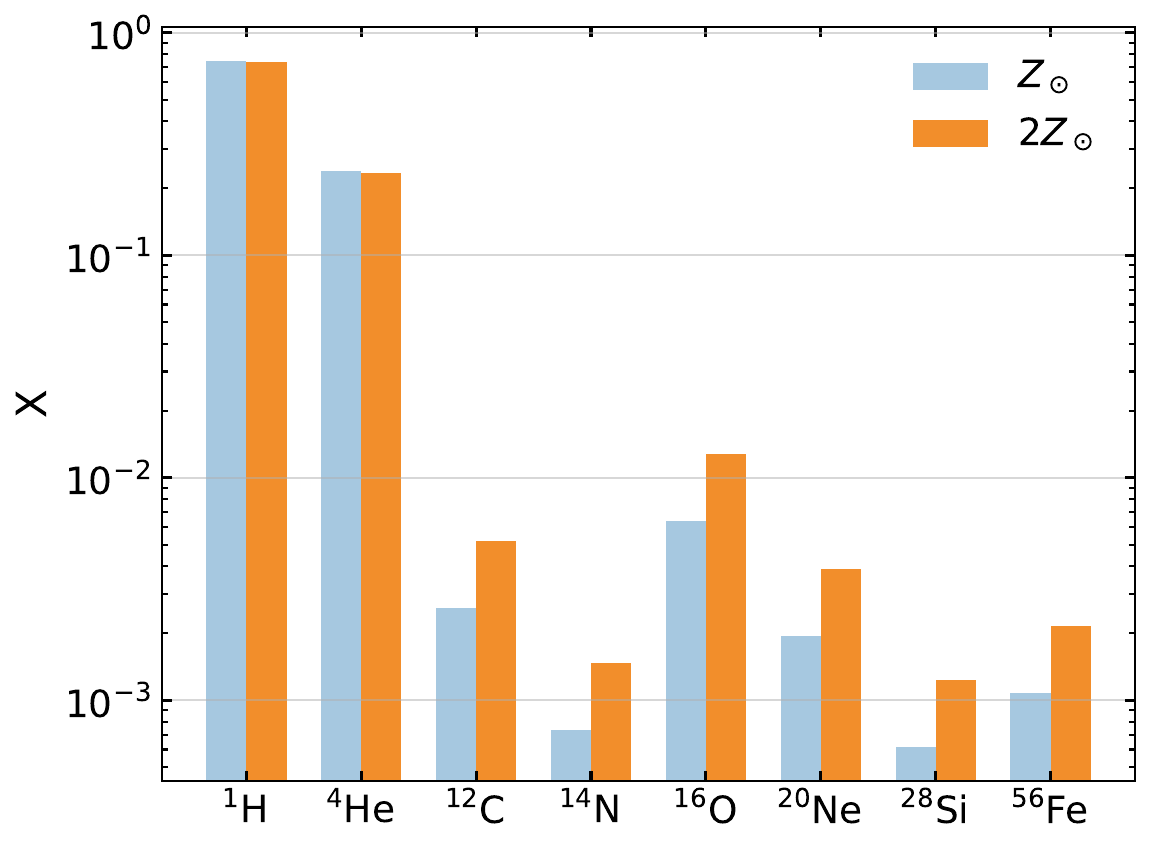}
    \caption{Comparison of the default solar composition with
      a composition scaled to twice the solar metallicity. 
      It shows the mass fractions of $\isotm{H}{1}$, $\isotm{He}{4}$, and
      the six most abundant metal isotopes in the solar mixture. 
      \supnote{solar\_composition.ipynb}}
    \label{fig:solar-comp}
\end{figure}

This class also supports rescaling the mixture to a
user-specified metallicity. When a target $Z$ is supplied, it scales
all metal isotopes together while scaling the combined $\mathrm{H}$
and $\mathrm{He}$ contribution so that the mass fractions continue to
sum to unity. This makes it possible to initialize networks with solar
relative abundances at a different total metallicity.
Figure~\ref{fig:solar-comp} illustrates this scaling behavior.

Since most reaction networks do not carry every isotope measured in \citet{lodders2020},
the resulting composition can be mapped onto a reduced set of
network nuclei using {\tt Composition.bin\_as()}.
This is done by mapping the original set of nuclei onto the
target nuclei with the closest mass and atomic number, $(A,Z)$.
The mapping prioritizes the mass number, followed by the atomic number, where both are chosen to be less or equal to those
of the original nuclei. This preserves the
total mass fraction while assigning the isotopic abundances to the
nuclei present in the chosen network.

\subsection{Importing Data}

\begin{figure*}[t]
\centering
\includegraphics[width=\linewidth]{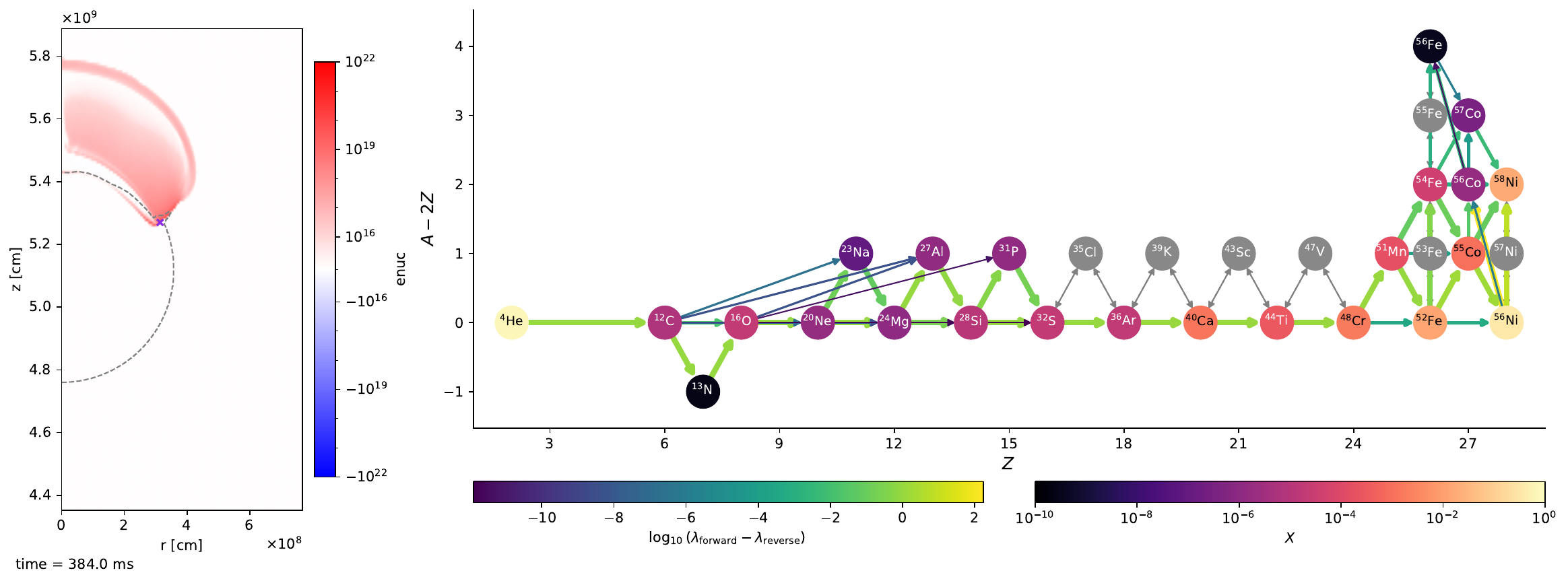}
\caption{\label{fig:yt-connect} An example of ``roundtripping'' a
  simulation.  A network was created in \pynucastro, exported via
  \amrexastrocxxnetwork to \cxx, used for a \castro simulation of a
  double detonation thermonuclear supernova (following the methodology
  of \citealt{Zingale:2024}), and then the plotfile was read in via a
  \yt-bridge to \pynucastro to visualize the flow through the network.
  In the slice plot on the left, the energy generation rate is
  visualized at a point where the surface helium detonation is about a
  third of the way around the surface of the white dwarf.  The gray
  contour indicates a density of $10^6~\gcc$, roughly indicating the
  base of the helium layer, and the transition to the underlying C/O
  white dwarf.  Finally, the purple-$\times$ indicates the location where
  the thermodynamic data was read for the network visualization.  This
  corresponds to the location of peak nuclear energy generation rate.}
\end{figure*}

One of the core design features of \pynucastro is the ability to
export networks in a form (usually \cxx) that an astrophysical
simulation code can use.  We've now added the ability for \pynucastro
to import data from simulations, to allow for the evaluation and
post-processing of nuclear reaction rates and more.  This capability
allows for ``roundtripping'' of the nucleosynthesis---we use
\pynucastro to generate a network for a simulation code, run the
simulation (with the \pynucastro network), and then read the
simulation output into python (leveraging the \yt package, \citealt{yt}).
This can be useful to explore which reaction rates are most important
in different stages of a simulation, and with \pynucastro's plotting
capabilities, we can visualize the nuclear flow through a reaction
network.

Figure~\ref{fig:yt-connect} shows an example---this is a helium
detonation propagating across the surface of a white dwarf, as a model
of a double-detonation Type Ia supernovae.  This simulation was run
with \castro, using the methodology from \citet{Zingale:2024}.  The
left panel of the figure shows the energy generation rate, visualized
with \yt, and the network is shown on the right, visualized with
\pynucastro.  The purple $\times$ in the energy generation plot indicates
the zone where we are reading in the simulation data for the network
visualization.  In the network visualization, we color the nodes by
composition and the edges by reaction rate.  Only the net rate is
shown (forward minus reverse) for each process, with the arrow head
indicating the net flow.  This example also highlights new plotting features: coloring the nodes by composition and
showing the net flow. Using
\yt to read the simulation data opens this capability up to a wide
variety of simulation codes.


\subsection{Thermodynamics}

Recent versions of \pynucastro\ have seen added more thermodynamic support, through
the inclusion of an equation of state (EOS).  Among the capabilities this enables is more advanced time-integration
within a Jupyter environment, by including an energy equation.  Here we describe these components.

\subsubsection{Stellar Equation of State}

The core stellar EOS consists of an ideal gas description of ions,
radiation (assuming local thermodynamic equilibrium), and an
electron-positron gas. The total thermodynamic quantities are obtained
by summing the contributions from each component. For example, the total pressure
is then given by
\begin{equation}
    P = P_{\mathrm{ion}} + P_{\mathrm{rad}} + P_{\mathrm{e^-}} + P_{\mathrm{e^+}} \enskip.
\end{equation}
The ion (ideal gas) and black body radiation pressures are given by
\begin{equation}
    P_{\mathrm{ion}} = \frac{\rho k_B T}{m_u \mu_I} \enskip , \quad P_{\mathrm{rad}} = \frac{aT^4}{3} \enskip,
\end{equation}
where $k_B$ is the Boltzmann constant, $m_u$ is the atomic mass unit,
$a$ is the radiation constant,
and $\mu_I$ is the mean molecular weight of the ions defined as
\begin{equation}
\dfrac{1}{\mu_I} = \sum_i Y_i \enskip.
\end{equation}

Electrons and positrons in a stellar plasma
can behave classically or as a degenerate Fermi gas, and their
energies can require a relativistic treatment.  
To compute their contribution to the EOS, we work in terms of
the Fermi-Dirac distribution function,
\begin{equation}
n(p) = \frac{2}{h^3} \frac{1}{e^{(\mathcal{E}(p)/k_BT - \eta)} + 1} \enskip,
\end{equation}
where $p$ is the momentum of a particle, $\eta$ is the degeneracy
parameter, defined in terms of the chemical potential, $\mu$,  electron
rest mass, $m_e c^2$, and temperature as
\begin{equation}
\eta = \frac{\mu - m_e c^2}{k_BT} \enskip,
\end{equation}
where $c$ is the speed of light,
and $\mathcal{E}(p)$ is the kinetic energy of a particle.  The
quantity $n(p) d^3x d^3p$ represents the number of particles with
momentum $p$ to $p+dp$ in a volume $d^3 x$.  The factor of two allows
for both a spin-up and spin-down Fermion to exist in the same phase
space location.  With this definition, the number density of electrons
is then just the integral over all momentum,
\begin{equation}
\label{eq:ne}
n_e = \int n(p) d^3 p = \frac{8\pi}{h^3} \int_0^\infty \frac{p^2}{e^{(\mathcal{E}(p)/k_BT - \eta)} + 1} dp \enskip,
\end{equation}
where we switch the integral over $d^3p$ to spherical coordinates (in momentum space),
$d^3p \rightarrow 4\pi p^2 dp$.

At high temperatures and low densities, electron-positron pairs become
important, and the distribution function for positrons has an energy
of $2m_e c^2$ more than the electrons, giving
\begin{equation}
\label{eq:nplus}
n_+(p) = \frac{8\pi}{h^3} \int_0^\infty \frac{p^2}{e^{(\mathcal{E}(p)/k_BT + 2m_e c^2/k_BT - \eta_+)} + 1} dp \enskip,
\end{equation}
where the degeneracy parameters are equal in magnitude and opposite in sign due to chemical equilibrium ($\eta_+ = -\eta$) \citep{CoxGiuli}.

If we express the kinetic energy of a particle as
\begin{equation}
\mathcal{E}(p) = m_e c^2 \left [ \left (1 + \left(\frac{p}{m_e c}\right )^2 \right )^{1/2} - 1 \right ]
\end{equation}
and define a dimensionless energy,
\begin{equation}
x = \frac{\mathcal{E}}{k_BT} \enskip ,
\end{equation}
then, with some algebra, we can write the number density of electrons in terms of $x$ as:
\begin{align}
n_e = \frac{8 \sqrt{2} \pi}{h^3} m_e^3 c^3 \beta^{3/2}
   \Biggl [ &\underbrace{\int_0^\infty \frac{x^{1/2} \left(1 + \frac{1}{2}\beta x \right )^{1/2}}{e^{x-\eta} + 1} dx}_{F_{1/2}(\eta, \beta)} + \nonumber \\
     \beta &\underbrace{\int_0^\infty \frac{x^{3/2} \left(1 + \frac{1}{2} \beta x \right )^{1/2}}{e^{x-\eta} + 1} dx}_{F_{3/2}(\eta, \beta)}
   \Biggr ] \enskip ,\label{eq:fermi}
\end{align}
where we define the Fermi-Dirac integral as
\begin{equation}
          F_k(\eta, \beta) = \int_0^\infty
           \frac{x^k \left (1 + \frac{1}{2}\beta x \right )^{1/2}}{e^{x-\eta} + 1} dx \enskip . \label{eq:fermi_integral}
\end{equation}
Here, the degree of relativity is represented by
$\beta$, defined as
\begin{equation}
\beta = \frac{k_BT}{m_e c^2} \enskip .
\end{equation}
For the case of positrons,
we can evaluate the Fermi integrals with an effective degeneracy parameter
$\tilde{\eta} = -\eta - 2 /\beta$, which makes Eq.~\ref{eq:nplus} look like
Eq.~\ref{eq:ne}, giving:
\begin{equation}
n_+ = \frac{8 \sqrt{2} \pi}{h^3} m_e^3 c^3 \beta^{3/2} \left [ F_{1/2}(\tilde{\eta},\beta) + \beta F_{3/2}(\tilde{\eta},\beta) \right ] \enskip .
\end{equation}
Similar expressions exist for momentum and internal energy for both electrons and positrons \citep{CoxGiuli,timmesarnett}.  

To support the electron-positron EOS, the \fermiintegral class
supports the integration of Fermi-Dirac integrals in the form of
Eq.~\ref{eq:fermi_integral} using the quadrature method of \citet{Gong:2001}, which
follows the ideas of \citet{Aparicio:1998} and combines Legendre and Laguerre methods in different regions of the momentum space.  The quadrature nodes and weights are computed via \sympy for 200 points.

The \electroneos class manages computing the Fermi integrals and assembling
the thermodynamic quantities for electrons and positrons.  It starts by first finding $\eta$ by root-finding on the constraint: 
\begin{equation}
   n_e - n_+ = \rho N_A Y_e \enskip ,
\end{equation}
where $Y_e$ is the electron fraction of the nuclei in the plasma.
First derivatives
with respect to $\rho$ and $T$ are also provided, using the derivatives of the Fermi integrals with respect to $\eta$ and $\beta$ given in \citet{Gong:2001}.  A current limitation 
of this class is that  \citet{Gong:2001} recommends 128-bit
precision, but \numpy does not robustly support for higher than 64-bit
precision.  This is something that will be addressed in 
future releases by leveraging the {\tt numpy\_quaddtype} package \citep{numpy-quaddtype}.  This is predominantly an issue at low temperatures.

In the stellar plasma, we need to account for the combined effects of ions, electrons and positrons, and radiation.
The \stellareos computes the thermodynamic state with
these three components, including the derivatives of number density, pressure,
and internal energy with respect to density and temperature.

\begin{figure}[t]
\centering
\plotone{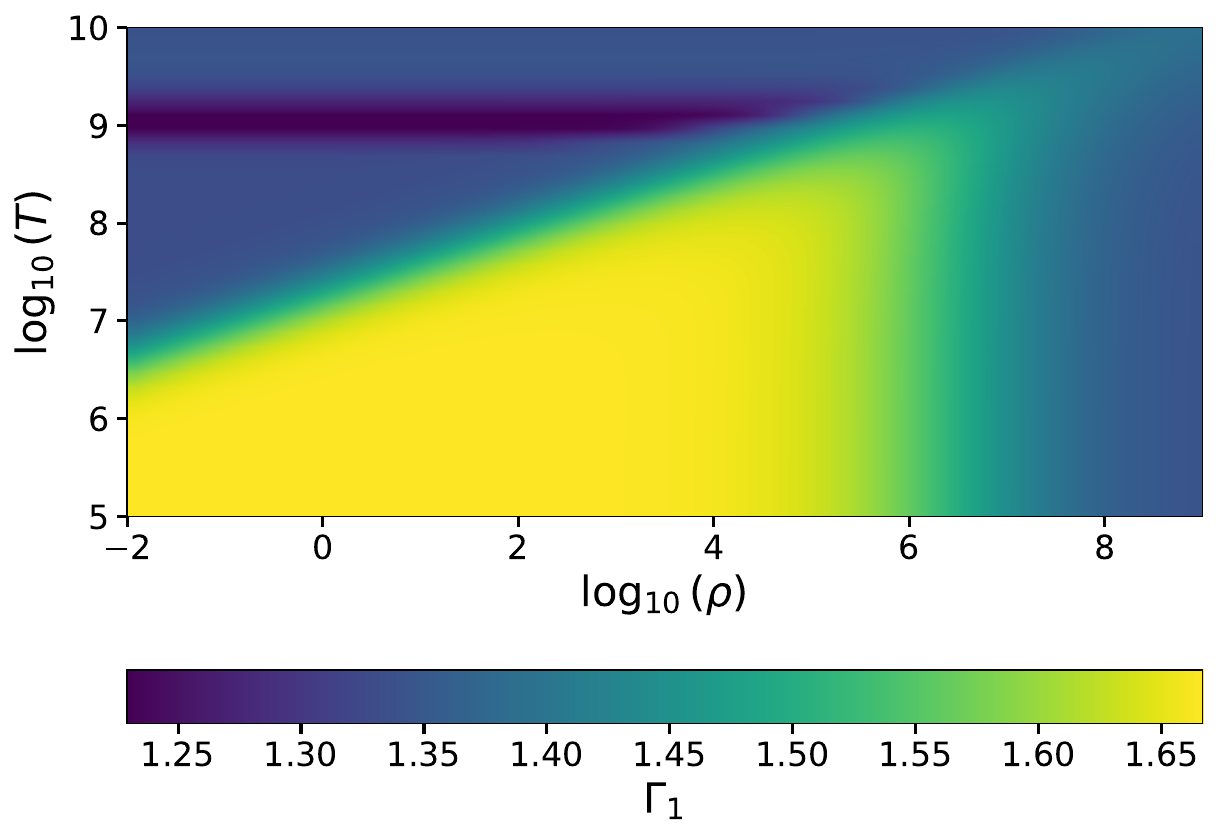}
\caption{\label{fig:eos} The adiabatic index $\Gamma_1$ in the $T$-$\rho$ plane\, computed with \stellareos.
\supnote{stellar-eos.ipynb}}
\end{figure}

Figure~\ref{fig:eos} shows a plot of the adiabatic index 
$\Gamma_1 \equiv d\log{P}/d\log{\rho} |_s$,
where $s$ is specific
entropy, constructed with \stellareos.  At low temperatures, we see the transition from
non-relativistic to relativistic degeneracy as $\Gamma_1$ shifts from
5/3 to 4/3.  At higher temperatures, we transition from the ideal gas
value of 5/3 to the radiation dominated value of 4/3.  Finally, at a
narrow range of temperatures around $10^9~\mathrm{K}$, we see the
effect of electron-position pairs, which drops $\Gamma_1$ below
4/3---a star in this region is unstable.

The \stellareos currently does not implement Coulomb effects for ions, so crystallization is not
supported.  For nuclear reactions, we are usually well outside of the thermodynamic conditions
where this is important.  Adding Coulomb corrections will be a future change.

\subsubsection{Thermal Neutrino Loss}

At the high temperatures reached during advanced stages of stellar
burning, thermal neutrino energy losses can become significant \citep{raffelt2012}. Unlike photons, neutrinos interact
only weakly with stellar matter and can usually escape directly from
the region where they are produced. Their emission therefore removes
energy locally from the stellar plasma, acting as a cooling term in
the stellar energy balance \citep{clayton:1968}.

\pynucastro evaluates these losses as a function of the local
thermodynamic state, separate from the nuclear reaction rates. They are
implemented in the
\texttt{neutrino\_cooling} module, which computes the total
neutrino cooling rate, $\epsilon_{\nu,\mathrm{therm}}$, in $\mathrm{erg~g^{-1}~s^{-1}}$
using the analytic fits of \citet{itoh:1996}. This includes
energy
losses from pair annihilation, plasma neutrino emission,
photoneutrino emission, bremsstrahlung, and recombination.

\begin{figure}
    \centering
    \plotone{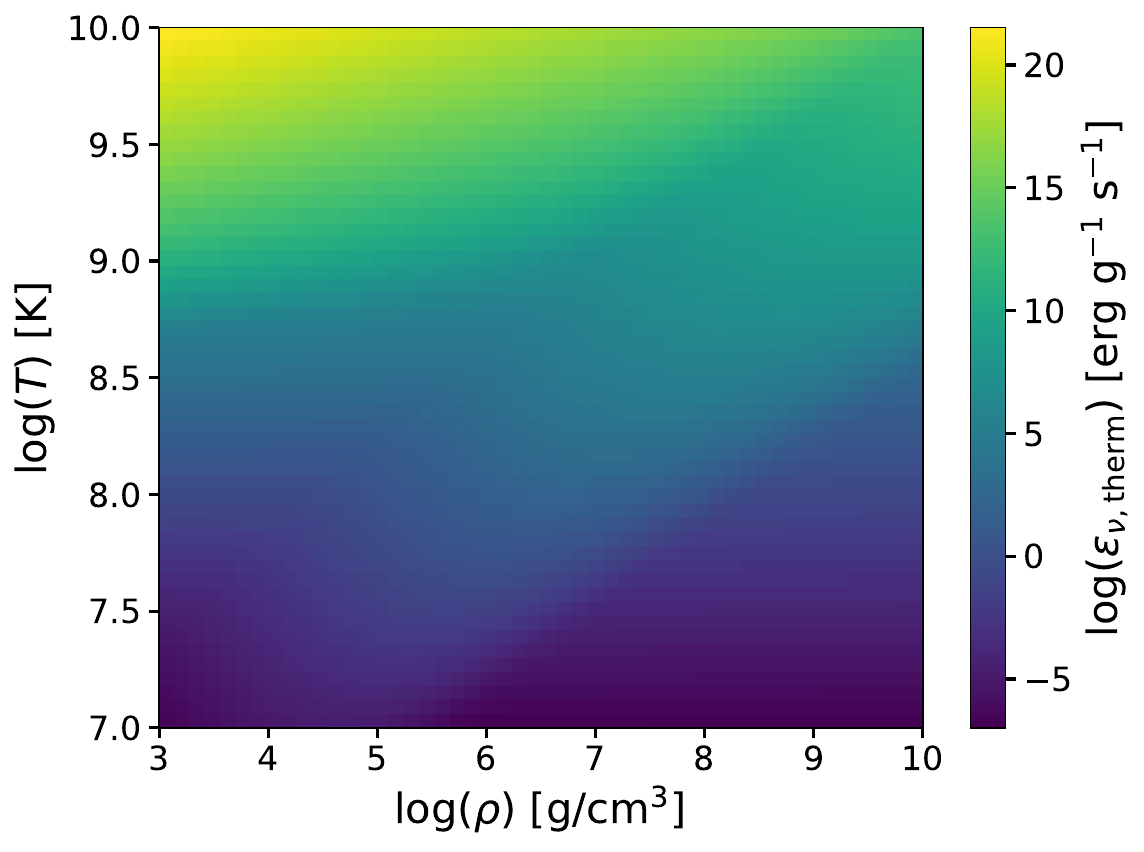}
    \caption{Thermal neutrino cooling rate in the $T$-$\rho$ plane computed with the {\tt NeutrinoCooling} interface. The color shows $\log_{10}(\epsilon_{\nu,\mathrm{therm}})$, where $\epsilon_{\nu,\mathrm{therm}}$ is in $\mathrm{erg~g^{-1}~s^{-1}}$, for fixed composition-averaged quantities {\tt abar} and {\tt zbar}. \supnote{neutrino-rho-T-plot.ipynb}}
    \label{fig:rho-T}
\end{figure}

For standalone calculations, neutrino losses can be evaluated directly with
a thermodynamic state.  Optionally, the individual
contributions from the different neutrino emission processes can be returned
in addition to the total cooling rate. The dependence of the
cooling rate on the thermodynamic state can also be visualized using
the {\tt NeutrinoCooling} class, as shown in Figure~\ref{fig:rho-T}.

\subsubsection{Self-heating burn}
\label{sec:self-heating-burn}
The addition of the equation of state allows for a self-heating burn,
where we evolve the composition together with a temperature equation:
\begin{subequations}
\begin{align}
\frac{d{\bf Y}}{dt} &= {\bf f}({\bf Y}) \\
\frac{dT}{dt} &= \frac{\epsilon_\mathrm{nuc} - \epsilon_{\nu,\mathrm{weak}}}{c_v} \enskip ,
\end{align}
\end{subequations}
where $\epsilon_\mathrm{nuc}$  is the nuclear energy generation rate computed from the network,
$\epsilon_{\nu,\mathrm{weak}}$ are the weak-rate neutrino losses,
and $c_v$ is the specific heat at constant volume, computed from the EOS, $c_v = c_v(\rho, T, {\bf Y})$.
Note that there is also a term related to the dependence of the EOS on the composition,
i.e. $\partial e/\partial Y_k |_{\rho, T, Y_{i,i \neq k}}$, contributing to the temperature evolution
(see e.g. \citet{cabezon_2004}). However, this term is known to be small compared to 
the nuclear energy release except at NSE (see e.g. \citet{calder:2007, chamulak_2007}) and is therefore ignored.
Currently there are no options for evolving density under a prescribed thermodynamic trajectory
or self-consistently from fluid equations. 
Therefore, density is assumed to be constant throughout the evolution to
focus on the response to the nuclear energy release, thereby
neglecting the contribution of hydrodynamic expansion or compression
and the dependence of the EOS on density to the temperature evolution.
When integrating this system,
the specific heat is recomputed from the current temperature and composition every time the rates are evaluated.
This builds on the {\tt integrate\_network} method described in Section~\ref{sec:helper}.
A self-heating burn can also include thermal neutrino losses,
using the same interface.
In this case, the temperature evolution equation becomes:
\begin{equation}
    \frac{dT}{dt} =  \frac{\epsilon_\mathrm{nuc} - \epsilon_{\nu,\mathrm{weak}} - \epsilon_{\nu,\mathrm{therm}}}{c_v} \enskip .
\end{equation}

\begin{figure}[t]
    \centering
    \plotone{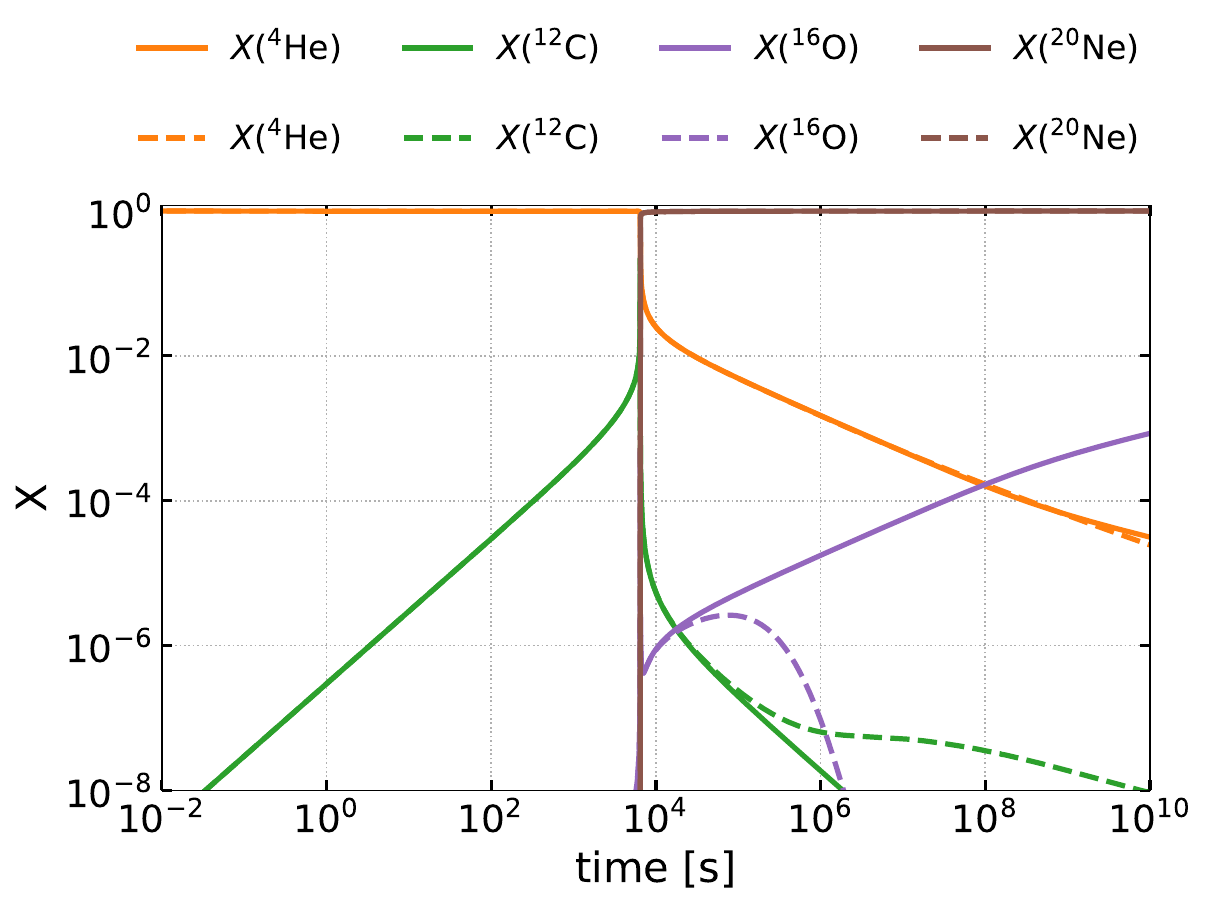}
    \caption{Comparison of composition evolution for the one-zone helium-burning calculation. The solid lines represent the case without effects of neutrino cooling. The dashed lines represent the evolution including effects of neutrino cooling. \supnote{neutrino-cooling.ipynb}}
    \label{fig:neutrino-cooling-composition}
\end{figure}
\begin{figure}[t]
    \centering
    \plotone{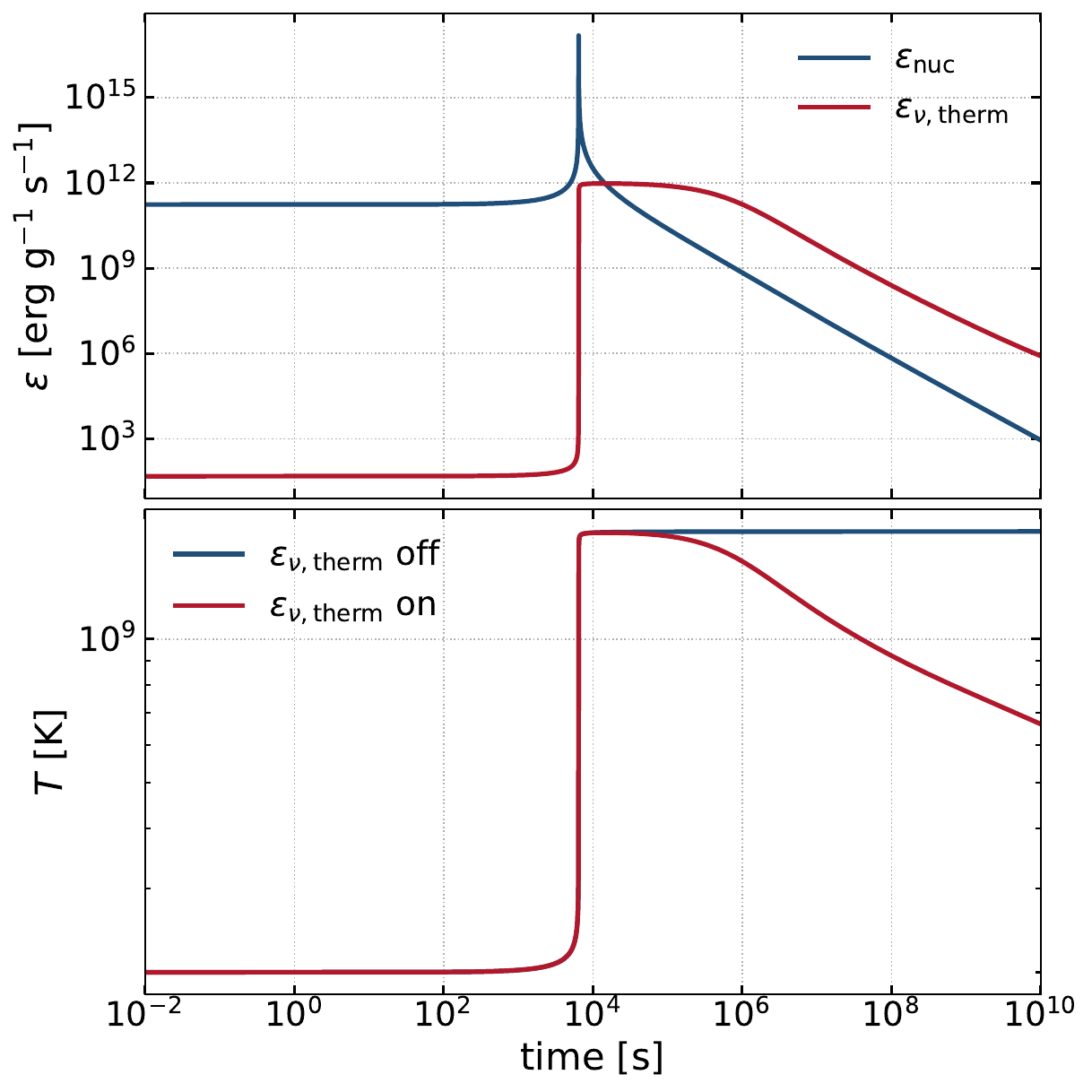}
    \caption{Thermal neutrino cooling in a one-zone helium-burning
      calculation. The upper panel compares the nuclear energy
      generation rate, $\epsilon_\mathrm{nuc}$, with the thermal
      neutrino loss rate, $\epsilon_{\nu,\mathrm{therm}}$. The lower panel compares the
      temperature evolution with and without the effects of neutrino
      cooling included in the evolution equation. \supnote{neutrino-cooling.ipynb}}
    \label{fig:neutrino-cooling}
\end{figure}

We illustrate this with a simple helium-burning network which includes
$\isotm{H}{1}$, $\isotm{He}{4}$, $\isotm{C}{12}$, $\isotm{N}{13}$,
$\isotm{O}{16}$ and $\isotm{Ne}{20}$. We initialize the zone with pure
$\isotm{He}{4}$, at $\rho = 10^{5}~\gcc$ and $T = 2 \times
10^{8}~\mathrm{K}$ and perform two self-heating integrations,
with and without thermal neutrino losses.
The corresponding composition evolution for both cases is shown in Figure~\ref{fig:neutrino-cooling-composition}. The two integrations follow nearly the same evolution through the initial helium-burning runaway. The effects of neutrino cooling become apparent in the late-time abundances of the intermediate species, mostly visible in carbon and oxygen mass fractions.  Figure~\ref{fig:neutrino-cooling} shows the resulting energy rates and
temperature evolution. In the calculation with thermal neutrino
cooling, $\epsilon_\mathrm{nuc}$ dominates during the initial
runaway, but after the peak, the nuclear energy generation rate decreases
while $\epsilon_{\nu,\mathrm{therm}}$ remains large for longer and eventually exceeds
$\epsilon_\mathrm{nuc}$. This produces the decline in temperature for
the cooled trajectory which is observed in the lower
panel. 

The inclusion of the EOS allows for other types of one-zone
burns, including constant-pressure burns appropriate to flames \citep{calder:2007} and Lagrangian particle trajectory post-processing.


\subsection{New Rate Approximations}

The cost of solving a reaction network with an implicit integrator
increases quickly with the number of nuclei explicitly carried in the network.
This is because the LU-decomposition routine associated with the implicit solve
can scale as $\mathcal{O}(N^3)$, where $N$ is the number of nuclei.
Therefore, multi-dimensional reactive flow simulations aim to use 
the smallest reaction network that can still accurately capture the energy release
to reduce the computational cost.
The hard-coded {\tt aprox}-family of networks achieves this by introducing
various types of rate approximations.
One example is the $(\alpha,p)(p,\gamma)$ approximation that combines the rate sequences
$A(\alpha, p)X(p,\gamma)$ and $A(\alpha,\gamma)B$ into a single
effective rate $A(\alpha,\gamma)B$ rate.
This assumes that the intermediate nucleus $X$ remains in equilibrium,
allowing it to be removed from the network.
However, these approximations are hardcoded into the {\tt aprox}-family network implementation,
making it difficult to update or add other rate-sequences with the same set of approximations.
In \citet{pynucastro2}, we demonstrated the \approximaterate
functionality, with the initial support for the $(\alpha,p)(p,\gamma)$ approximation.
\approximaterate provides a flexible framework for constructing these approximated rates
as well as extending it with new approximations, and now \pynucastro 3 has added several new rate approximations, which we describe below.

\subsubsection{Double n-capture}
\label{sec:double-n-capture}

The {\tt aprox} networks make extensive use of the assumption that the
intermediate species, $X$, are in equilibrium.  In addition to its use
in $A(\alpha, p)X(p,\gamma)B$ rate sequence, 
the same concept can also be used in $A(n,\gamma)X(n,\gamma)B$
rate sequence. The overall effect of approximating $dY_X/dt \approx 0$
is that this rate sequence can now be treated as a single double
neutron capture process on species, $A$, i.e.
\begin{equation}
    A(n,\gamma)X(n,\gamma)B \xrightarrow{dY_X/dt \ = \ 0} A(nn,\gamma)B \enskip .
\end{equation}

The evolution equation for $A(n,\gamma)X(n,\gamma)B$, including both the forward and reverse,
can be written as the following:
\begin{subequations}
\label{eq:ngng_evolution}
\begin{align}
\frac{d Y_A}{dt} =& -\rho Y_A Y_n \lambda_{A(n,\gamma)X} + Y_X \lambda_{X(\gamma,n)A} \label{eq:ngng_evolution_1}\\
\frac{d Y_B}{dt} =& +\rho Y_X Y_n \lambda_{X(n,\gamma)B} - Y_B \lambda_{B(\gamma,n)X} \label{eq:ngng_evolution_2}\\
\frac{d Y_X}{dt} =& +\rho Y_A Y_n \lambda_{A(n,\gamma)X} - Y_X \lambda_{X(\gamma,n)A} \\ \nonumber
                  & -\rho Y_X Y_n \lambda_{X(n,\gamma)B} + Y_B \lambda_{B(\gamma,n)X} \\
\frac{d Y_n}{dt} =& -\rho Y_A Y_n \lambda_{A(n,\gamma)X} + Y_X \lambda_{X(\gamma,n)A} \\ \nonumber
                  & -\rho Y_X Y_n \lambda_{X(n,\gamma)B} + Y_B \lambda_{B(\gamma,n)X} \enskip .
\end{align}
\end{subequations}
The equilibrium condition for species $X$, $dY_X/dt \approx 0$,
imposes a constraint on the expression of its molar abundance to be
\begin{equation}
\label{eq:Y_X}
    Y_X = \frac{Y_B \lambda_{B(\gamma,n)X} + \rho Y_A Y_n \lambda_{A(n,\gamma)X}}{\rho Y_n \lambda_{X(n,\gamma)B} + \lambda_{X(\gamma,n)A}} \enskip .
\end{equation}
Substituting Eq.~\ref{eq:Y_X} into Eq.~\ref{eq:ngng_evolution_1} and Eq.~\ref{eq:ngng_evolution_2},
the system of evolution equations described by two-step reaction sequence, Eq.~\ref{eq:ngng_evolution},
now reduces to a simplified system described by a single double neutron capture process
\begin{equation}
\label{eq:ngng_evolution_new}
\begin{aligned}
\frac{1}{2} \frac{d Y_n}{dt}
    &= \frac{d Y_A}{dt}
     = - \frac{d Y_B}{dt} \\
    &= - \frac{1}{2}\rho^2 Y_A Y_n^2
       \lambda_{A(nn,\gamma)B}^{\mathrm{eff}}
     + Y_B
       \lambda_{B(\gamma,nn)A}^{\mathrm{eff}} \enskip,
\end{aligned}
\end{equation}
where the forward and reverse rate for the double neutron capture rate is defined as 
\begin{subequations}
    \begin{align}
    \lambda_{A(nn,\gamma)B}^{\mathrm{eff}} =& \frac{2\lambda_{A(n,\gamma)X} \lambda_{X(n,\gamma)B}}{\rho Y_n \lambda_{X(n,\gamma)B} + \lambda_{X(\gamma,n)A}} \\
    \lambda_{B(\gamma,nn)A}^{\mathrm{eff}} =& \frac{\lambda_{B(\gamma,n)X} \lambda_{X(\gamma,n)A}}{\rho Y_n \lambda_{X(n,\gamma)B} + \lambda_{X(\gamma,n)A}} \enskip .
    \end{align}
\end{subequations}

\pynucastro can now do this set of approximations automatically via
\approximaterate.  It is important to note that since
\approximaterate is derived from the balance of
forward and reverse rates in the rate sequence, it is
intrinsically compatible with nuclear statistical
equilibrium provided that the reverse rates in this
multi-step rate sequence are derived from detailed-balance.

\begin{figure}[t]
    \centering
    \plottwo{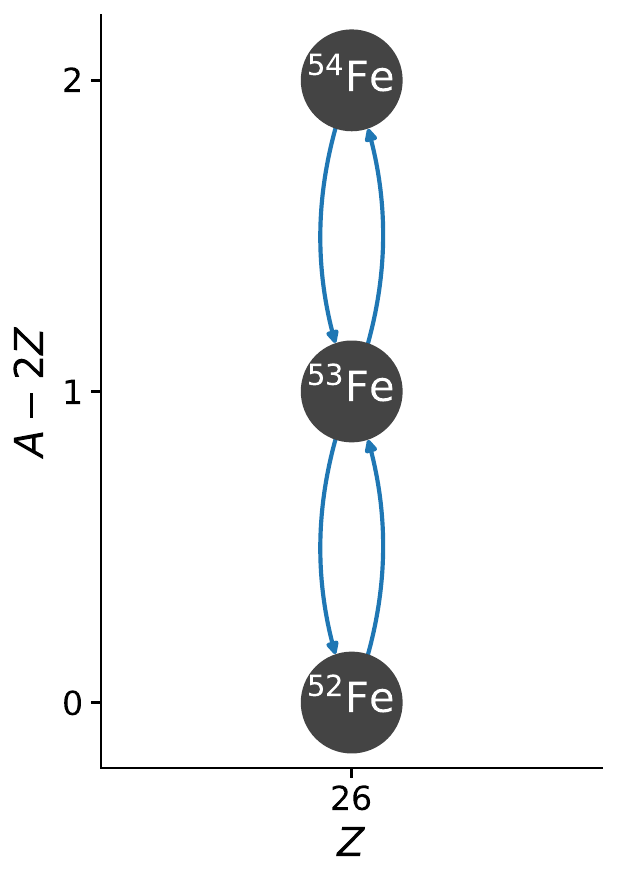}{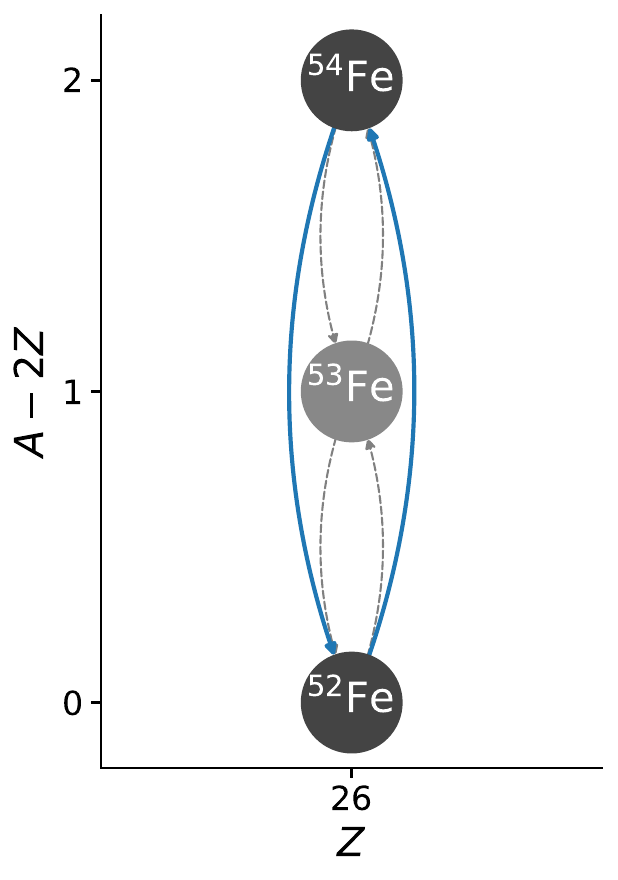}
    \caption{{\it Left}: flow diagram for full reaction sequence, ${}^{52}\mathrm{Fe}(n,\gamma){}^{53}\mathrm{Fe}(n,\gamma){}^{54}\mathrm{Fe}$. {\it Right}: flow diagram for ${}^{52}\mathrm{Fe}(nn,\gamma){}^{54}\mathrm{Fe}$, an approximated version of ${}^{52}\mathrm{Fe}(n,\gamma){}^{53}\mathrm{Fe}(n,\gamma){}^{54}\mathrm{Fe}$. The $\isotm{Fe}{53}$ node is grayed out to indicate that it is removed from the network but is still connected by the hidden rates represented by the dotted lines. \supnote{double-neutron-capture.ipynb}}
    \label{fig:double-neutron-capture}
\end{figure}

\begin{figure}[t]
    \centering
    \plotone{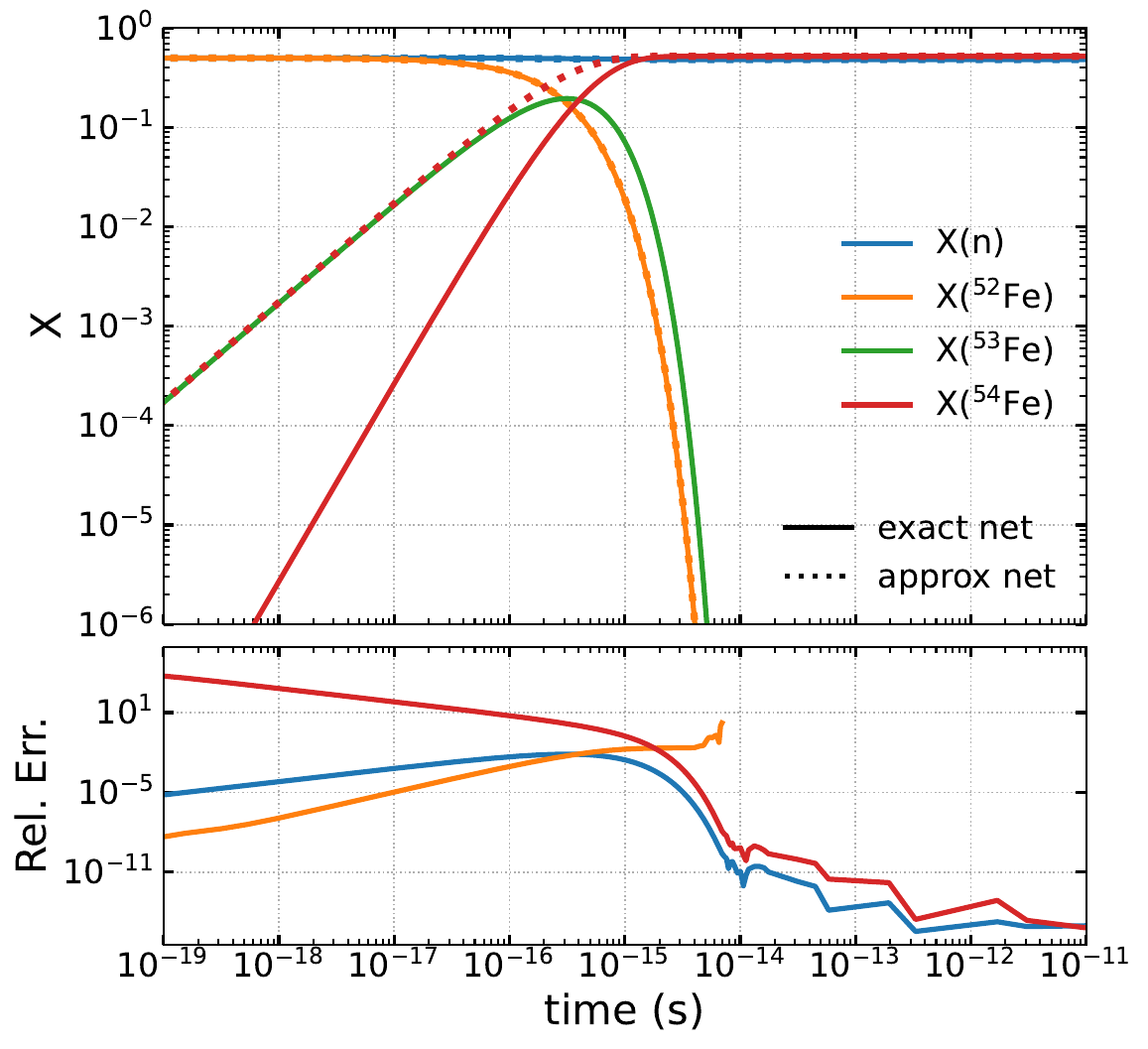}
    \caption{{\it Top}: Time evolution of the two networks shown in Figure \ref{fig:double-neutron-capture}. {\it Bot}: Relative error in the mass fraction of nuclei common to both networks.
    Relative error for ${}^{52}\mathrm{Fe}$ terminates as the mass fraction drops
    below the integration tolerance.
    Both network show the same nucleosynthesis yields after a short timescale
    of $\sim 10^{-15}$ s. \supnote{double-neutron-capture.ipynb}}
    \label{fig:double-neutron-capture-integration}
\end{figure}

An example use case for the double neutron capture approximation
is to approximate 
${}^{52}\mathrm{Fe}(n,\gamma){}^{53}\mathrm{Fe}(n,\gamma){}^{54}\mathrm{Fe}$
into ${}^{52}\mathrm{Fe}(nn,\gamma){}^{54}\mathrm{Fe}$, 
which is shown graphically in Figure~\ref{fig:double-neutron-capture}.
A sample code that does this approximation is:

\begin{lstlisting}
nuc_list = ["fe52", "fe53", "fe54", "n"]
net = pyna.network_helper(nuc_list) 
net.make_nn_g_approx(intermediate_nuclei="fe53")
net.remove_nuclei(["fe53"])
\end{lstlisting}

The evolution of the exact network and the approximate network under the same initial conditions is shown in
Figure~\ref{fig:double-neutron-capture-integration}. 
We see that ${}^{52}\mathrm{Fe}$ is consumed at nearly the same rate for both networks. 
The early production of ${}^{54}\mathrm{Fe}$ in the approximate network
closely follows the production of ${}^{53}\mathrm{Fe}$ in the exact network.  
The depletion of ${}^{53}\mathrm{Fe}$ occurs under a short timescale of 
$\sim 10^{-15}$ s, which causes the subsequent production of 
${}^{54}\mathrm{Fe}$ from the approximate network to become
consistent with the exact network.

\subsubsection{Approximating C/O burning}
\label{sec:co-burning}

Carbon and oxygen burning has many branches that may need to be accounted for.
The dominant branches are:
\begin{subequations}
\begin{gather}
\isotm{C}{12} + \isotm{C}{12} \rightarrow
\begin{cases}
    \isotm{Ne}{20} + \alpha \\
    \isotm{Na}{23} + p \\
    \isotm{Mg}{23} + n
\end{cases}
\\[1ex]
\isotm{O}{16} + \isotm{C}{12} \rightarrow
\begin{cases}
    \isotm{Mg}{24} + \alpha \\
    \isotm{Al}{27} + p \\
    \isotm{Si}{27} + n
\end{cases}
\\[1ex]
\isotm{O}{16} + \isotm{O}{16} \rightarrow
\begin{cases}
    \isotm{Si}{28} + \alpha \\
    \isotm{P}{31} + p \\
    \isotm{S}{31} + n
\end{cases}
\enskip .
\end{gather}
\end{subequations}
Networks like {\tt aprox13} leave out the neutron-emission branch and
approximate the alpha and proton emission branches to eliminate the
intermediate nucleus.  This allows the process to be described using
only nuclei along the $\alpha$-chain. 
\pynucastro\ can implement this approximation for each of these reactions.
Note that the neutron-emission branch is ignored in this approximation
and is not considered in the subsequent discussion.

\begin{figure}[t]
\centering
\plotone{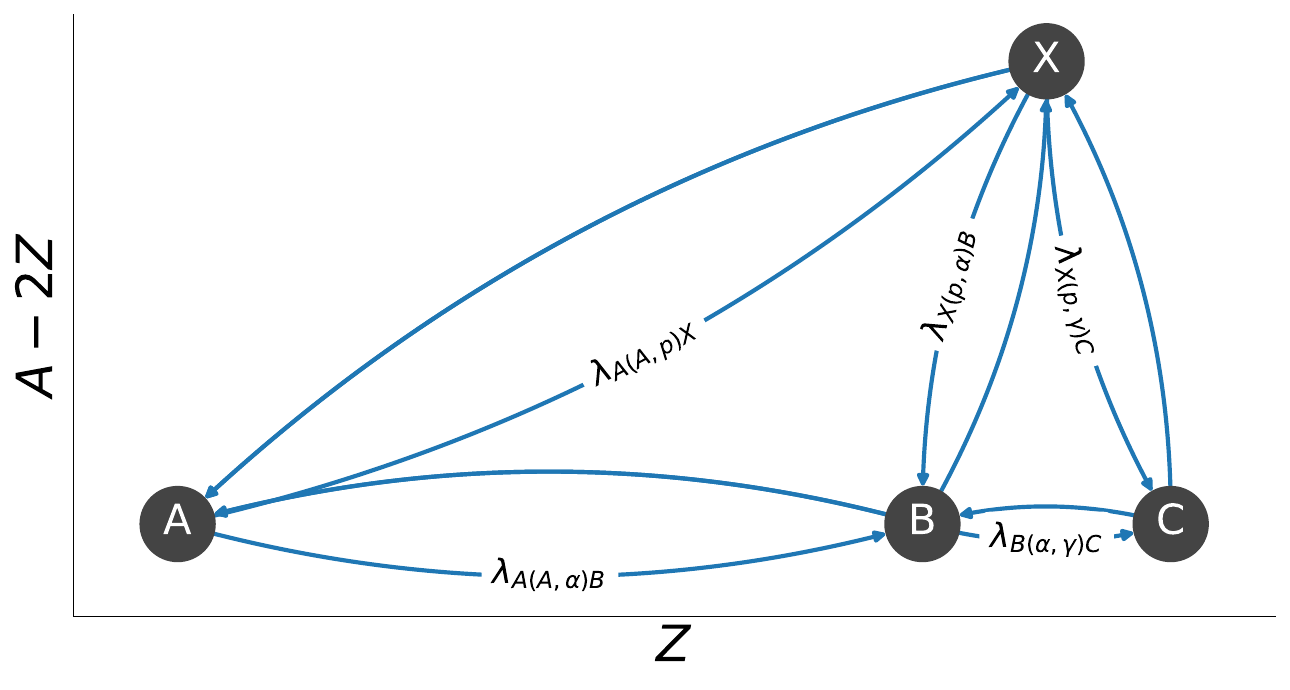}
\caption{\label{fig:co} A schematic showing the nuclei and rates involved in approximating carbon / oxygen burning.  Our goal is to eliminate nucleus $X$ from the system will keeping the flow through it assuming equilibrium.}
\end{figure}

We will describe the generic process.  We will call the starting nucleus $A$ (even in the
case where it is $\isotm{C}{12} + \isotm{O}{16}$), the nucleus in the alpha-emission branch $B$,
the nucleus in the proton-emission branch $X$, and the nucleus that
would return if the neutron combined with the nucleus in the
neutron-emission branch $C$.  For $\isotm{C}{12}+\isotm{C}{12}$, we have $A = \isotm{C}{12}$,
$B = \isotm{Ne}{20}$, $C = \isotm{Mg}{24}$, and $X = \isotm{Na}{23}$.
This is shown graphically in Figure~\ref{fig:co}.

Our goal is to eliminate $X$, and create 3 approximate rates:
\begin{itemize}
\item $\lambda_{A(A,\alpha)B}^\mathrm{eff}$ : combining $A(A,\alpha)B$ and the sequence $A(A,p)X(p,\alpha)B$
\item $\lambda_{A(A,\gamma)C}^\mathrm{eff}$ : representing $A(A,p)X(p,\gamma)C$
\item $\lambda_{B(\alpha,\gamma)C}^\mathrm{eff}$ : combining $B(\alpha,\gamma)C$ and the sequence $B(\alpha,p)X(p,\gamma)C$
\end{itemize}
as well as their inverses.
We will write this for the case of $C+C$ or $O+O$, in which
case there is an identical particle factor of $1/2$
in the rate and we have to account for 2 nuclei $A$
consumed or produced.  But this will not matter in the end,
and the expressions will be equally valid for the $C+O$ case.

Starting with nucleus $X$, the evolution equation is:
\begin{align}
   \frac{dY_X}{dt} = &+ \tfrac{1}{2}\rho Y_A^2 \lambda_{A(A,p)X} - \rho Y_p Y_X \lambda_{X(p,A)A} \nonumber \\
                      &- \rho Y_p Y_X \lambda_{X(p,\alpha)B} + \rho Y_\alpha Y_B \lambda_{B(\alpha,p)X} \nonumber \\
                      &- \rho Y_p Y_X \lambda_{X(p,\gamma)C} + Y_C\lambda_{C(\gamma,p)X} \enskip .
\end{align}
Setting this to zero to find the equilibrium solution, we find:
\begin{equation}
  \rho Y_p Y_X = \frac{\tfrac{1}{2} \rho Y_A^2 \lambda_{A(A,p)X} + \rho Y_\alpha Y_B \lambda_{B(\alpha,p)X} + Y_C \lambda_{C(\gamma,p)X}}{\lambda_{X(p,A)A} + \lambda_{X(p,\alpha)B} + \lambda_{X(p,\gamma)C}} \enskip .
\end{equation}
Notice that the denominator is basically normalizing by all the
possible decay paths from nucleus $X$.  This also shows that the
evolution of protons is the same as for nucleus $X$.

We now write the evolution for nuclei $A$, $B$, and $C$.    The evolution of $A$ is given as:
\begin{align}
   \frac{1}{2}\frac{dY_A}{dt} = &-\tfrac{1}{2} \rho Y_A^2 \lambda_{A(A,p)X} - \tfrac{1}{2} \rho Y_A^2 \lambda_{A(A,\alpha)B}  \nonumber \\
                                &+ \rho Y_p Y_X \lambda_{X(p,A)A} + \rho Y_\alpha Y_B \lambda_{B(\alpha,A)A} \enskip .
\end{align}
The evolution of $\alpha$ is the same as that of nucleus $B$ and is:
\begin{align}
   \frac{dY_\alpha}{dt} = \frac{dY_B}{dt} = &- \rho Y_\alpha Y_B \lambda_{B(\alpha,A)A} - \rho Y_\alpha Y_B \lambda_{B(\alpha,p)X} \nonumber \\
                                            &- \rho Y_\alpha Y_B \lambda_{B(\alpha,\gamma)C}\nonumber + \tfrac{1}{2} \rho Y_A^2 \lambda_{A(A,\alpha)B} \nonumber \\
                                            &+ \rho Y_p Y_X \lambda_{X(p,\alpha)B}  + Y_C \lambda_{C(\gamma,\alpha)B} \enskip .
\end{align}
Finally, the evolution of $C$ is:
\begin{align}
   \frac{dY_C}{dt} = &- Y_C \lambda_{C(\gamma,p)X} - Y_C \lambda_{C(\gamma,\alpha)B} \nonumber \\
                     &+ \rho Y_p Y_X \lambda_{X(p,\gamma)C} + \rho Y_\alpha Y_B \lambda_{B(\alpha,\gamma)C} \enskip .
\end{align}

Now substituting in our equilibrium value for $\rho Y_pY_X$, and defining
the branching normalization as
\begin{equation}
    f \equiv \frac{1}{\lambda_{X(p,A)A} + \lambda_{X(p,\alpha)B} + \lambda_{X(p,\gamma)C}}
\end{equation}
and
grouping terms, we find:
\begin{align}
  \frac{1}{2} \frac{dY_A}{dt} = &-\tfrac{1}{2}\rho Y_A^2 \underbrace{\left(\lambda_{A(A,\alpha)B} + f\lambda_{A(A,p)X} \lambda_{X(p,\alpha)B} \right)}_{\lambda^\mathrm{eff}_{A(A,\alpha)B}} \nonumber \\ 
                                &+ \rho Y_\alpha Y_B \underbrace{\left (\lambda_{B(\alpha,A)A} + f\lambda_{X(p,A)A} \lambda_{B(\alpha,p)X} \right )}_{\lambda^\mathrm{eff}_{B(\alpha,A)A}} \nonumber\\
                                &+ Y_C \underbrace{f\lambda_{X(p,A)A} \lambda_{C(\gamma,p)X}}_{\lambda^\mathrm{eff}_{C(\gamma,A)A}} \nonumber \\
                                &-\tfrac{1}{2}\rho Y_A^2 \underbrace{f\lambda_{A(A,p)X} \lambda_{X(p,\gamma)C}}_{\lambda^\mathrm{eff}_{A(A,\gamma)C}}
\end{align}
for the evolution of $A$, where we've identified the effective rates for each process.  Similarly, we find:
\begin{align}
  \frac{dY_B}{dt} = &+ \tfrac{1}{2} \rho Y_A^2 \underbrace{\left (\lambda_{A(A,\alpha)B} + f\lambda_{A(A,p)X} \lambda_{X(p,\alpha)B} \right )}_{\lambda^\mathrm{eff}_{A(A,\alpha)B}} \nonumber \\
                    &- \rho Y_\alpha Y_B \underbrace{\left (\lambda_{B(\alpha,A)A} + f\lambda_{X(p,A)A}\lambda_{B(\alpha,p)X} \right)}_{\lambda^{\mathrm{eff}}_{B(\alpha,A)A}}  \nonumber \\
                    &- \rho Y_\alpha Y_B \underbrace{\left (\lambda_{B(\alpha,\gamma)C} + f\lambda_{B(\alpha,p)X} \lambda_{X(p,\gamma)C} \right)}_{\lambda^\mathrm{eff}_{B(\alpha,\gamma)C}} \nonumber \\
                    &+Y_C \underbrace{\left (\lambda_{C(\gamma,\alpha)B} + f\lambda_{X(p,\alpha)B} \lambda_{C(\gamma,p)X} \right)}_{\lambda^{\mathrm{eff}}_{C(\gamma,\alpha)B}}
\end{align}
for $B$, again identifying the effective rates, and finally,
\begin{align}
  \frac{dY_C}{dt} = &+ \tfrac{1}{2} \rho Y_A^2 \underbrace{f\lambda_{A(A,p)X} \lambda_{X(p,\gamma)C} }_{\lambda^\mathrm{eff}_{A(A,\gamma)C}}
                    - Y_C \underbrace{f\lambda_{X(p,A)A} \lambda_{C(\gamma,p)X} }_{\lambda^\mathrm{eff}_{C(\gamma,A)A}} \nonumber \\
                    &+ \rho Y_\alpha Y_B \underbrace{\left (\lambda_{B(\alpha,\gamma)C} + f\lambda_{B(\alpha,p)X} \lambda_{X(p,\gamma)C} \right)}_{\lambda^\mathrm{eff}_{B(\alpha,\gamma)C}} \nonumber \\
                    &- Y_C \underbrace{\left ( \lambda_{C(\gamma,\alpha)B} + f\lambda_{X(p,\alpha)B} \lambda_{C(\gamma,p)X} \right )}_{\lambda^\mathrm{eff}_{C(\gamma,\alpha)B}}
\end{align}
for the evolution of $C$.
From these expressions, we can read off the effective forward rates:
\begin{subequations}
\begin{align}
  \lambda^\mathrm{eff}_{A(A,\alpha)B} &= \lambda_{A(A,\alpha)B} + f\lambda_{A(A,p)X} \lambda_{X(p,\alpha)B}
  \\
  \lambda^\mathrm{eff}_{A(A,\gamma)C} &= f\lambda_{A(A,p)X} \lambda_{X(p,\gamma)C}
  \\
  \lambda^\mathrm{eff}_{B(\alpha,\gamma)C} &= \lambda_{B(\alpha,\gamma)C} + f\lambda_{B(\alpha,p)X} \lambda_{X(p,\gamma)C} \enskip ,
\end{align}
\end{subequations}
and reverse rates:
\begin{subequations}
\begin{align}
 \lambda^\mathrm{eff}_{B(\alpha,A)A} &= \lambda_{B(\alpha,A)A} + f\lambda_{X(p,A)A} \lambda_{B(\alpha,p)X}
  \\
  \lambda^\mathrm{eff}_{C(\gamma,A)A} &= f\lambda_{X(p,A)A} \lambda_{C(\gamma,p)X}
  \\
  \lambda^\mathrm{eff}_{C(\gamma,\alpha)B} &= \lambda_{C(\gamma,\alpha)B} + f\lambda_{X(p,\alpha)B} \lambda_{C(\gamma,p)X} \enskip .
\end{align}
\end{subequations}

The form of these makes sense---anytime you pass through nucleus $X$, you need to normalize
by all the possible decay paths from it.
We note that the last rate is almost the same as the $A(\alpha,\gamma)B$ described
in \citet{pynucastro2}, except the normalization, $f$, is different, since there is
a third nucleus present which presents an alternate branching from $X$.  In \pynucastro 3,
we allow for another branching in \approximaterate.  The current \approximaterate is written in a way that makes it easy to add new approximations in the future.

To test this, we build two networks. The first includes $\alpha$-chain
and all of the intermediate nuclei that are needed, and is created as
\begin{lstlisting}
nuclei = ["p", "he4",
          "c12", "o16", "ne20", "na23",
          "mg24", "al27", "si28", "p31", "s32",
          "cl35", "ar36", "k39", "ca40",
          "sc43", "ti44", "v47", "cr48",
          "mn51", "fe52", "co55", "ni56"]
net = pyna.network_helper(nuclei)
\end{lstlisting}
This has 23 nuclei, linked with 92 rates (46 from ReacLib and
46 rederived via detailed balance).  We'll call this the full network.

\begin{figure*}[t]
\centering
\plotone{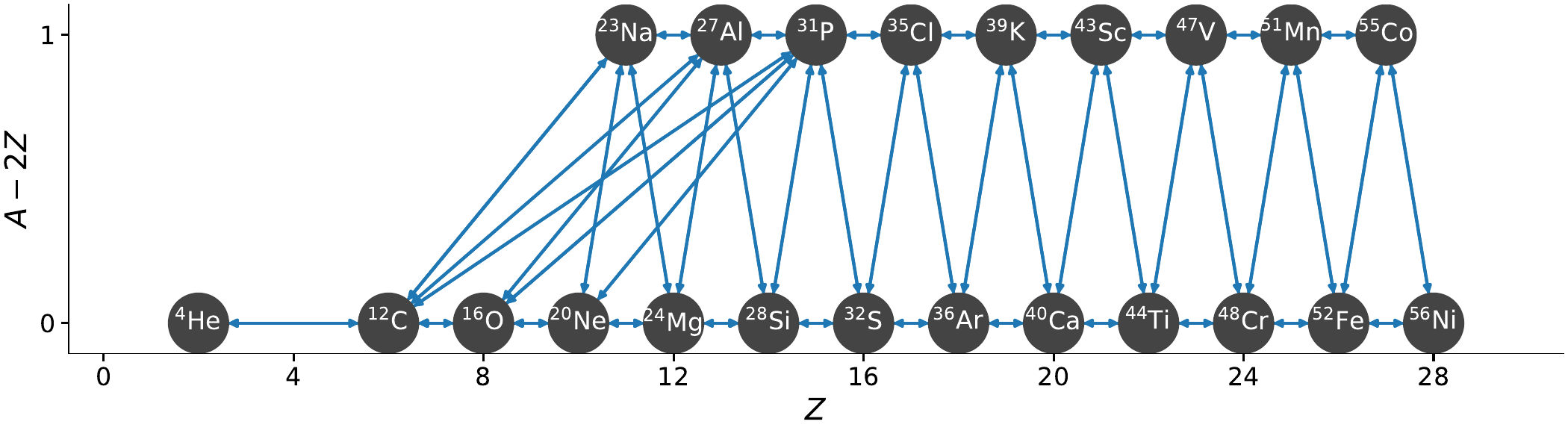}
\plotone{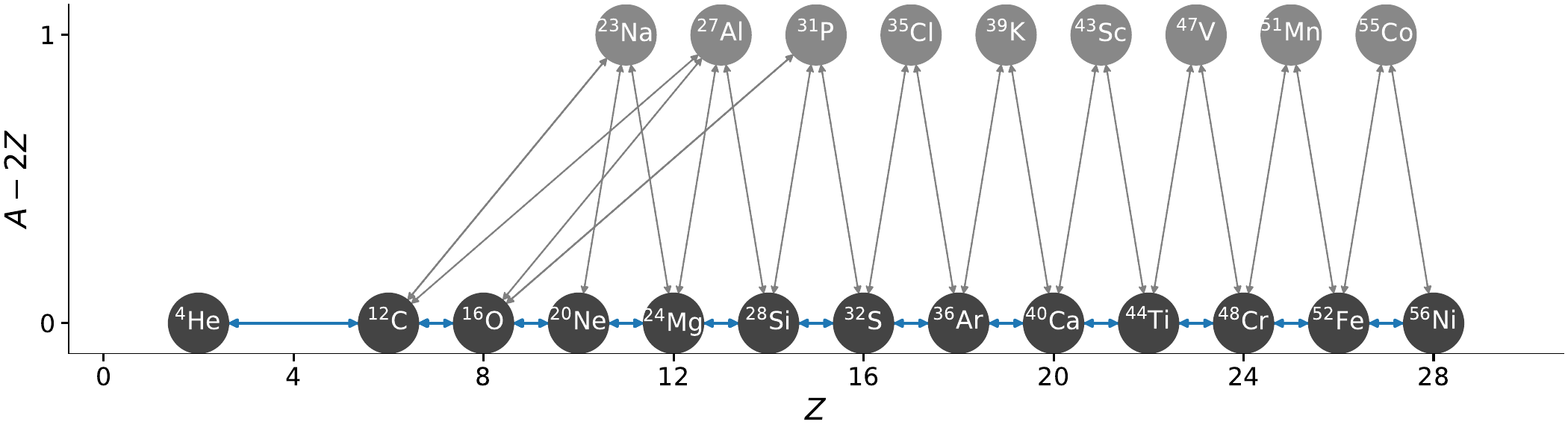}
\caption{\label{fig:aprox13-nets} A comparison of the original, unapproximated network (top) and the {\tt aprox13}-style network created from this using ``make\_CO\_burning\_approx'' (bottom).  In the approximate network, the nuclei in gray are removed, but the rates flowing through them are kept, and evaluated via the \approximaterate objects. \supnote{aprox13.ipynb}}
\end{figure*}

Next we create the approximate network.  For each of $\isotm{C}{12} +
\isotm{C}{12}$, $\isotm{O}{16} + \isotm{C}{12}$, and $\isotm{O}{16} +
\isotm{O}{16}$, we use the approximation derived in this section to
replace the rates, and remove the intermediate nucleus.  In
\pynucastro, this is done using the {\tt make\_CO\_burning\_approx}
function. As a result of removing $\isotm{P}{31}$, we also drop
$\isotm{Ne}{20}(\isotm{C}{12}, p)\isotm{P}{31}$. We don't expect this
rate to be significant, given the high Coulomb barrier.  For
consistency, we explicitly remove $\isotm{Ne}{20}(\isotm{C}{12},
\alpha)\isotm{Si}{28}$ and its inverse.  We also approximate out
\nucleilist{Cl35, L39, Sc43, V47, Mn51, Co55} using the $(\alpha,
p)(p, \gamma)$ approximation.  The process for generating the
approximate network starting with the network above is:
\begin{lstlisting}
# create a new net with the rates from the
# full net
rates = net.get_rates()
apnet = pyna.PythonNetwork(rates=rates)

# do the C+C, C+O, O+O approximations
apnet.make_CO_burning_approx("C")
apnet.remove_nuclei(["na23"])

apnet.make_CO_burning_approx("CO")
apnet.remove_nuclei(["al27"])

apnet.make_CO_burning_approx("O")
apnet.remove_nuclei(["p31"])

# remove some additional heavy nuclei rates
rne = apnet.get_rate_by_name("Ne20(C12,a)Si28")
rne_r = apnet.get_rate_by_name("Si28(a,C12)Ne20")
apnet.remove_rates([rne, rne_r])

# (a,p)(p,g) approximation
intermediate_nucs = ["cl35", "k39", "sc43",
                     "v47", "mn51", "co55"]
apnet.make_ap_pg_approx(intermediate_nucs)
apnet.remove_nuclei(intermediate_nucs)
\end{lstlisting}
The result is that the number of nuclei
in the net is reduced from 23 to 13, and the total number of rates is
102, with 36 ReacLib rates, 36 derived reverse rates, and 30
approximate rates (only 36 of these rates explicitly connect nuclei).
This network is the \pynucastro version of the classic {\tt aprox13}
network.  Figure~\ref{fig:aprox13-nets} shows the structure of the two
networks.

\begin{figure}[t]
\plotone{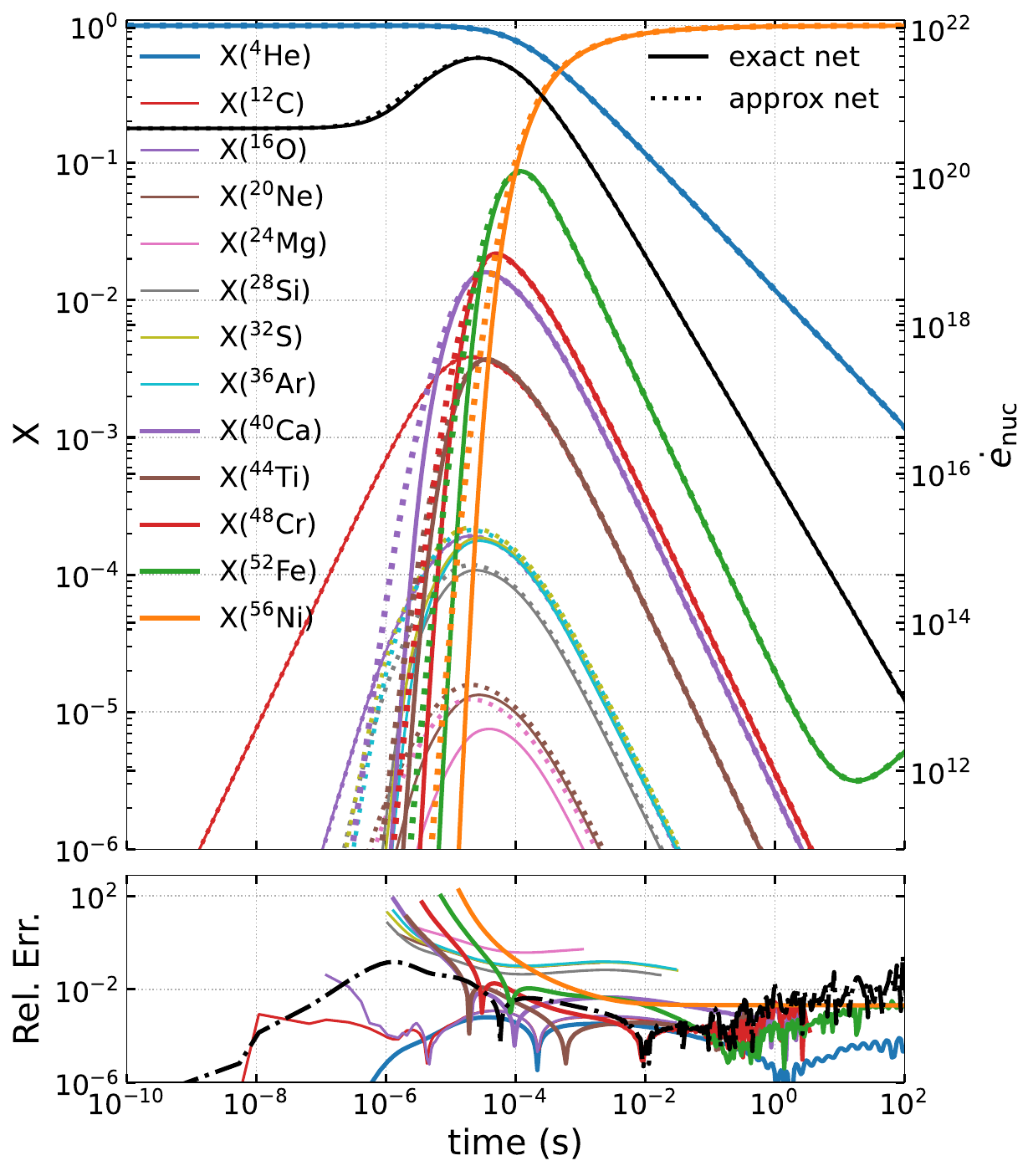}
\caption{\label{fig:aprox-integration} {\it Top}: Integrating the {\tt
    aprox13}-style network compared to the unapproximated / full
  network.  The solid lines show the mass fractions for the full network
  and the dotted lines show the approximate network.
  The black lines represent the instantaneous nuclear energy generation rate, 
  $\dot{e}_{\mathrm{nuc}}$, evaluated from the full and approximate network.
  {\it Bot}: Relative error in the mass fraction of nuclei common to
  both networks. Relative errors are shown when the mass fractions are above $10^{-6}$.
  We see that for the most abundant nuclei, indicated by the thicker lines,
  the relative error for the approximate network is $10^{-2}\sim10^{-6}$.
  The maximum relative error in $\dot{e}_{\mathrm{nuc}}$, 
  indicated by the dot-dash black line, is $\sim 0.26$.
  Differences are only
  apparently during the early rise of some nuclei because the full network
  has additional nuclei that the reaction flow passes through before winding up at
  the same nuclei in the approximate network. \supnote{aprox13.ipynb}}
\end{figure}

To test how well the approximation works, we next integrate each
network with the same starting conditions: pure $\isotm{He}{4}$ with a
density of $\rho = 10^7~\gcc$ and temperature of $3\times
10^9~\mathrm{K}$.  Figure~\ref{fig:aprox-integration} shows the
results.  Overall, we see that the approximate network captures the
behavior of the full network quite well, demonstrating both that this
\pynucastro is able to perform the approximation and that this
approximation is useful for science applications.

\subsubsection{Modified rates}
\label{sec:modified_rates}


In \citet{pynucastro2}, we introduced the {\tt modify\_products()}
method that allows the products of the rate to be modified along with
its Q-value updated to reflect the new mass difference between the
reactants and products.  It was intended to approximate a multi-step
reaction rate sequence into a single effective reaction rate.  Assume
that the initial reaction of the sequence is the slowest step compared
to the subsequent reactions, then the intermediate nuclei produced can
be treated as being in the steady state.  Under these assumptions, the
overall effective rate of the full sequence is approximately the rate
of the first reaction, while the reactants and products are updated
accordingly to reflect the net outcome of the rate sequence.  

As an example, consider the following {\it forward} rate sequence
approximation introduced by \citet{pynucastro2}.
\begin{equation}
\label{eq:modify_rate_ex1}
{}^{12}\mathrm{C}({}^{12}\mathrm{C},n){}^{23}\mathrm{Mg}(n,\gamma){}^{24}\mathrm{Mg} \xrightarrow{dY_{{}^{23}\mathrm{Mg}}/dt\, = \, 0} {}^{12}\mathrm{C}({}^{12}\mathrm{C},\gamma){}^{24}\mathrm{Mg} \enskip .
\end{equation}
The approximation assumes that the intermediate nuclei, $\isotm{Mg}{23}$ and $n$,
remain in local equilibrium due to balance of the two {\it forward} rates, i.e.
\begin{equation}
\begin{aligned}
\frac{d}{dt}Y_{{}^{23}\mathrm{Mg}} =\frac{d}{dt} Y_{n}
=& +\frac{1}{2} \rho Y_{{}^{12}\mathrm{C}}^2 \lambda_{{}^{12}\mathrm{C}({}^{12}\mathrm{C},n){}^{23}\mathrm{Mg}} \\
 & - \rho Y_{{}^{23}\mathrm{Mg}} Y_n \lambda_{{}^{23}\mathrm{Mg}(n,\gamma){}^{24}\mathrm{Mg}} \\
 =& \ 0 \enskip .
\end{aligned}
\end{equation}
Under this assumption, the two-step reaction sequence behaves as a single
effective reaction that converts two ${}^{12}\mathrm{C}$ into ${}^{24}\mathrm{Mg}$,
which then evolves according to
\begin{equation}
    \frac{1}{2}\frac{d}{dt}Y_{{}^{12}\mathrm{C}} = -\frac{d}{dt}Y_{{}^{24}\mathrm{Mg}} = -\frac{1}{2} \rho Y_{{}^{12}\mathrm{C}}^2 \underbrace{\lambda_{{}^{12}\mathrm{C}({}^{12}\mathrm{C},\gamma){}^{24}\mathrm{Mg}}}_{\lambda_{{}^{12}\mathrm{C}({}^{12}\mathrm{C},n){}^{23}\mathrm{Mg}}} \enskip ,
\end{equation}
where the effective rate is simply the rate of the first reaction in the sequence
as the subsequent neutron capture occurs right after the ${}^{23}\mathrm{Mg}$ and $n$
are produced.

While {\tt modify\_products()} is sufficient when only the products of
a given reaction need to be modified, it cannot be used for reactions
with modified reactants or non-standard stoichiometry.
To address this limitation, \pynucastro 3 introduces a new class,
\modifiedrate, that allows both the reactants and products of a
given rate to be modified, as well as supporting a custom
stoichiometry.  The \modifiedrate with a custom stoichiometry has
a rate equation following
Eq.~\ref{eq:molar_fraction_stoichiometry_ode}, which is slightly
different from the rate equation (Eq.~\ref{eq:molar_fraction_ode}) of
a generic reaction.

In order to motivate and illustrate the usage of
\modifiedrate, consider the following reaction rate sequence from
CNO-II cycle and reduce it to a single effective rate:
\begin{equation}
    {}^{16}\mathrm{O} (p, \gamma) {}^{17}\mathrm{F} (, e^+ \nu) {}^{17}\mathrm{O} (p,\alpha) {}^{14}\mathrm{N}  
    \rightarrow {}^{16}\mathrm{O} (pp, \alpha e^+ \nu) {}^{14}\mathrm{N} \enskip .
\end{equation}
Assume that ${}^{16}\mathrm{O} (p,\gamma){} {}^{17}\mathrm{F}$ is the rate limiting step,
so that ${}^{17}\mathrm{F}$ is in a steady state, balanced by the the proton-capture rate and 
the subsequent $\beta^+$ decay.
If the reaction network does not explicitly carry ${}^{17}\mathrm{O}$, 
equivalently $X_{{}^{17}\mathrm{O}} \sim 0$,
then ${}^{17}\mathrm{O}$ can also be treated as being in a steady state.
With these assumptions, the rate equation for the above multi-step reaction
sequence reduces to

\begin{equation}
\label{eq:modifiedrate_ex}
\begin{aligned}
    \frac{dY_{{}^{16}\mathrm{O}}}{dt} &= \frac{1}{2}\frac{dY_{p}}{dt} = - \frac{dY_{{}^{4}\mathrm{He}}}{dt} = - \frac{dY_{{}^{14}\mathrm{N}}} {dt} \\
    &= - \rho Y_{{}^{16}\mathrm{O}} Y_p \underbrace{\lambda_{{}^{16}\mathrm{O} (pp, \alpha e^+ \nu) {}^{14}\mathrm{N}}}_{\lambda_{{}^{16}\mathrm{O} (p, \gamma) {}^{17}\mathrm{F}}} \enskip .
\end{aligned}
\end{equation}

The above rate equation can be interpreted as a single effective reaction,
whose reaction rate is taken to be the rate of the first reaction.
At first glance, Eq.~\ref{eq:modifiedrate_ex} does not have the proper density and
molar abundance dependency for a direct double-proton capture, which would normally scale as
$\propto \rho^2 Y_{{}^{16}\mathrm{O}} Y_p^2$. This is because 
Eq.~\ref{eq:modifiedrate_ex} represents the process of two {\it consecutive} proton captures
rather than a direct double-proton capture on ${}^{16}\mathrm{O}$.
As a result, the net consumption of species reflects the full sequence, 
but the rate equation still has the density and molar abundance dependency of the individual steps. 
This situation is well-suited for \modifiedrate and corresponds to the form of rate equation shown in Eq.~\ref{eq:molar_fraction_stoichiometry_ode}.
Note that this specific example is utilized in the classic {\tt aprox19} network.
To incorporate this rate approximation, we can modify the product of
${}^{16}\mathrm{O} (p, \gamma) {}^{17}\mathrm{F}$ to be $[{}^{4}\mathrm{He}, {}^{14}\mathrm{N}]$
and set the stoichiometric coefficient for proton to be 2, while keeping the reactants unchanged.
A sample code that does the above approximation is shown below

\begin{lstlisting}
rl = pyna.ReacLibLibrary()
O16p_gF17 = rl.get_rate_by_name("o16(p,g)f17")
stoichiometry = {pyna.Nucleus("p"): 2}
O16pp_aN14 = pyna.ModifiedRate(O16p_gF17,
                 new_reactants=["p", "o16"],
                 new_products=["n14", "he4"],
                 stoichiometry=stoichiometry)
\end{lstlisting}

\begin{figure}
    \centering
    \plotone{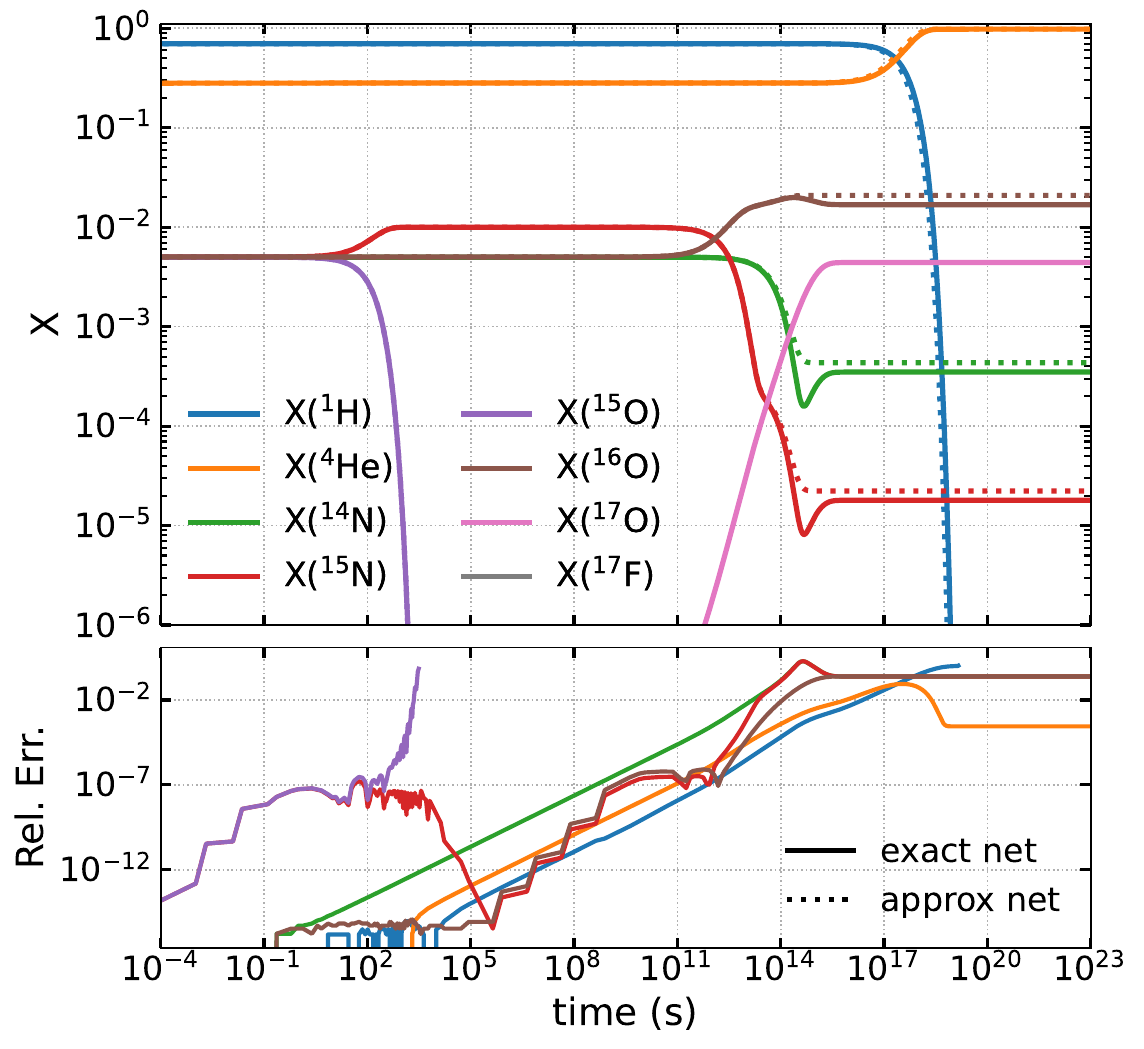}
    \caption{{\it Top}: Time evolution of the CNO-II cycle. The approximate
    network replaces the reaction sequence ${}^{16}\mathrm{O} (p, \gamma) {}^{17}\mathrm{F} (, e^+ \ \nu) {}^{17}\mathrm{O} (p,\alpha) {}^{14}\mathrm{N}$
    to ${}^{16}\mathrm{O} (pp, \alpha) {}^{14}\mathrm{N}$ using the \modifiedrate.
    {\it Bot}: Relative error in the mass fraction of nuclei common to both networks.
    Relative errors are shown when the mass fractions are above $10^{-10}$.
    \supnote{modified-rates.ipynb}}
    \label{fig:modified-rate-integrate}
\end{figure}

To demonstrate the above rate approximation, Figure~\ref{fig:modified-rate-integrate}
shows the time evolution of the CNO-II cycle under $\rho = 100\ \mathrm{g} \ \mathrm{cm}^{-3}$ and $T = 2\times 10^7$ K with a solar-like initial composition.
The nuclei removed by the \modifiedrate approximation, 
${}^{17}\mathrm{O}$ and ${}^{17}\mathrm{F}$,
are initialized with zero mass fraction so that both the exact and approximate
networks have the same initial conditions.
We see that the overall evolution is nearly identical until $t \sim 10^{12}$ s,
where the abundance of $\isotm{O}{17}$ begins to increase significantly.
This increase in the mass fraction indicates that the equilibrium assumption made
for the \modifiedrate approximation is no longer valid.
As a result, the final abundance of $\isotm{O}{17}$ in the exact network is
redistributed among the remaining species in the approximate network. 

The use of custom {\tt stoichiometry} is also not limited to integer stoichiometric
coefficients. For instance, the use of custom {\tt stoichiometry} allows for a ``3/2-$\alpha$'' capture,
like used in {\tt aprox19} or the MESA {\tt basic.net} network, modifying $\isotm{N}{14}(\alpha,\gamma)\isotm{F}{18}$ to
$\isotm{N}{14}(1.5\alpha,\gamma)\isotm{Ne}{20}$---this is useful for
bridging odd-Z and even-Z nuclei in small networks (where $Z$ is the
proton number).

It is important to note that reaction rates with a custom stoichiometry,
which the stoichiometric coefficient of any reactants or products
differs from the corresponding exponent in the molar abundance dependency {\it cannot} 
construct a physically meaningful corresponding inverse rate from detailed balance
and is therefore incompatible with the NSE description (also see Section~\ref{sec:nse}).
The detailed balance relation requires that the
ratio between the forward and reverse reaction rates
to be $\propto \exp{\left(-Q_{\mathrm{for}}/k_B T\right)}$,
where $Q_{\mathrm{for}}$ is the Q-value of the forward rate defined by Eq.~\ref{eq:q-value}.
See Appendix~\ref{appendix:modifiedrate} and \ref{appendix:nse} for more details.
Since detailed balance relation is derived from assuming that the multiplicities
of the reactants and products are given by the exponents in the molar abundance dependency,
the same multiplicities must be used when computing $Q_{\mathrm{for}}$ in Eq.~\ref{eq:q-value}.
However, if the custom stoichiometric coefficient differs from the molar abundance dependency,
then computing $Q_{\mathrm{for}}$ from the molar abundance dependency can
violate baryon conservation, which then has no physical meaning.
On the other hand, computing $Q_{\mathrm{for}}$ from the stoichiometric coefficient given by
the custom stoichiometry is physically meaningful, but is incompatible
with the true detailed-balance description.
This is the case for both ${}^{16}\mathrm{O} (pp, \alpha) {}^{14}\mathrm{N}$ and 
$\isotm{N}{14}(1.5\alpha,\gamma)\isotm{Ne}{20}$ shown earlier.
This behavior is expected since in these approximations,
the rate of creation and destruction of the
intermediate nuclei is assumed to be balanced between the successive forward reactions in the
sequence, whereas detailed balance describes the balance between the forward and reverse reaction.

\subsubsection{Branched rates}
\label{sec:branched_rate}

Consider the CNO-I cycle---we need 6 nuclei in addition to $p$ and $\alpha$ to represent
the burning.  We'd like to use a reduced approximation that still captures the effective
rate of burning.  Our approach reproduces how the {\tt basic.net} in {\sf MESA} works---that network is
used for simple main-sequence and He-burning phases of stellar evolution.

We start by writing the CNO cycle as two sequences:
\begin{subequations}
\begin{align}
\isotm{C}{12}(p,\gamma)\isotm{N}{13}(,e^+\nu)\isotm{C}{13}(p,\gamma)\isotm{N}{14} \label{eq:cno-c12} \enskip ,\\
\isotm{N}{14}(p,\gamma)\isotm{O}{15}(,e^+\nu)\isotm{N}{15}(p,\alpha)\isotm{C}{12} \enskip . \label{eq:cno-n14}
\end{align}
\end{subequations}
We can represent the first part of the sequence (Eq.~\ref{eq:cno-c12}) using
\modifiedrate, but for the
second part of the sequence (Eq.~\ref{eq:cno-n14}), we also need to consider the branching
$\isotm{N}{15}(p,\gamma)\isotm{O}{16}$ that starts the CNO-II cycle:
\begin{equation}
\isotm{N}{14}(p,\gamma)\isotm{O}{15}(,e^+\nu)\isotm{N}{15}(p,\gamma)\isotm{O}{16} \label{eq:cno-o16} \enskip .
\end{equation}
The \branchedrate class provides the necessary capability.  It takes
an underlying rate which acts as the rate-limiting step for the
sequence (as we'll see in a moment, the first rate,
$\isotm{N}{14}(p,\gamma)\isotm{O}{15}$ is the slowest).  It also takes
the two branches which define the end-point of the sequence.  In our
case, these are $\isotm{N}{15}(p,\alpha)\isotm{C}{12}$ and
$\isotm{N}{15}(p,\gamma)\isotm{O}{16}$.  Finally, we need to modify
the stoichiometry, since two protons are consumed in the sequence.
The effective rates for sequences \ref{eq:cno-n14} and \ref{eq:cno-o16} are then computed as:
\begin{align}
\lambda_{{}^{14}\mathrm{N}(pp,e^+\nu\alpha){}^{12}\mathrm{C}} &= \lambda_{{}^{14}\mathrm{N}(p,\gamma){}^{15}\mathrm{O}} \frac{\lambda_{{}^{15}\mathrm{N}(p,\alpha){}^{12}\mathrm{C}}}{\lambda_{{}^{15}\mathrm{N}(p,\alpha){}^{12}\mathrm{C}} + \lambda_{{}^{15}\mathrm{N}(p,\gamma){}^{16}\mathrm{O}}} \\
\lambda_{{}^{14}\mathrm{N}(pp,e^+\nu){}^{16}\mathrm{O}} &= \lambda_{{}^{14}\mathrm{N}(p,\gamma){}^{15}\mathrm{O}} \frac{\lambda_{{}^{15}\mathrm{N}(p,\gamma){}^{16}\mathrm{O}}}{\lambda_{{}^{15}\mathrm{N}(p,\alpha){}^{12}\mathrm{C}} + \lambda_{{}^{15}\mathrm{N}(p,\gamma){}^{16}\mathrm{O}}} \enskip .
\end{align}

\begin{figure}[t]
\centering
\plotone{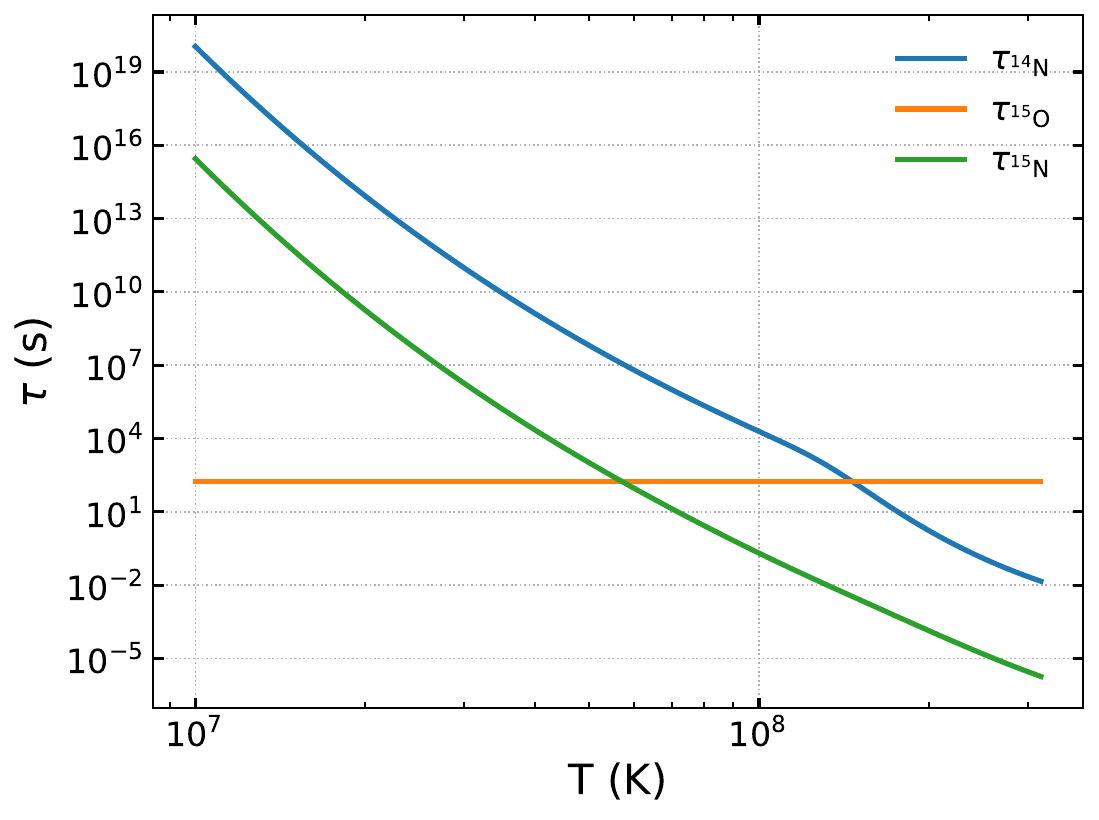}
\caption{\label{fig:cno-timescales} Timescales for destroying a
  nucleus for the 3 reactions in sequence Eq.~\ref{eq:cno-n14}.  These
  were computed using $\rho = 100~\gcc$ and a solar composition.  We
  see that the $\isotm{N}{14}$ destruction timescale is the slowest
  until about $2\times
  10^8~\mathrm{K}$. \supnote{cno-rate-comparison.ipynb}}
\end{figure}

We can construct these effective sequences using \branchedrate as:
\begin{lstlisting}
rl = pyna.ReacLibLibrary()
rn14pg = rl.get_rate_by_name("n14(p,g)o15")
rn15pa = rl.get_rate_by_name("n15(p,a)c12")
rn15pg = rl.get_rate_by_name("n15(p,g)o16")
stoichiometry = {pyna.Nucleus("p"): 2}

rn14_2p_c12 = pyna.BranchedRate(rn14pg,
                   primary_branch=rn15pa,
                   other_branch=rn15pg,
                   stoichiometry=stoichiometry)
rn14_2p_o16 = pyna.BranchedRate(rn14pg,
                   primary_branch=rn15pg,
                   other_branch=rn15pa,
                   stoichiometry=stoichiometry)
\end{lstlisting}
In this construction, the {\tt primary\_branch} is the branch that is taken
by the \branchedrate we are constructing.  To understand why we can
use the $\isotm{N}{14}(p,\gamma)\mathrm{O}{15}$ rate as the underlying / limiting
rate, we can look at the timescales of all the rates in the sequence Eq.~\ref{eq:cno-n14}.
We define the timescale as:
\begin{equation}
\tau_\xi = \frac{Y(\xi)}{|dY(\xi)/dt|} \enskip,
\end{equation}
where $dY(\xi)/dt$ is just the single-rate contribution to the
evolution of the molar fraction.  These timescales are shown in
Figure~\ref{fig:cno-timescales}.  We see that the first rate in the
sequence, $\isotm{N}{14}(p,\gamma)\isotm{O}{15}$, is the slowest rate
until about $2\times 10^8~\mathrm{K}$.  Above that temperature, the
CNO cycle transitions to the hot-CNO cycle and $\beta$-limiting of the
rate sequence would need to be done (this is something that is done in
the common {\tt aprox19} and {\tt aprox21} networks).  We will only
consider the cooler regime where regular CNO takes place here.

A similar timescale analysis shows that the $\isotm{C}{12}(p,\gamma)\isotm{N}{13}$ rate
is the limiter for sequence Eq.~\ref{eq:cno-n14} over about the same temperature range.
That sequence can be approximated then as:
\begin{lstlisting}
rc12pg = rl.get_rate_by_name("c12(p,g)n13")
stoichiometry = {pyna.Nucleus("p"): 2}
rc12_2p_n14 = pyna.ModifiedRate(rc12pg,
                  new_products=[Nucleus("n14")],
                  stoichiometry=stoichiometry)
\end{lstlisting}

Finally, to complete the cycle, we should consider the sequence:
\begin{equation}
\isotm{O}{16}(p,\gamma)\isotm{F}{17}(,e^+\nu)\isotm{O}{17}(p,\alpha)\isotm{N}{14} \enskip ,
\end{equation}
which we explored as a \modifiedrate in Section \ref{sec:modified_rates}.

\begin{figure}[t]
\centering
\plotone{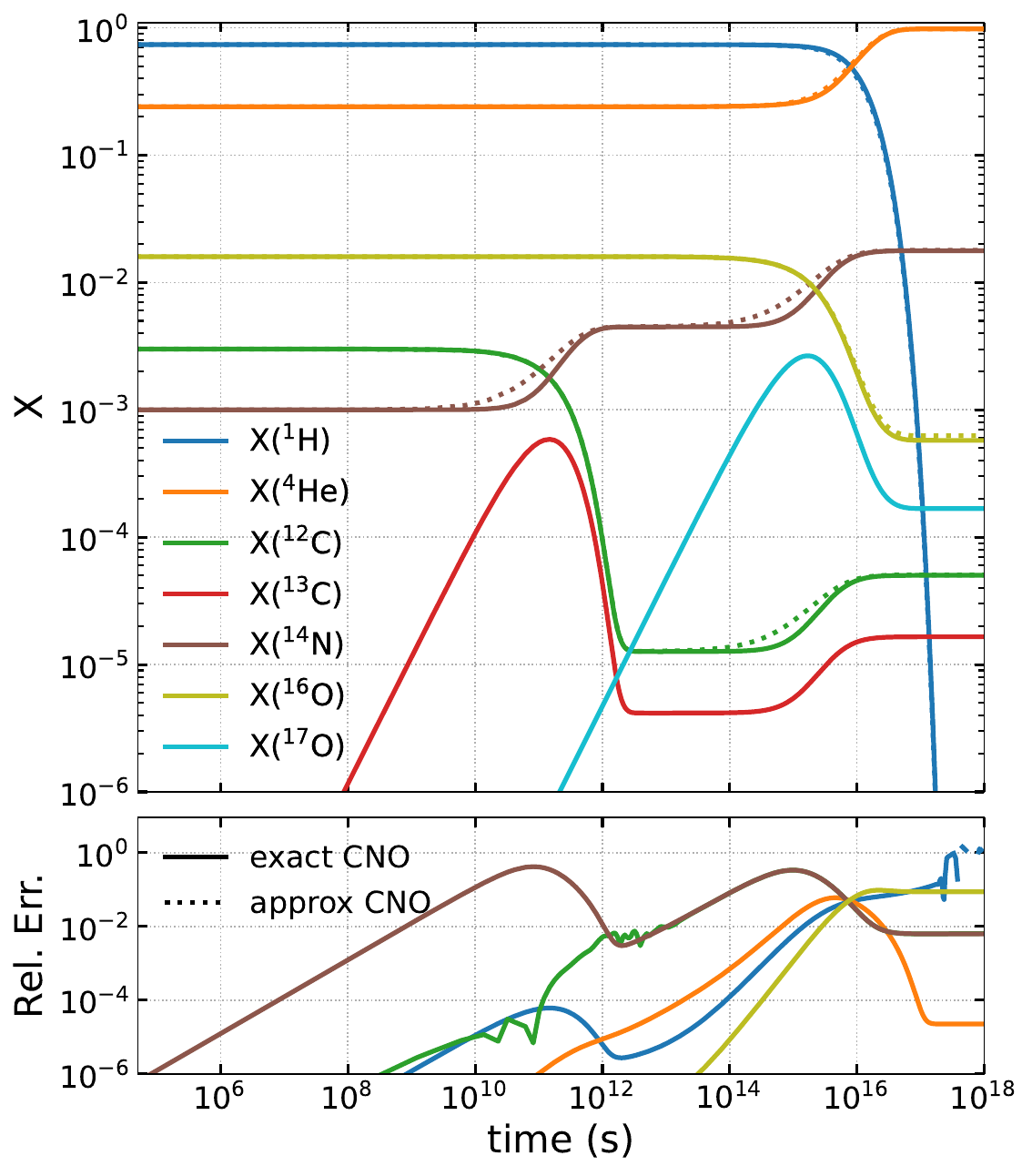}
\caption{\label{fig:cno-compare} Comparison of the full CNO network (solid lines) and reduced-CNO
approximate network (dotted lines) for a constant-$T$ burn with $\rho
= 100~\gcc$ and $T = 2\times 10^7~\mathrm{K}$.  
We see good agreement between the two networks indicated by
the relative error panel shown at the bottom. \supnote{branched-rates.ipynb}}
\end{figure}

Together, these four approximate rates model the CNO-I \& II cycle as:
\begin{align}
\isotm{C}{12} + 2 p &\rightarrow \isotm{N}{14} + e^+ + \nu \nonumber \\
\isotm{N}{14} + 2 p &\rightarrow \isotm{C}{12} + \isotm{He}{4} + e^+ + \nu \nonumber \\
\isotm{N}{14} + 2 p &\rightarrow \isotm{O}{16} + e^+ + \nu \nonumber \\
\isotm{O}{16} + 2 p &\rightarrow \isotm{N}{14} + \isotm{He}{4} + e^+ + \nu \enskip .\nonumber
\end{align}
We can test the accuracy of this reduced CNO network by comparing to a full / unapproximated
CNO network, constructed as:
\begin{lstlisting}
nuclei = ["p", "he4", "c12", "c13",
          "n13", "n14", "n15",
          "o15", "o16", "o17", "f17"],
net = pyna.network_helper(nuclei
                tabular_ordering=["ffn", "oda"],
                with_reverse=False)
\end{lstlisting}
Figure~\ref{fig:cno-compare} shows the comparison.  We see that the reduced CNO network
agrees really well with the full network with a relative error
ranging from $\sim 10^{-1}$--$10^{-5}$ at steady state.
Overall, the full network has 11 nuclei, while
the reduced network has only 5 (including $p$ and $\alpha$).  This approximation can
be built upon further by adding $\beta$-limiting to the underlying rates in the future
to broaden its applicability.

\subsection{New rates}

\subsubsection{StarLib}
\label{sec:starlib}

StarLib \citep{starlib} is a library of thermonuclear reaction rates and weak interaction rates.
Its distinguishing feature is that it provides rate probability densities at temperatures relevant to nucleosynthesis studies for each rate. 
These rate probability densities obey lognormal distributions and can thus be described through two parameters, $\mu$ and $\sigma$:
\begin{equation}
f(x;\mu,\sigma) = \frac{1}{\sigma \sqrt{2\pi}} 
                  \frac{1}{x} \exp\!\left(-\frac{(\ln x-\mu)^2}{2\sigma^2} \right),
                  \enskip 0 < x < \infty.
\end{equation}
The recommended value for a rate is the median of this
distribution, i.e., $x_{med} = e^{\mu}$, with a factor uncertainty of $f.u. = e^{\sigma}$. 
Thus, for each rate, StarLib's tabular format provides $x_{med}$ and $f.u.$ across a grid of temperature points ranging from 0.001 GK to 10 GK. 
\citet{starlib} recommends that rates be sampled as:
\begin{equation}
    x_{i} = e^{\mu+p_{i}\sigma} = x_{med}(f.u.)^{p_i},
\end{equation}
where $p_{i}$ is a standard normal deviate, and $i$ is an index over rates. Note that a single deviate applies to all tabulated data for a given rate.

In \pynucastro, rates that are tabulated as a function of temperature
are made available through \temperaturetabularrate.  An instance of
this class holds two lists: one for a temperature grid and the other
for corresponding rate values.  For smooth interpolation, both lists
are stored in log-space.  Rate evaluations bounded by a rate's
temperature grid employ monotone cubic Hermite interpolation.
Evaluations beyond the grid employ linear interpolation.  The
functionality of \temperaturetabularrate is expanded by \starlibrate,
a subclass capable of sampling its rate data.  An instance of
\starlibrate holds lists for log median rate values $(\ln{x_{med}})$,
log factor uncertainties $(\ln{f.u.})$, and temperature grid points as
immutable attributes.  It also holds another list for sampled rate
data initialized with log median rate values.  This attribute is
dynamically mutated whenever the rate is asked to be sampled.  The
sampling for individual instances of \starlibrate is handled by the
{\tt sample\_rates()} method.  The newly sampled rate data is then
used to inform future rate evaluations.  Note that
{\temperaturetabularrate} has applications in \pynucastro beyond
StarLib (see Section~\ref{sec:alternate-rates}).

The intended way to access StarLib rates in \pynucastro is via
\starliblibrary.  The constructor of \starliblibrary accepts an
integer seed as an optional argument.  If a seed is provided, then it
is used to initialize a {\tt numpy.random.Generator} to generate
Gaussian deviates and thus sample rates within that instance of
\starliblibrary \citep{numpy}.  If no seed is provided, then all rates
within the library are initialized with their rate data at median
values.  An instance of \starliblibrary may also be resampled via the
{\tt resample()} method, which too accepts an integer seed as an
optional argument.  If no seed is provided to {\tt resample()}, then a
random seed is used.  The {\tt unsample()} method restores the rate
data of all rates in an instance of \starliblibrary to their
respective median values.  The {\tt resample()} and {\tt unsample()}
methods are also available for \ratecollection.  For a \ratecollection, the methods only apply
to instances of \starlibrate present within the \ratecollection.  The following code demostrates an application of StarLib in \pynucastro by repeatedly sampling and integrating a {\tt PythonNetwork} (a subclass of \ratecollection): 
\begin{lstlisting}
# assuming rho, T, and comp are previously 
# initialized and net is a PythonNetwork
# with one or more instances of StarLibRate

tmax = 1.e20
nsamples = 25
sampled_sols = []
for seed in np.random.randint(0, 10000, 
                              size=nsamples):
    net.resample(seed=seed)
    sol = net.integrate_network(tmax, rho, T, 
                                comp,atol=1.e-8)
    sampled_sols.append(sol)
\end{lstlisting}

\begin{figure}
    \centering
    \plotone{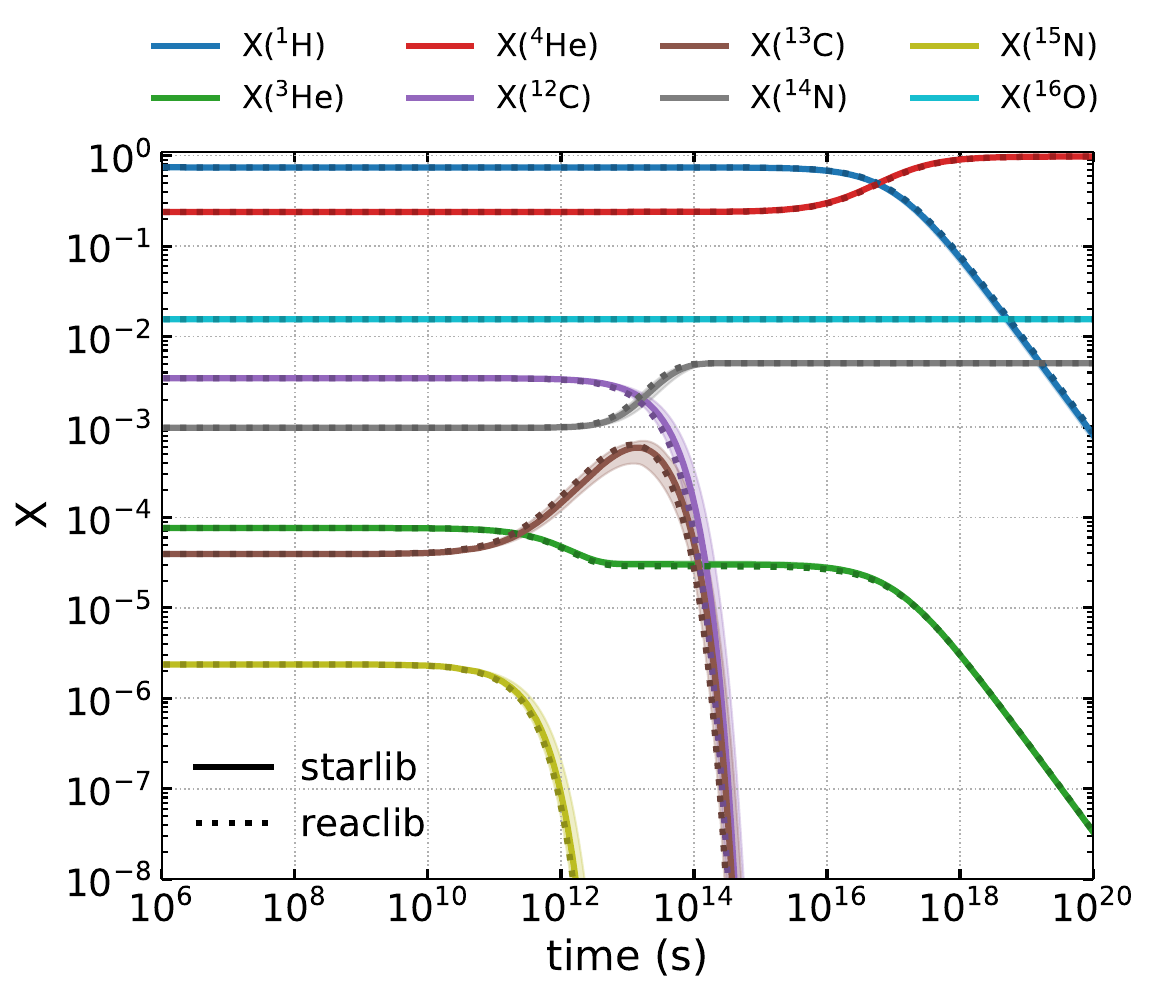}
    \caption{Integration of a H-burning network (pp chain + CNO) with StarLib rates (at median as well as sampled values) against that with Reaclib rates. Initial composition is set to be the Lodders solar composition, as described in Section \ref{sec:lodders}, binned into the nuclei listed in the legend. The solid lines depict integration with StarLib rates at median values and the darker dotted lines depict integration with Reaclib rates. The shaded regions encompass a total of 25 integration solutions each obtained with a unique sampling of StarLib rates. The boundaries of the shaded region represent the highest and the lowest value for each mass fraction across all solutions with sampled rates, evaluated at consistent time stamps. \supnote{starlib-rates.ipynb}}
    \label{fig:starlib}
\end{figure}

Figure~\ref{fig:starlib} is generated by employing this code on an H-burning (pp-chain + CNO) reaction network. 
The network is resampled and re-integrated $25+1$ times ($+1$ accounts for the median case).
Across all runs, initial composition is set to be the Lodders solar composition (see Section~\ref{sec:lodders}) binned into nuclei most relevant for H-burning.
The solutions with sampled rates are evaluated at consistent time stamps to obtain the spread in mass fractions over time as depicted in Figure~\ref{fig:starlib}.
For comparison, the integration of a reaction network with ReacLib rates is also presented. 
The effects of StarLib sampling seem most apparent for ${}^{12}\mathrm{C}$, ${}^{13}\mathrm{C}$, and ${}^{15}\mathrm{N}$. 
We also observe differences between the evolution of mass fractions as described by StarLib rates and that by ReacLib rates. 

Sampling of the StarLib rate uncertainties is also available in the
\amrexastrocxxnetwork exported networks, where the random seed can be
specified at runtime.  This allows for suites of full multidimensional
reacting hydrodynamics simulations that explore how rate uncertainties
affects nucleosynthesis and explosion mechanisms.

\subsubsection{Alternate rates}
\label{sec:alternate-rates}

Aside from major rate compilations such as ReacLib and StarLib, there are more
recent up-to-date rates that have not yet been included in these
collections. To incorporate these reaction rates, \pynucastro provides
a set of alternate rates that store these updated reaction rates from
various sources as an alternative to the existing rates from the
standard rate libraries.  There are currently two alternate rates
implemented and more rates can be added conveniently in the future.

\begin{enumerate}
    \item ${}^{12}\mathrm{C}(\alpha,\gamma){}^{16}\mathrm{O}$ reaction rate from \citet{deBoer:2017} given
    in the ReacLib format as \reaclibrate.
    \item ${}^{16}\mathrm{O}(p,\gamma){}^{17}\mathrm{F}$ reaction rate from \citet{iliadis:2022} given in
    temperature tabular form as \temperaturetabularrate.
\end{enumerate}
These two rates were chosen to demonstrate how
anyone in the community can add new rates to a \pynucastro by simply subclassing an existing rate class and passing in the new data.

\begin{figure}
    \centering
    \plotone{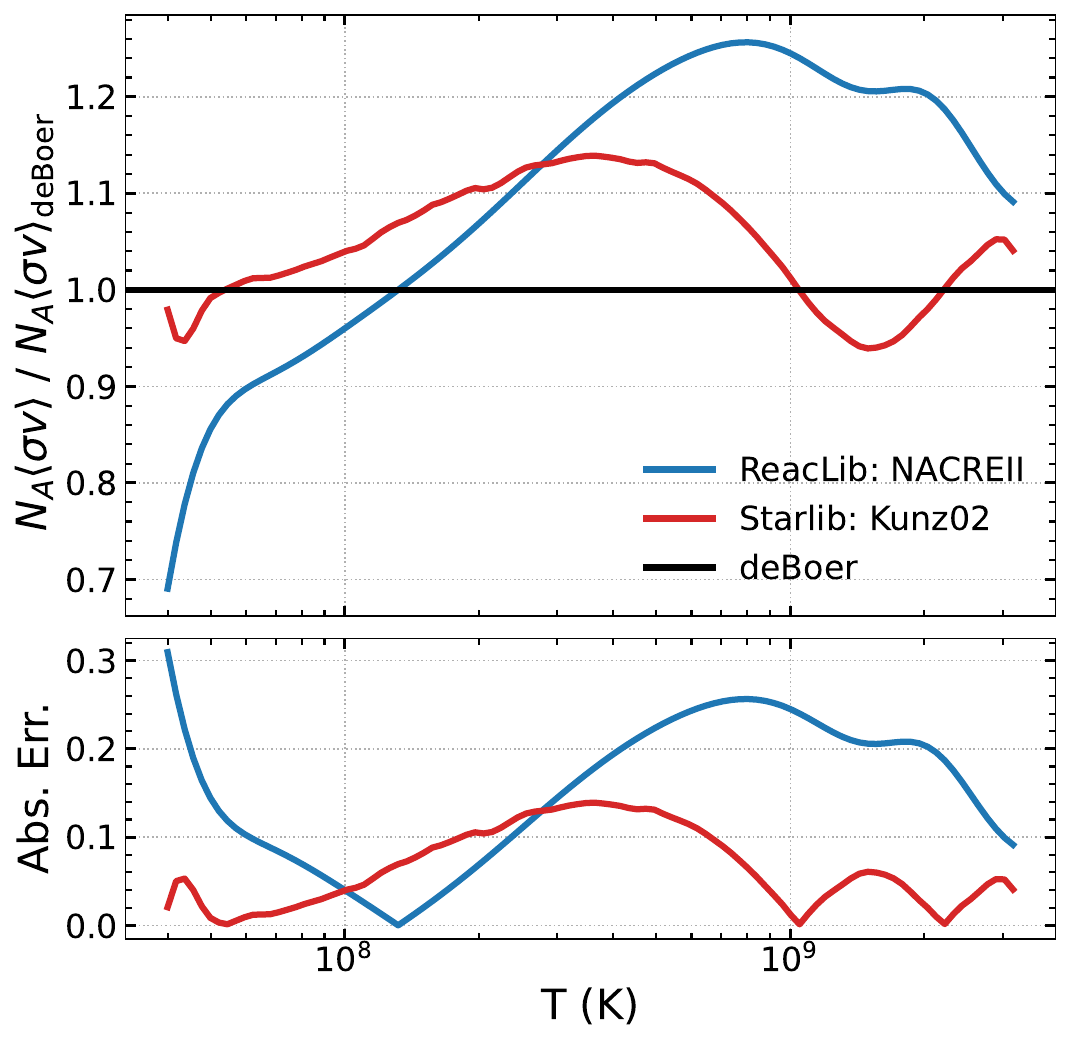}
    \caption{{\it Top}: ${}^{12}\mathrm{C}(\alpha,\gamma){}^{16}\mathrm{O}$ reaction rate
    from various sources normalized by the rate given in \citet{deBoer:2017}. 
    {\it Bot}: Absolute error of the normalized reaction rate relative to the \citet{deBoer:2017} rate. \supnote{alternate-rates.ipynb}}
    \label{fig:deboer}
\end{figure}

The ${}^{12}\mathrm{C}(\alpha,\gamma){}^{16}\mathrm{O}$ reaction is a
critical link in Helium burning that connects the triple-alpha process
to the rest of the $\alpha$-chain.  \citet{deBoer:2017} demonstrated
that the updated R-matrix calculations for
${}^{12}\mathrm{C}(\alpha,\gamma){}^{16}\mathrm{O}$ has significantly
smaller uncertainties at lower temperatures compared to the rate from
\citet{kunz_astrophysical_2002}, which was used in the StarLib
Library. It also showed that the results were similar compared to the
NACR2 rate \citep{xu_nacre_2013} computed from the potential model
(PM) approach, which was used in the most recent version of the
ReacLib Library.  Even though tabulated version of the rate is given
in Table XXV of \citet{deBoer:2017}, we implemented the rate following
the ReacLib parameterization given in Table XXVI.  A comparison of all
the available ${}^{12}\mathrm{C}(\alpha,\gamma){}^{16}\mathrm{O}$
rates in \pynucastro is given in Figure~\ref{fig:deboer}.

\begin{figure}
    \centering
    \plotone{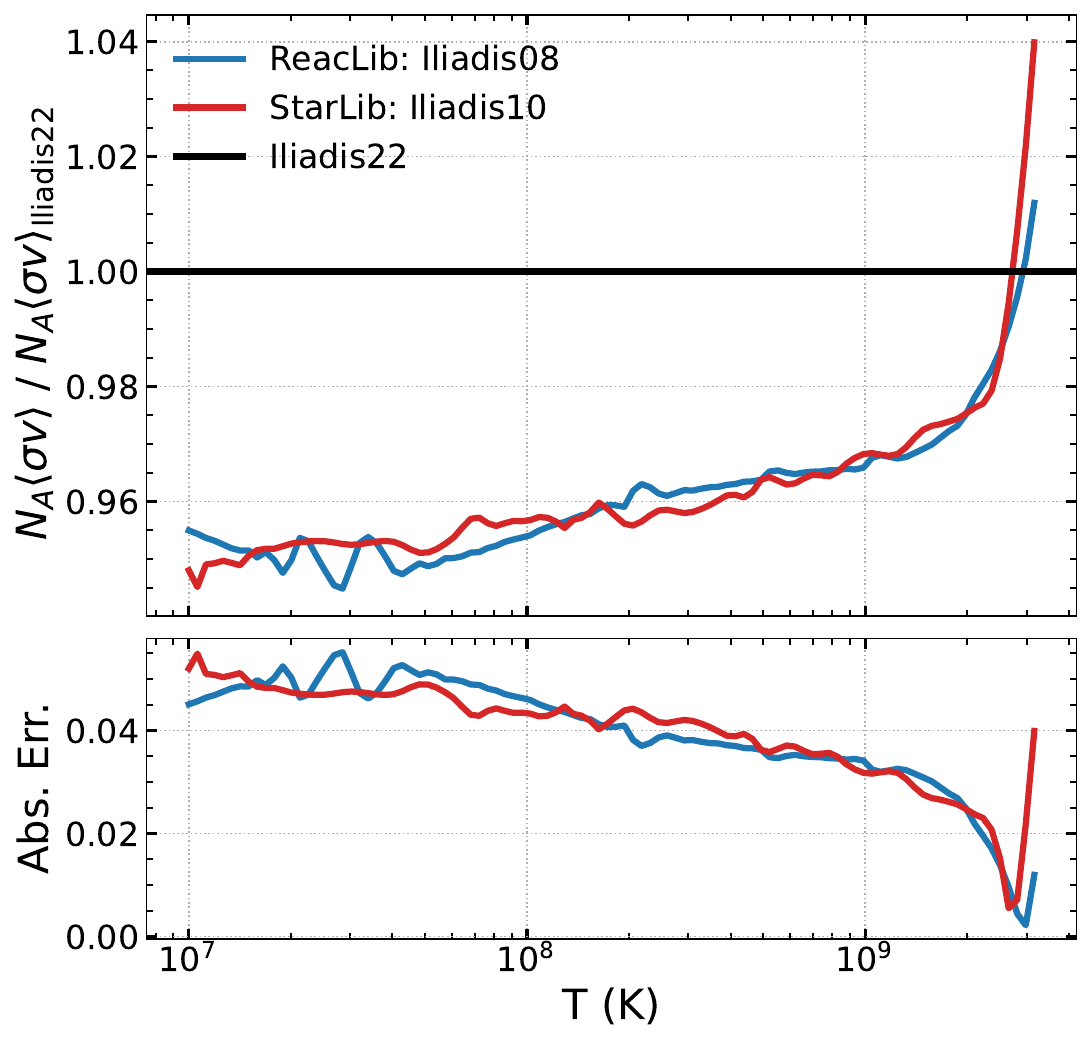}
    \caption{{\it Top}: ${}^{16}\mathrm{O}(p,\gamma){}^{17}\mathrm{F}$ reaction rate
    from various sources normalized by the rate given in \citet{iliadis:2022}. 
    {\it Bot}: Absolute error of the normalized reaction rate relative to the \citet{iliadis:2022} rate. \supnote{alternate-rates.ipynb}}
    \label{fig:iliadis}
\end{figure}

${}^{16}\mathrm{O}(p,\gamma){}^{17}\mathrm{F}$ plays an important role in hydrogen burning 
via the CNO-cycle. As we noted earlier in Section~\ref{sec:modified_rates},
this is the rate-limiting step for the rate sequence 
${}^{16}\mathrm{O} (p, \gamma) {}^{17}\mathrm{F} (, e^+ \nu) {}^{17}\mathrm{O}$.
This means that the sensitivity of this rate can severely impact the isotopic ratio
of ${}^{17}\mathrm{O}/{}^{16}\mathrm{O}$. 
\citet{iliadis:2022} provided an updated version of this reaction
based on the Markov chain Monte Carlo sampling.
This result showed nearly half the uncertainties compared to \citet{iliadis_2008},
which was used in the ReacLib library. 
This rate was implemented using the tabulated data for temperature and the median rate
through the use of \temperaturetabularrate.
A comparison of all the available ${}^{16}\mathrm{O}(p,\gamma){}^{17}\mathrm{F}$
rates in \pynucastro is given in Figure~\ref{fig:iliadis}.

\subsubsection{More weak rate support}
\label{sec:weaktab}

Reactions mediated by
weak interactions are responsible for the evolution of the electron fraction, $Y_e$, of the composition.
Changes in $Y_e$ directly affects the electron-positron EOS (especially in degenerate matter),
strength of the electron screening effect, and the composition in nuclear statistical equilibrium.
Moreover, since weak rates emit neutrinos and antineutrinos,
they act as an active cooling source as they are assumed to freely
escape from the stellar environment and carry away energy.
Due to these unique characteristics of weak reactions, they play an important role in
both nucleosynthesis and stellar evolution,
especially in the structure and evolution of the massive star prior to core collapse \citep{heger_2001},
and the convective Urca process in simmering white dwarfs \citep{martinez-rodriguez_neutronization_2016, piersanti_pre-explosive_2022}.

Since weak reactions have nonlinear dependence on $\rho Y_e$ and temperature,
theoretical weak rates have adopted a tabulation of the rates on the $(\rho Y_e,  T)$ grid \citep{ffn}.
In \citet{pynucastro2}, we've introduced the {\tt TabularRate} class to store the relevant rate data and
to evaluate the rate using a bilinear interpolation scheme on $(\log{\rho Y_e}, \ \log{T})$ grid.  
{\tt TabularRate} is now called \tabularrate
to better distinguish from \temperaturetabularrate.
In the latest version of \pynucastro, we have expanded the coverage of the 
theoretical weak rates to include:

\begin{enumerate}
    \item FFN rates from \citet{ffn}, covering sd-shell and fp-shell nuclei with $21 \leq A \leq 60$
    \item Oda rates from \citet{oda:1994}, covering sd-shell nuclei with $17 \leq A \leq 39$
    \item Langanke rates from \citet{langanke:2001} covering fp-shell nuclei with $45 \leq A \leq 65$ (added in \citealt{pynucastro2.1})
    \item Pruet \& Fuller rates from \citet{pruetfuller:2003}, covering fp-shell nuclei with $65 \leq A \leq 80$
    \item Suzuki rates from \citet{suzuki:2016}, covering sd-shell nuclei with $17 \leq A \leq 28$ (added in \citealt{pynucastro2})
\end{enumerate}

\begin{figure*}
    \centering
    \plotone{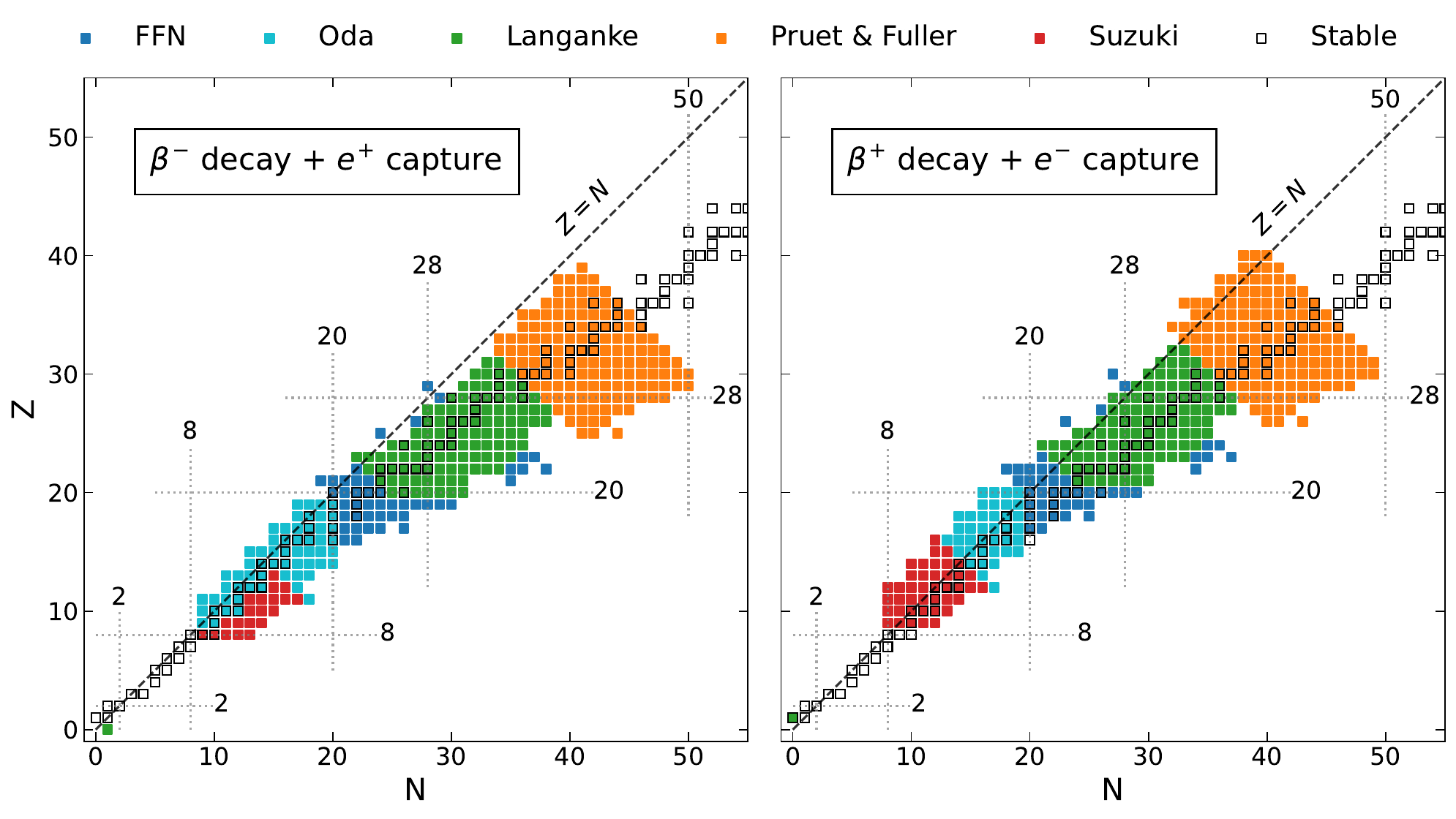}
    \caption{Tabular weak rate coverage map. Rates are overloaded by the 
    default priorities: FFN, Oda, Langanke, Pruet \& Fuller, and Suzuki.
    {\it Left}: $\beta^-$ decay and $e^+$ capture coverage map.
    {\it Right}: $\beta^+$ decay and $e^-$ capture coverage map. \supnote{weak-rates.ipynb}}
    \label{fig:weak-rate-coverage}
\end{figure*}

Because there are overlaps in the coverage between these sources, 
an order of precedence is required when loading the \tabularlibrary, which loads
all available tabular weak rates. The default ordering is listed above, with later entries
taking higher priority, i.e. Suzuki rates have the highest priority while FFN rates have the lowest.
The current tabular weak rate coverage of \pynucastro, following the default ordering,
can be visualized with the weak rate coverage map 
(similar to Fig.~10 shown in \citealt{winnet})
shown in Figure~\ref{fig:weak-rate-coverage}.

\subsection{Nuclear Statistical Equilibrium}
\label{sec:nse}

Nuclear statistical equilibrium (NSE) describes a state of global chemical equilibrium
among all species in the network established by nuclear reactions mediated by the strong nuclear force.
The molar abundance of the $i$-th species under NSE is given by
\begin{align}
\begin{split}
\label{eq:nse_eq}
    Y^{\mathrm{NSE}}_i =& \frac{1}{\rho}\frac{m_i}{A_i}(2J_i + 1)G_i \left(\frac{m_i k_B T}{2\pi \hbar^2}\right)^{3/2} \\
    &\times \exp{\Biggl(\frac{Z_i (\overbrace{\mu^{\mathrm{id}}_p + \mu^C_p}^{\bar{\mu}^{\mathrm{id}}_p})+ N_i \mu^{\mathrm{id}}_n + B_i - \mu^C_i}{k_B T}\Biggr)} \enskip ,
\end{split}
\end{align}
where $m_i$ is the mass of $i$-th species, $\mu^C_i$ is the coulomb
correction of the chemical potential, $B_i$ is the binding energy,
$J_i$ is the ground state spin, and $G_i$ is the internal nuclear
partition function describing the distribution of the particle
population at different internal energy levels.
Given the input triplet of $(\rho, T, Y_e)$, the only two unknowns in the equation are
$\mu^{\mathrm{id}}_p$ and $\mu^\mathrm{id}_n$, the kinetic portion of the chemical potential 
for proton and neutron, where $\mu_i^{\mathrm{id}} = \mu_i - m_i c^2 - \mu^C_i$.
Without any loss of generality, it is more convenient to solve for
$\bar{\mu}^{\mathrm{id}}_p$ (defined in Eq.~\ref{eq:nse_eq} as $\bar{\mu}^\mathrm{id}_p \equiv \mu^{\mathrm{id}}_p + \mu^C_p$) and $\mu^\mathrm{id}_n$ subject to
two constraints
%
\begin{subequations}
\label{eq:nse_constraint_eq}
\begin{align}
    \sum_i A_i Y_i - 1 &= 0 \label{eq:nse_constraint_eq1}\\
    \sum_i Z_i Y_i - Y_e &= 0 \label{eq:nse_constraint_eq2} \enskip.
\end{align}
\end{subequations}
Eq.~\ref{eq:nse_constraint_eq} represents the constraints on the baryon density, $\rho$,
and the electron fraction, $Y_e$, which both remain constant under strong interactions. 
Detailed derivation of Eq.~\ref{eq:nse_eq} can be found in various literature, e.g.\ \citet{pynucastro2, skynet}.
Here we just emphasize 3 important assumptions made when deriving Eq.~\ref{eq:nse_eq}.

\begin{enumerate}
    \item \label{nse_assump:1} Only strong-mediated nuclear reactions are considered such that the 
    electron fraction, $Y_e$, remains unchanged, i.e.\ Eq.~\ref{eq:nse_constraint_eq2} holds.
    
    \item \label{nse_assump:2} Ions are described by Maxwell-Boltzmann statistics so that the molar abundance of 
    the $i$-th species in the network is given by
    \begin{equation}
    \label{eq:mb_massfrac}
    \begin{aligned}
    Y^{\mathrm{MB}}_i =& \frac{1}{\rho}\frac{m_i}{A_i}(2J_i + 1)G_i \left(\frac{m_i k_B T}{2\pi \hbar^2}\right)^{3/2}\\
    &\times \exp{\left(\frac{\mu_i - m_i c^2 - \mu^C_i}{k_B T}\right)} \enskip ,
    \end{aligned}
    \end{equation}
    where $\mu_i$ is the chemical potential of species $i$.
    
    \item \label{nse_assump:3} Under local equilibrium, each strong reaction rate pair 
    imposes a a constraint on the chemical potentials of the reactants and the products.
    Assume that all the species are {\it sufficiently} connected via a series of strong reactions,
    then the chemical potential of the $i$-th species in the network, $\mu_i$, can be expressed in terms of the
    chemical potentials of proton and neutrons, $\mu_{\mathrm{p}}$ and $\mu_{\mathrm{n}}$, i.e.
    \begin{equation}
        \label{eq:nse_constraint}
    \mu_i = Z_i \mu_{\mathrm{p}} + N_i \mu_{\mathrm{n}} \enskip ,
    \end{equation}
    where $Z_i$ and $N_i$ are the atomic and neutron number of the $i$-th species.
    With Eq.~\ref{eq:nse_constraint}, a reaction network with $N$ number
    of distinct species, each characterized by its own chemical potential, is reduced to a system 
    described by only 2 independent chemical potentials, $\mu_{\mathrm{p}}$ and $\mu_{\mathrm{n}}$.
    Even though it is common to choose $\{\mu_{\mathrm{p}}, \mu_{\mathrm{n}}\}$ as 
    the basis to express $\mu_i$, a linear combination of any two independent chemical potentials can be used.
    Therefore, it is completely general to work with $\{\mu_{\mathrm{p}}, \mu_{\mathrm{n}}\}$ even if
    protons and/or neutrons are not explicitly included in the network.
    See Appendix B in \citet{skynet} for related discussions.
    
\end{enumerate}

In \citet{pynucastro2}, we introduced the 
{\tt get\_comp\_nse()} method that solves Eq.~\ref{eq:nse_eq}
using SciPy {\tt fsolve()} subject to Eq.~\ref{eq:nse_constraint_eq}
and validated the solutions against the results of \citet{Seitenzahl_2008}.
We also demonstrated a comparison between the mass abundances obtained from 
integrating the reaction network to the steady state and the solution computed from Eq.~\ref{eq:nse_eq}.
On average, we showed a relative error of $\sim 10^{-2}$--$10^{-3}$,
which are small enough to demonstrate that our overall implementation was correct.
However, since they do not agree to machine precision or to the tolerances of the
numerical integrator and the root-finding solver, it suggests minor systematic errors in our implementation.
To address this issue, the 3 assumptions used in deriving Eq.~\ref{eq:nse_eq}
must be incorporated into the reaction network itself.
Here we address the them individually.

\begin{enumerate}
    \item Assumption \ref{nse_assump:1} is trivially satisfied if weak rates are excluded from the network.
    If weak rates are included, then the electron fraction, $Y_e$, evolves according to 
    \begin{equation}
    \label{eq:ye_evolution}
        \dot{Y}_e = \sum_i Z_i \dot{Y}_i^{\mathrm{weak}} \enskip ,
    \end{equation}
    where $\dot{Y}_i^{\mathrm{weak}}$ is the rate of change of the molar abundance for the $i$-th species
    in the network due to weak rates.
    In this case, Eq.~\ref{eq:nse_eq} still describes the correct composition at steady state 
    for the instantaneous value of $Y_e$ provided that the the timescale for the strong reaction
    to establish NSE is significantly shorter than the timescale over which $Y_e$ evolves according to $\dot{Y}_e$.
    
    \item To address assumption \ref{nse_assump:2}, the strong reaction rates must be consistent with the 
    explicit form of the molar abundances derived under Maxwell-Boltzmann statistics, i.e. Eq.~\ref{eq:mb_massfrac}.
    This requirement can be enforced by ensuring that every strong reaction rate is paired with its
    inverse rate, where either the forward or the reverse must be derived from the other as a source
    rate using detailed balance. This procedure then explicitly uses the molar abundance expression
    described in Eq.~\ref{eq:mb_massfrac}.
    
    In \pynucastro, rates constructed via detailed balance are done via the \derivedrate class,
    which is first introduced in \citet{pynucastro2}.
    However, we found a slight inconsistency in the expression used for the molar abundance
    derived from Maxwell-Boltzmann statistic when constructing \derivedrate
    compared to what was used to derive the NSE abundance equation, Eq.~\ref{eq:nse_eq}.
    More details of this can be found in Appendix \ref{appendix:nse}.
    This inconsistency led to an average relative error of $\sim 10^{-2}$--$10^{-3}$ shown in \citet{pynucastro2}
    instead of a machine precision level of agreement when comparing the NSE solution to the abundances
    obtained from integrating the network to steady state.

    In the recent version of \pynucastro, this issue is fixed where \derivedrate is now constructed
    following Eq.~\ref{eq:equilibrium_ratio_full}. In addition, \derivedrate is now extended
    from only supporting \reaclibrate to all currently available rate types in \pynucastro 
    including \temperaturetabularrate and \starlibrate. However, rates with custom {\tt stoichiometry}
    are not supported (see Section \ref{sec:modified_rates} for detailed discussions).

    \item To address assumption \ref{nse_assump:3}, consider a general reaction network with
    $M$ number of strong reaction rate pairs and $N$ number of distinct species, 
    with a corresponding set of chemical potentials for each species, $\{\mu_j\}_{j=1}^N$.
    Consider a strong reaction rate pair following the generic form of Eq.~\ref{eq:rate_pair_form},
    then the net difference between the chemical potentials of the reactants and the products
    for the $i$-th reaction pair, $\delta_i$, can be written as
    \begin{equation}
        \sum_j S_{i,j} \mu_j = \delta_i \enskip ,
    \end{equation}
    where $\mathbf{S}$ is the stoichiometric matrix with shape $(M,N)$, and $S_{i,j}$ is
    the stoichiometric coefficient for the $j$-th species, participating in the $i$-th reaction
    pair with reactants and products having opposite signs. The overall sign of $\delta_i$ determines whether
    the net forward or reverse rate is favored.
    Global equilibrium is then reached when $\mu_j = \mu^{\mathrm{eq}}_j$ such that
    $\delta_i=0$ for all reaction pairs. In other words, $\mu^{\mathrm{eq}}$ is the kernel of $\mathbf{S}$ or 
    \begin{equation}
        \mu^{\mathrm{eq}} = \mathrm{Ker}(\mathbf{S}) \enskip .
    \end{equation}
    The number of NSE constraint equations then determines the maximum allowed dimension,
    $\mathcal{D}$, for the vector space spanned by the set of equilibrium chemical potentials, $\{\mu^{\mathrm{eq}}_j\}$.
    In general, since there are 2 possible constraint equations (Eq.~\ref{eq:nse_constraint_eq})
    on $\rho$ and $Y_e$, $\{\mu^{\mathrm{eq}}_j\}$ spans at most
    a two-dimensional vector space with a commonly chosen basis $\{\mu_p,\mu_n\}$.
    However, if all species in the network have the same $Z/A$ ratio,
    e.g.\ $\alpha$-chain network, then the constraint on $Y_e$ (Eq.~\ref{eq:nse_constraint_eq2})
    becomes redundant since $Y_e = 0.5$ always. 
    In this case, $\mathcal{D}$ reduces to 1.
    From a physical standpoint, $\{\mu^{\mathrm{eq}}_j\}$ for an $\alpha$-chain network
    can be written in terms of the chemical potential of ${}^{4}\mathrm{He}$.
    With the Rank-nullity theorem \citep{friedberg_linear_2014}, generally we have
    \begin{equation}
    \label{eq:nse-connection}
        \underbrace{\overbrace{\dim{(\mu)}}^{N} - \mathrm{Rank}(\mathbf{S})}_{\dim(\mu^{\mathrm{eq}})} = \mathcal{D} \enskip .
    \end{equation}
\end{enumerate}

The requirement of a {\it sufficiently connected} network is generally trivial for
large networks with hundreds of nuclei, which are commonly used for
one- or multi-zone simulations. However, constraint Eq.~\ref{eq:nse-connection} is important in expensive
multi-dimensional simulations, where the networks use less than few tens of nuclei
while still requiring evolution of the network under NSE description in extreme thermodynamic conditions.
To address this, we've implemented the {\tt is\_NSE\_Compatible()} method in 
\ratecollection, which determines the intrinsic compatibility of the network to NSE
by checking every rate in the network has a corresponding inverse rate
computed from detailed balance and that the nuclei are {\it sufficiently connected}
following Eq.~\ref{eq:nse-connection}.

\begin{figure}[t]
    \centering
    \plotone{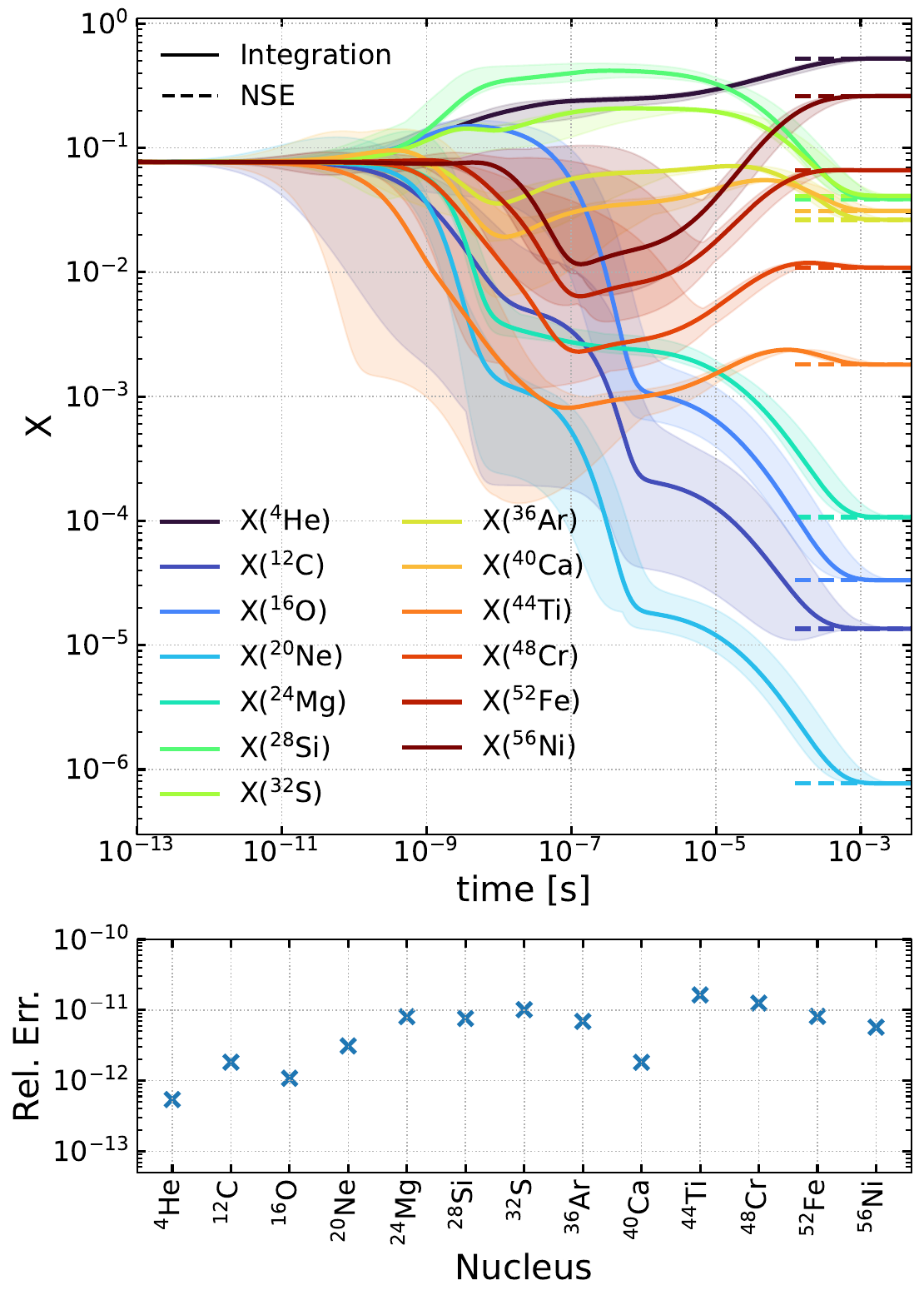}
    \caption{{\it Top}: Time integration of {\tt aprox13} network to steady state at
    $\rho=10^7~\gcc$ and $T = 6\times 10^9~\mathrm{K}$ with StarLib rates.
    Solid lines show the results obtained using the median StarLib rates,
    while the shaded regions span the range of the mass fractions
    obtained from 15 different random samples of the \starliblibrary.
    The horizontal dashed lines show the equilibrium abundances obtained
    by solving Eq.~\ref{eq:nse_eq}.
    {\it Bot}: Relative error between the final steady-state from obtained
    from network integration and those obtained by solving the Eq.~\ref{eq:nse_eq}.
    \supnote{nse.ipynb}}
    \label{fig:nse}
\end{figure}

To demonstrate the updated NSE module,
Figure~\ref{fig:nse} shows the time evolution of the {\tt aprox13} network at
$\rho=10^7~\gcc$ and $T = 6\times 10^9~\mathrm{K}$, 
starting with an uniform initial composition.
Unlike Figure~\ref{fig:aprox-integration}, this example uses StarLib rates together
with 15 different random samples from the \starliblibrary instead of ReacLib rates.
As a result, the network follows a wide spread of nucleosynthesis pathways during the non-equilibrium
evolution indicated by the shaded regions.
Nonetheless, they all converge to the same equilibrium composition after $t \sim 10^{-3}~\mathrm{s}$.
This is expected since the ratio between the forward and reverse reaction rates is fixed
by detailed-balance and only depends on the nuclear properties of the reactants and products.
The dashed lines show the theoretical equilibrium abundance obtained by
solving Eq.~\ref{eq:nse_eq}, while the bottom panel shows the relative error
between the final steady-state abundances from integration and the corresponding NSE solution.
On average, both solutions agree up to $\sim 10^{-11}$--$10^{-12}$,
where the relative tolerance used for integrating the network and solving the NSE equations (Eq.~\ref{eq:nse_eq})
is $10^{-12}$.

\subsection{Generating NSE Tables}
\label{sec:nse-table}

When a region in a simulation enters NSE, directly enforcing the
conditions of NSE and computing the composition from
Eq.~\ref{eq:nse_eq} is often more reliable and less expensive than
computing the change due to reaction rates (where the forward and
backward strong rates need cancel to machine precision). It can even
be made more accurate by computing an NSE state from many more nuclei
($\mathcal{O}(100)$) than are advected in the simulation.  In this
way, a simulation code can choose to carry a reduced, representative
reaction network for most of the domain, to avoid the heavy computational penalty of reacting
many nuclei together in an implicit solver.  The state from
the larger collection of nuclei can be used in the regions in NSE,
and can provide the weak-rate evolution via
$\dot{Y_e}$, resulting in an accurate determination of nucleosynthesis
and the electron fraction evolution.

The computational burden of computing NSE can further be vastly
reduced in-simulation by precomputing and tabulating the NSE states.
This has been done, for example, in the Type Ia supernova simulations by \citet{townsley:2007,seitenzahl:2009,ma:2013}.
\pynucastro provides the capability to create an NSE table with \nsenetwork,
and makes it easy to utilize extensive nuclear data for a large and 
accurate basis of the computation. 

An NSE table is computed over a user-specified range of temperature, density, and electron fraction, $Y_e$. The temperature and density ranges should not exceed the valid domain of the weak rate tables, typically:
\begin{subequations}
\begin{align}
    1 \leq \log_{10}\left(\rho Y_e\right) \leq 11 \\
    7 \leq \log_{10}\left(T\right) \leq 11 \enskip .
\end{align}
\end{subequations}
The rates for low-mass nuclei included from \citet{oda:1994} or
\citet{suzuki:2016} have smaller limits, but these rates are not
typically needed for NSE.

The precomputation of the NSE states is done as follows. Beginning
with an initial guess for the chemical potentials of protons and
neutrons, $\mu_p=-3.5$ and $\mu_n=-15$, the constraint equations
\ref{eq:nse_constraint_eq1} and \ref{eq:nse_constraint_eq2} are solved
iteratively with SciPy {\tt fsolve()} until we reach an error of
$10^{-11}$. The composition is calculated via
Eq.~\ref{eq:mb_massfrac}. The change in composition, $\dot{\bf Y}$, and
loss of energy via neutrinos, $\epsilon_{\nu,\mathrm{weak}}$, arising
from weak rates are recorded. We then compute
the mean molecular weight and its time-derivative:
\begin{subequations}
    \begin{align}
        \bar{A} &= \left (\sum_i Y_i\right)^{-1}\\
        \frac{d}{dt} \bar{A} &= -\bar{A}^2 \sum_i \dot{Y}_i \enskip ,
    \end{align}
\end{subequations}
the average
binding energy per nucleon, $\langle{B}/{A}\rangle$ and its time-derivative,
\begin{subequations}
    \begin{align}
        \left\langle\frac{B}{A}\right\rangle &= \sum_i B_i Y_i \\
        \frac{d}{dt}\left\langle\frac{B}{A}\right\rangle &= \sum_i B_i\dot{Y}_i \enskip ,
    \end{align}
\end{subequations}
and $\dot{Y_e}$,
\begin{equation}
        \dot{Y_e} = \sum_i Z_i \dot{Y}_i \enskip .
\end{equation}
These are output to a table alongside $\epsilon_{\nu,\mathrm{weak}}$ and optionally, the composition of the NSE state. These quantities can be used in calculations with the equation of state and energy generation rates (see, e.g., \citealt{Zingale:2024b}).

Tabulated compositions are useful for simulating regions that transition out of NSE. As such, the most useful set of nuclei to use as a basis for the composition is the set that matches the tracked nuclei outside of NSE regions. When constructing an NSE table with \pynucastro, you may specify a reduced set of nuclei to bin the composition into and subsequently output as part of the table. 

While the reduced set of tracked nuclei is often chosen in application codes to strike a balance between computational cost and accuracy, the full set of nuclei used to tabulate NSE has no in-simulation relation between computational cost and size of network. Thus, best practice is to construct the table using the largest possible reaction network, constrained only by available nuclear data. 

The NSE table will only accurately describe an evolution under nuclear statistical equilibrium conditions. Assumption 1 of Eq.~\ref{eq:mb_massfrac} is satisfied if matter is at sufficient density and temperature that the strong nuclear reactions occur on a much quicker scale than the weak reactions. In this case we may treat $Y_e$ as instantaneously constant, and calculate its change over a longer timescale than that of strong reactions. The choice of sufficient density and temperature is the responsibility of application codes. Assumption 2 is satisfied in the construction of reaction networks. Assumption 3 is satisfied for a network that {\tt is\_NSE\_Compatible()}, for which it is the user's responsibility to construct a sufficiently large network. 

\nsenetwork\ can be used to create tables similar to those in the literature, e.g., \citet{seitenzahl:2009}.  In particular, it is easy to create updated versions of these tables, with new and expanded mass measurements, spins, partition functions, and weak rates when new data is published. \pynucastro notably features nuclear masses and spins from NuBase 2020 \citep{nubase:2020}, and will continue to implement new nuclear data as it is published.

A table generated in this fashion was used for the evolution of
massive stars with the core in NSE in \cite{Zingale:2024b}. The table
was generated from data about 96 nuclei and the composition was binned
down to correspond to a 19-isotope network. The table was invoked
under conditions of $\rho>10^7~\gcc$, $T>3\times10^9~\mathrm{K}$,
sufficiently heavy average nuclei, and sufficiently low concentration
of silicon. Since then, more accurate tables using over 200 nuclei
have been generated using \pynucastro.


\subsection{New tools}

\subsubsection{Exporting to \networkx}

\pynucastro uses \networkx \citep{networkx} to make plots of the
network, treating the nuclei as nodes and the rates as the edges in a graph.
Beyond visualizations, \networkx has a large variety of
graph theory algorithms that can be applied to a \pynucastro network.
As part of the construction of a plot, a \networkx\ {\tt MultiDiGraph} is created.
This graph object can be exported from
the network via the {\tt create\_network\_graph} method.  This takes
all the same options as the plotting methods, allowing for filtering
of rates, nuclei, and more.

\begin{figure*}[t]
\centering
\plottwo{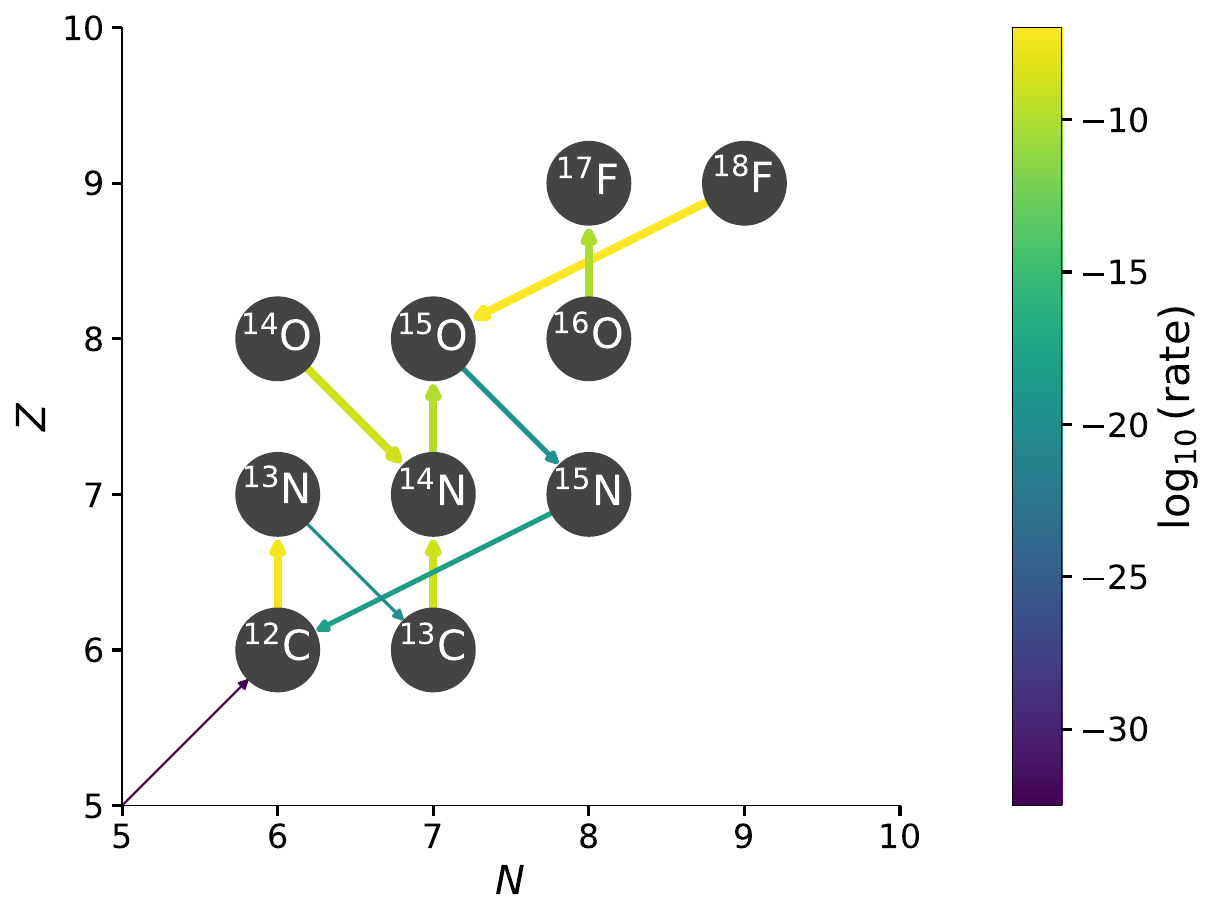}{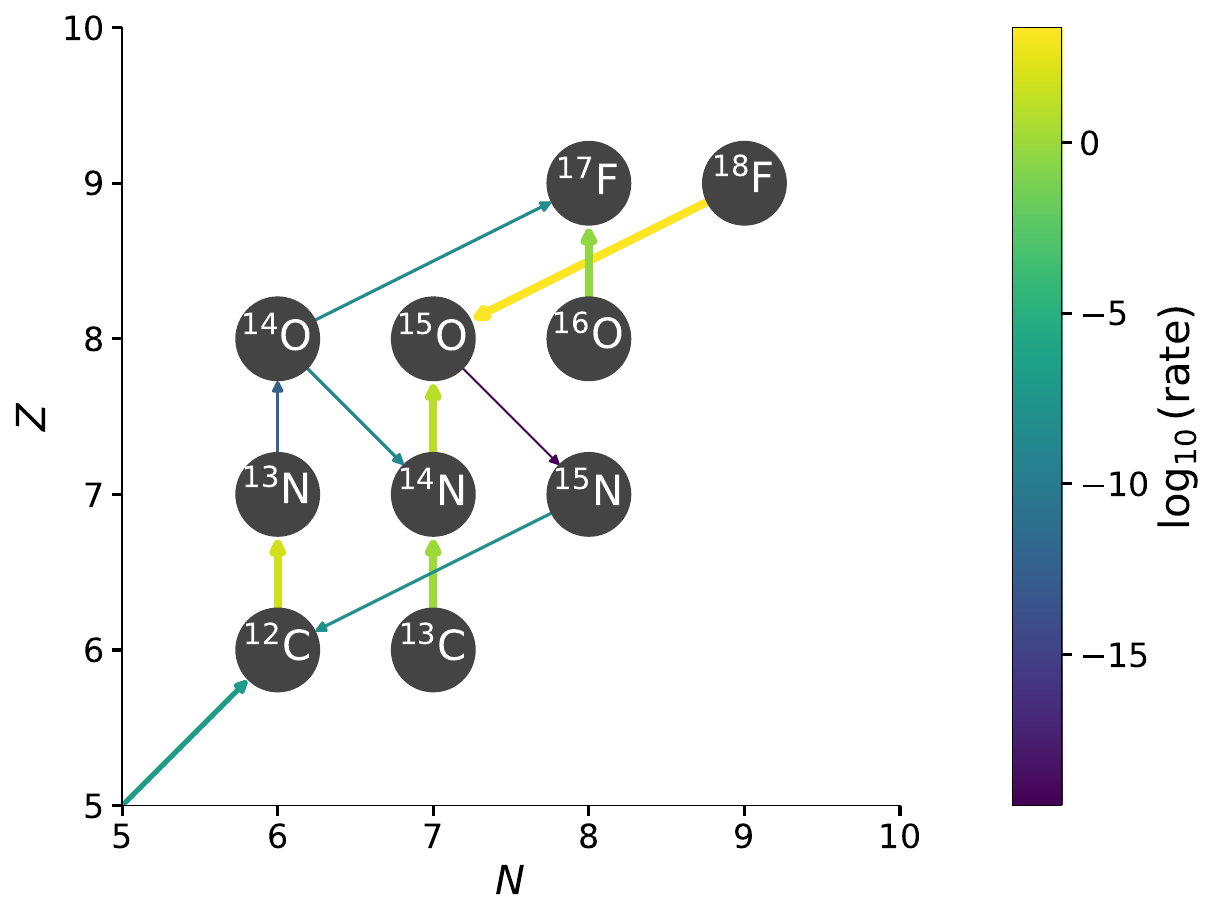}
\caption{\label{fig:networkx} A CNO network showing branching.  For
  each nucleus, only the rates within 10\% of the fastest rate are
  shown.  For both panels, the density is $10^4~\gcc$ and the
  composition of solar.  On the left, the temperature is $5\times
  10^7~\mathrm{K}$ and we see the traditional CNO cycle connecting the
  nuclei.  For the right panel, the temperature is $5\times
  10^8~\mathrm{K}$, and now we see the hot-CNO cycle. \supnote{network-cycles.ipynb}}
\end{figure*}

As an example of using \networkx to analyze a network, we can look at
CNO burning.  We start by creating a simple CNO network with some
breakout nuclei like Fluorine.  We'll ignore reverse rates for this:
\begin{lstlisting}
nuclei = ["p", "he4",
          "c12", "c13",
          "n13", "n14", "n15",
          "o14", "o15", "o16",
          "f17", "f18"]
rl = pyna.ReacLibLibrary()
lib = rl.linking_nuclei(nuclei,
                        with_reverse=False)
rc = pyna.RateCollection(libraries=lib)
\end{lstlisting}
Now we'll create a thermodynamic state.  To begin, we'll choose
a temperature of $5\times 10^7~\mathrm{K}$, density of $10^4~\gcc$, and solar composition.
\begin{lstlisting}
comp = pyna.Composition(rc.unique_nuclei,
                        init="solar")
state = pyna.ThermoState(rho=1e4, T=5e7,
                         comp=comp)
\end{lstlisting}
The left panel of Figure~\ref{fig:networkx} shows the network with
the rates evaluated for these conditions.  We use the {\tt
  consuming\_rate\_threshold} option to limit which rates are added to
the graph.  Setting it to {\tt 0.1} means that only those rates that
are within 10\% of the fastest rate consuming a nucleus are added to
the graph.

Now we export this to a \networkx graph:
\begin{lstlisting}
rates = rc.evaluate_rates(state)
G = rc.create_network_graph(
       rc.unique_nuclei,
       rate_ydots=rates,
       consuming_rate_threshold=0.1)
\end{lstlisting}
The weights on the edges of this graph are set according to the reaction rate evaluation for our
thermodynamic state.  The \networkx function {\tt simple\_cycles} can
find cycles in our reaction network by operating on the graph:
\begin{lstlisting}
import networkx as nx
cycles = nx.simple_cycles(G)
for n, c in enumerate(cycles):
    print(n, c)
\end{lstlisting}
This returns simply
\begin{verbatim}
0 [C12, N13, C13, N14, O15, N15]
\end{verbatim}
Here the {\tt 0} is the cycle index---it finds only a single cycle.  The
progression of nuclei in this case is the standard CNO cycle.  \networkx
does this with no knowledge of astrophysics,  just by evaluating the
weights assigned to the edges.  

The right panel of Figure~\ref{fig:networkx} uses a higher temperature,
$5\times 10^8~\mathrm{K}$, and now we see visually that for $\isotm{N}{13}$,
the proton-capture dominates and the $\beta$-decay is not included (for our threshold).
Reevaluating the rates with this temperature, exporting the graph, and finding
the cycles with {\tt simple\_cycles} now gives:
\begin{verbatim}
0 [C12, N13, O14, N14, O15, N15]
\end{verbatim}
which is the hot-CNO cycle.  We see this difference visually in the plots
in Figure~\ref{fig:networkx}---at the cooler temperature, we have the
$\beta$-decay of $\isotm{N}{13}$, while at the hotter temperature, we
have a proton-capturing onto $\isotm{N}{13}$.

Beyond just cycles, \networkx has a large suite of functions that can
assess the complexity of a graph.  By exporting a reaction network to
a \networkx graph object, it becomes easy to explore reaction networks
from a graph-theory perspective.  It also
opens the door to using graph methods for
network reduction \citep{hashemi:2024}.

\subsubsection{Stiffness Assessment}

Nuclear reaction networks governed by Eq.~\ref{eq:reaction_ydot} are
generally unsuitable for explicit time integrators, such as the
Forward Euler method.  The standard numerical stability criteria for
Forward Euler method \citep{leveque_finite_2007} requires that
\begin{equation}
\label{eq:forward_euler_stability}
\Delta t \leq \frac{2}{\max{(|\lambda_i|)}} \enskip ,
\end{equation}
where $\max{(|\lambda_i|)}$, also known as the {\it spectral radius},
is the largest eigenvalue in magnitude of the Jacobian of the
reaction network, Eq.~\ref{eq:net_jacobian}, 
under the assumption that the eigenvalues are real and negative.
The stability criteria,
Eq.~\ref{eq:forward_euler_stability}, sets the upper limit on the
timestep, $\Delta t$, for the explicit time integrator, which only
depends on the fastest varying timescale of the system, $\sim
1/\max{|\lambda_i|}$.  To characterize the range of the timescales
present in the system, it is convenient to define the stiffness ratio
\begin{equation}
\label{eq:stiff_ratio}
S \equiv \frac{\max{(|\lambda_i|)}}{\min{(|\lambda_i|)}} \enskip ,
\end{equation}
where a system is considered to be stiff when $S \gg 1$.  Due to the
highly nonlinear dependence on temperature for the reaction rates and
the wide range of species abundances in the nuclear reaction network
\citep{timmes_integration_1999}, it is not uncommon to see $S
>10^{15}$ \citep{hixmeyer} in stellar environments.  In order to avoid
taking extremely small timestep to ensure stability with explicit
methods, nuclear reaction networks are better suited with implicit
time integrator, like Backward Euler method, which is unconditionally
stable \citep{leveque_finite_2007} or higher order methods, like the
variable order VODE solver \citep{vode}.  Stabilized-explicit methods,
like Runge-Kutta-Chebyshev \citep{rkc-parker,rkc} have also been shown
to work well with nuclear reaction networks in some instances.
Regardless of the integrator, being able to assess the stiffness of
the network is an important diagnostic.

In \pynucastro, we've included convenient methods for evaluating
spectral radius of the network, utilizing
\scipy's linear algebra routines to find the eigenvalues of the Jacobian.
Furthermore, the method {\tt find\_stiffest\_rate} can automate the process of
finding which reaction rate is most responsible for the stiff-nature of the
network by removing one rate at a time from the network and recomputing the spectral radius.
As an example, consider an H-burning network for classical nova.  We can build the network as:
\begin{lstlisting}
all_nuclei = ["p", "h2", "he3", "he4",
              "li6", "li7", "be7", "be8", "b8",
              "c12", "c13", "n13", "n14", "n15",
              "o14", "o15", "o16", "o17", "o18",
              "f17", "f18", "f19", 
              "ne18", "ne19", "ne20", "ne21"]
              
net = pyna.network_helper(all_nuclei,
      tabular_ordering=["ffn", "oda"])
\end{lstlisting}
and then get the spectral radius, by first creating a thermodynamic state and
then calling {\tt spectral\_radius}:
\begin{lstlisting}
>>> comp = pyna.Composition(net.unique_nuclei,
                            init="solar")
>>> state = pyna.ThermoState(rho=1.7e3, T=7e7,
                             comp=comp)
>>> sprad = net.spectral_radius(state)
>>> sprad
np.float64(366628.7133838029)
\end{lstlisting}
For this network, it yields a spectral radius of $3.66\times 10^5$.
We can next ask if there is a rate that, when removed, would reduce this significantly, making our network less stiff, and the integration easier:
\begin{lstlisting}[mathescape=true]
>>> r = net.find_stiffest_rate(state)
>>> r
Li6 + p $\rightarrow$ He4 + He3
\end{lstlisting}
This returns the rate $\isotm{Li}{6}(p,\isotm{He}{3})\isotm{He}{4}$.  We can
then measure how much influence this one rate has on the stiffness of the
network by recomputing the spectral radius with it removed:
\begin{lstlisting}
>>> sprad = net.spectral_radius(state,
                    exclude_rates=[r])
>>> sprad
np.float64(7523.704931433871)
\end{lstlisting}
This gives the new spectral radius as $7.5\times 10^3$---almost a factor of 50 smaller.
This information can then be used with other methods in \pynucastro to decide
if this rate is needed, for example, by evaluating the various rates that produce
and consume $\isotm{Li}{6}$.   If it can be removed from a simulation, then the network's
stiffness can be decreased dramatically, potentially allowing explicit integrators
like Runge-Kutta-Chebyshev \citep{rkc} to be used.


\subsection{Exporting networks}

\cxx output for networks was originally added to support the
AMReX-Astrophysics suite of simulation codes \citep{amrex-astro}, with
the networks written in a fashion that supported GPU offloading.
Since then, a simpler \cxx network type was created, \simplecxxnetwork, to support
other simulation codes.  Both of these use the \cxx{}20 standard.  Finally, a set of Fortran wrappers (provided by \fortrannetwork) was added to interface to \simplecxxnetwork.
Here we discuss some of the features of these network types.

\subsubsection{GPU performance of \amrexastrocxxnetwork}

An important capability of \pynucastro\ is outputting networks for the
\amrex-Astrophysics suite of simulation codes \citep{amrex-astro}.
These networks are managed by the \microphysics\ project
\citep{Microphysics,microphysics-joss}, which provides ODE integrators (including VODE,
\citealt{vode}, Runge-Kutta-Chebyshev, \citealt{rkc}, and a 5th order
stiffly-accurate Rosenbrock solver, \citealt{rodas5p}), that work with
our simulation codes both via operator splitting
\citep{strang:1968,strang_rnaas} and spectral deferred correction
\citep{minion:2003, castro-sdc, castro-simple-sdc} coupling methods.

The \amrex\ library \citep{amrex} is an adaptive mesh refinement
framework that decomposes the computational domain into a nested
hierarchy of boxes containing the state data describing a simulation.
Importantly, \amrex\ provides a \cxx lambda-capturing mechanism for
launching GPU kernels \citep{amrex-gpu}.  In both
\castro\ \citep{castro} and \maestroex \citep{maestroex}, the data is
allocated on the GPU at the start of a simulation and kept there for
the duration of the simulation.  All physics, including the reactions,
are run on GPUs.  \pynucastro's \amrexastrocxxnetwork\ outputs the
right-hand side function and Jacobian for a network in a manner that
can be run on the GPU.  The entire ODE integration is then launched as
a GPU kernel, spreading the zones in an \amrex\ box across the GPU
cores.  Small differences in the thermodynamics can lead to large
differences in the difficulty of burning a zone, due to the strong
nonlinearity of the reaction rates.  This means that some zones will
take very few steps in the ODE integration while others might require
thousands or more.  This thread-divergence can make it difficult to
get good performance on GPUs.

Reaction network GPU kernels can be large and are known to cause
issues with compilers and register allocation
\citep{exascale-readiness}. We address these challenges using continuous integration (CI) workflows to test the performance of exported networks with the latest compiler suites. We also invest significant development effort into reducing the memory requirements and associated register pressure of our reaction kernels.  Since \pynucastro\ 2.0, we have focused on
optimizing the \amrexastrocxxnetwork\ kernels by caching table
indexing for partition functions and tabulated weak rates, removing or
combining expensive exponentials as well as using CUDA/HIP library implementations for exponentiation,
reducing memory usage, 
and experimenting with single-precision Jacobians.

\begin{figure}[t]
\centering
\plotone{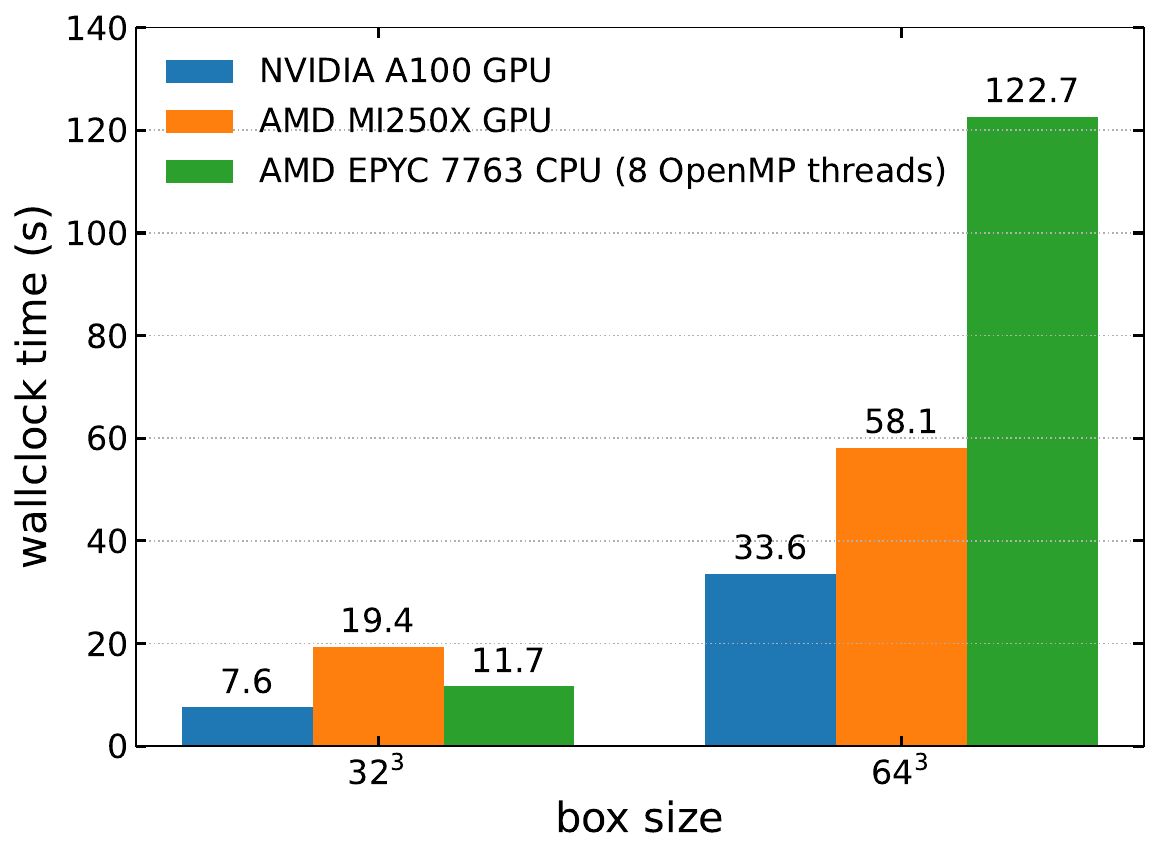}
\caption{\label{fig:gpu} Comparison of CPU and GPU performance for
network integration using {\tt AmrexAstroCxxNetwork} with
VODE ODE integrator and Helmholtz EOS provided in \microphysics.
This test measures the wallclock time for $32^3$ and $64^3$ zones
with varying density, temperature and composition.}
\end{figure}

We perform a test of the network integration on CPUs and GPUs.  Our
test creates a cube of data (a single AMR box), with density varying
along one axis (logarithmically-spaced from $10^6$ to $10^8~\gcc$),
temperature varying along a second axis (logarithmically-spaced from
$5\times 10^8$ to $2\times 10^9~\mathrm{K}$), and composition along the
last axis (setting the initial composition ranging from 20\% to 90\%
$\isotm{He}{4}$ with the remaining fraction equally split across the
nuclei).  This wide range in density and temperature will make the
difficulty of the burn vary greatly throughout the cube, with the 
number of right-hand side calls by the integrator ranging from about 70 to 7000.  This will
lead to thread-divergence on GPUs, which makes this a difficult (but
real-world) test.

In the \amrex-Astrophysics simulation codes, when doing operator-splitting, we work in terms
of mass fractions and solve:
\begin{subequations}
\begin{align}
\frac{dX_k}{dt} &= \dot{\omega}_k \\
\frac{de}{dt} &= \epsilon_\mathrm{nuc} - \epsilon_{\nu,\mathrm{weak}} - \epsilon_{\nu,\mathrm{therm}} \enskip ,
\end{align}
\end{subequations}
where $\dot{\omega}_k$ are the species creation and destruction rates (the mass fraction
equivalent to ${\bf f}$ in Eq.~\ref{eq:net_Y}), and as before,
$\epsilon_{\mathrm{nuc}}$ is the energy generation rate from the network, 
$\epsilon_{\nu, \mathrm{weak}}$ is the neutrino loss term from weak reactions,
and $\epsilon_{\nu, \mathrm{therm}}$ is the thermal neutrino loss term. 
The effect of electron screening is also included when evaluating $\dot{\omega}_k$, using the formulation of \citet{graboske:1973,alastuey:1978,itoh:1979}.
This system is closed via the equation of state (we use the Helmholtz EOS of
\citealt{timmes_swesty:2000}), which is called each time we evaluate
the right-hand side of the ODE system.  We use the VODE ODE integrator,
and integrate for $10^{-5}~\mathrm{s}$.  When running on GPUs, the
entire ODE integration is performed as a single kernel launch on the
GPU.

We use a box of either $32^3$ or $64^3$ zones---these are
representative of the sizes used for production science simulations.
When running on CPUs, we use OpenMP with logical tiling \citep{tiling}
to divide the box across threads, and compile with GCC 13.3.  When running on GPUs, the kernel
launch puts a single zone on each GPU core.  With AMD GPUs we use ROCm
7.2.0 (this gives much better performance on our kernels than earlier
versions) and with NVIDIA GPUs we use CUDA 12.9.  
While both Frontier and Perlmutter have multiple GPUs / node, 
for this test we run only on a single GPU (or single ``graphics compute
die'' for the AMD MI250X), since we are only using a single box of data. 
This is representative of how a large simulation would run, where each GPU
on a node would be given its own box in our AMR hierarchy to work on. 
Figure~\ref{fig:gpu} shows a comparison of the {\tt ase} network 
(shown in Appendix~\ref{appendix:nse})
on the CPUs (AMD EPYC 3 processors) and GPUs (AMD
MI250X) of the OLCF Frontier machine and the GPUs on the NERSC
Perlmutter machine (NVIDIA A100).  
The wallclock time includes only the integration kernel, 
and not any initialization.
On the smaller box ($32^3)$, the
NVIDIA A100 GPU gives the best performance, with the AMD MI250X about
$2.5\times$ slower.  The CPU run, with 8 threads is faster than the
AMD GPU.  On the larger box ($64^3$), we again see that the NVIDIA
A100 GPU leads the performance, but now the AMD MI250X GPU is not far
behind.  The CPU run (even with 8 threads) is much longer.
Overall, this test shows that the exported \amrexastrocxxnetwork gets
excellent GPU performance.

\subsection{\simplecxxnetwork and \fortrannetwork}
\label{sec:new_network_types}

The \simplecxxnetwork has added based on community needs.  The goal of
this network is to output the \cxx code required to evaluate the
right-hand side, ${\bf f({\bf y}})$ and Jacobian on the reaction ODE
system without needing to use the \amrex library.  To facilitate this,
\simplecxxnetwork uses a small header that implements the few
basic data types we rely on from \amrex (fixed-size arrays).  The
\simplecxxnetwork class outputs a header-only \cxx implementation of
the network, and supports all of the major rate types of \pynucastro,
except for \starlibrate currently.  It also includes a screening
function from \citet{chugunov:2007} and a basic driver that evaluates
the rates for a given thermodynamic state.  A user of this network
needs to provide their own integrator, since the integrator often
depends on how a network is coupled to the simulation code, and decide
if they want to include energy evolution.

The \fortrannetwork is a lightweight wrapper to \simplecxxnetwork that
uses Fortran 2003 C-bindings to pass information between the
languages.  A Fortran code can use these interfaces to access the
information computed in the \cxx functions.  We decided on this path
instead of outputting Fortran code directly for ease of maintenance.

\subsection{Testing}
\label{sec:testing}

Substantial unit tests (currently > 900 with > 85\%
code coverage\footnote{{\tt pytest}
{\tt coverage} misses lines in functions that are wrapped with
\numba JIT compilation, so the actual coverage is higher.}) are run on
every pull request, including tests to ensure external simulations
using generated \cxx code give the same answers with the changes.  The
Jupyter notebooks that make up the documentation are also run as tests
using the {\tt pytest} {\tt nbval} plugin. Several recent versions of
python are tested on both Linux and Mac GitHub runners, and a single python
version is tested with Windows.

A particularly important test is to compare the various network
backends, to make sure that they all agree.  To facilitate this
testing, \pynucastro\ provides a {\tt NetworkCompare} class that takes
a \library and then directly evaluates the rates and $d\mathbf{Y}/dt$ terms
from a \ratecollection, the python module written by \pythonnetwork,
and the \cxx outputs from \simplecxxnetwork and \amrexastrocxxnetwork.
The output of these four backends are then compared to a relative
tolerance of $10^{-11}$ when screening is not used and $10^{-6}$ with
screening (using the \citealt{chugunov:2007} screening formulation).
The looser tolerance with screening is due to optimizations done in
the \cxx implementation that create slightly larger floating point
differences with the python implementation.  This comparison is done
on several different network types, including rate approximations,
derived reverse rates, modified rates, and tabular weak rates, and
ensures that we get agreement across all backends.  This is run
automatically via {\tt pytest} through the GitHub CI infrastructure
on every pull request.

\section{Summary and Future Directions}

\pynucastro 3 is a major update to the \pynucastro library, and brings
many new tools for analyzing networks and new network types for
broader community use.  Enhanced thermodynamic support throughout the
library allows for much more realistic one-zone burning calculations
and greatly enhances the utility of interactive sessions exploring
concepts in nuclear astrophysics.  The addition of \branchedrate and
\modifiedrate, and the enhancements to \approximaterate allow for the
creation of complex, efficient reaction networks, as demonstrated with
the CNO and C/O-burning approximations.  The ability to sample
uncertainty in StarLib rates opens up new possibilities for
understanding uncertainty in nucleosynthetic yields, both within a
python environment and in exported networks.  New tools, including
interfacing with \networkx and assessing the stiffness of a network
add to the large capabilities \pynucastro provides for analyzing
networks.  Finally, the new export types for networks will enable more
simulation codes to use \pynucastro-generated networks.

The current features of \pynucastro have been
driven mostly by the science needs of the developers.  Looking
forward, there are a number of new developments we can imagine for a
4.0 release, which we briefly describe here.  These ideas for future
features also serve as an entry-point for new contributors from the
community.

Presently, \pynucastro assumes CGS units throughout, except in a few
places (like $Q$ values) where energies are given in MeV.  Units used
in nuclear astrophysics can be confusing, and a great way to fix this
is to adopt a unit system throughout the code.  We envision using the
\unyt library \citep{unyt} for this purpose---attaching units to every
quantity used in a python workflow with \pynucastro.

Support for an equation of state was one of the major additions in
\pynucastro 3.  Presently, the EOS computes the Fermi integrals as
needed, which is a major expense.  We will instead switch to a
tabulation of the Helmholtz free energy, following the ideas of
\citet{timmes_swesty:2000}, computing the table at 128 or 256-bit
precision for increased accuracy using \citet{electron_positron_cxx}.
We will also explore the possibility of adding nuclear equations of
state.

The current capabilities of \pynucastro are largely
driven by the needs of the developers.  We can imagine in the future 
expanding to other areas of nuclear physics, e.g., include nuclear
energy level structures and transition lifetimes to allow for weak
rate on-the-fly calculations (like done in \citealt{mesa:2015}).

The biggest future opportunity is supporting other astrophysics
simulation codes, outside of the \amrex community.  A lot of progress
has been made on this front already via \simplecxxnetwork and
\fortrannetwork.  Presently, only \amrexastrocxxnetwork runs on GPUs,
but if there is community interest, we can explore what is needed to
allow \simplecxxnetwork to utilize GPUs through a Kokkos interface.

Enhancing our interactions with \mesa
\citep{mesa:2011,mesa:2013,mesa:2015,mesa:2018,mesa:2019} is another
area to focus on.  Many multi-dimensional simulations of stellar
convection and explosions begin by mapping a 1D model from a stellar
evolution code onto the multi-d grid.  \pynucastro can already work
with data from a \mesa simulation using {\tt mesa\_reader}
\citep{mesa_reader}, producing a simple container class of \pynucastro
thermodynamic quantities indexed by \mesa zone. This then allows for a
network to be built and analysis of the reactive flow to be performed
using \pynucastro.  A potential use case for this would be to test if
the reaction network used in \mesa was sufficient, or if important
rates were missing.  At the moment, this works best for \mesa problem
setups that used a general network built from a list of isotopes.  But
many \mesa models use rate approximations, especially to pp and CNO
burning.  So a first area to focus on for improvement is to implement
remainder of the approximate rates that \mesa commonly uses, using the
\approximaterate, \branchedrate, and \modifiedrate classes.  The combined \mesa + \pynucastro workflow
would then allow a user to explore whether any additional rates may
have been important in a \mesa simulation or to interactively (using
Jupyter widgets) explore the flow through the reaction network in
different regions of the \mesa model.
Completing the pp and CNO rate approximations will also enable \pynucastro
to recreate the widely-used {\tt aprox19} and {\tt aprox21} networks,
just like we now do with {\tt aprox13}.

For H-burning, we do not currently account for the neutrino losses
from the weak reactions in the pp-chain and CNO cycle.  We already
have the mechanism to do this with the tabular weak rates, but the
weak rates used in H-burning are typically those from ReacLib or
StarLib.  To finish this, we need a data source for the neutrino energies for these weak rates.

Finally, \citet{pynucastro2.1} demonstrated the network-reduction
algorithms from \pynucastro 2, including the directed relation graph
with error propagation (DRGEP) algorithm \citep{drgep,niemeyersung}.
These methods have not been widely utilized since then, but the need
still exists for automated methods to take a large network (100s of
nuclei) and reduce it down to the size that can be run in
multidimensional simulations.  These reduction methods will see
renewed focus in \pynucastro in the near future.

\begin{acknowledgments}
All of the code need to reproduce these plots is available on Zenodo
at \citet{pynucastro3_zenodo}.  The work at Stony Brook was supported by
DOE/Office of Nuclear Physics grant DE-FG02-87ER40317. 
ASC was supported by NASA grant 80NSSC25K7213.
This research used resources of the National Energy Research Scientific Computing Center (NERSC), a Department of Energy Office of Science User Facility using NERSC award NP-ERCAP0041237.  Computer time at the Oak Ridge Leadership Computing Facility was provided via a Director's Discretionary award.

We thank anonymous referees for their careful reviews
and helpful comments, which helped to improve the clarity of this manuscript.
\end{acknowledgments}

\software{AMReX \citep{amrex,amrex-gpu},
          AMReX-Astro Microphysics \citep{Microphysics,microphysics-joss},
          autodiff \citep{autodiff},
          Jupyter \citep{jupyter},
          matplotlib \citep{Hunter:2007},
          NetworkX \citep{networkx},
          numba \citep{numba},
          numpy \citep{numpy,numpy2020},
          pynucastro \citep{pynucastro,pynucastro2},
          pytest \citep{pytest},
          SciPy \citep{scipy},
          SymPy \citep{sympy},
          VODE \citep{vode},
          yt \citep{yt}}

\appendix

\renewcommand{\theHequation}{\thesection.\arabic{equation}}

\section{Modified Rate}
\label{appendix:modifiedrate}

In this section, we discuss the form of the rate equation for \modifiedrate
involving custom stoichiometry.
Let's first consider a general nuclear reaction rate without a custom stoichiometry.
Consider a reaction involving a set of $\widetilde{r}$ number of distinct reactants $\{\widetilde{C}_i\}_{i=1}^{\widetilde{r}}$ 
and $\widetilde{p}$ number of distinct products $\{\widetilde{B}_j\}_{j=1}^{\widetilde{p}}$
with their corresponding multiplicity counts, $c_\xi$ for each species  
$\xi \in \{\widetilde{C}_i\}_{i=1}^{\widetilde{r}} \cup \{\widetilde{B}_j\}_{j=1}^{\widetilde{p}}$.
Or equivalently, a set of $r$  number of total reactants $\{C_i\}_{i=1}^r$
and $p$ number of total products $\{B_j\}_{j=1}^p$, 
where $r = \sum_{i=1}^{\widetilde{r}} c_{\widetilde{C}_i}$ and $p = \sum_{j=1}^{\widetilde{p}} c_{\widetilde{B}_j}$.
Any class that inherits from the base \rate class, including \modifiedrate,
defines {\tt reactants} and {\tt products}, which store the lists of nuclei participating in the reaction rate
and correspond to the sets $\{C_i\}_{i=1}^r$ and $\{B_j\}_{j=1}^p$.
For example, consider the $\isotm{C}{12}(\isotm{C}{12},\alpha)\isotm{Ne}{20}$ rate.
This reaction can be represented in two equivalent ways.
The first described by the non-tilde notation so that the reactant and product lists
are $\{C_i\}_{i=1}^{r=2} = \{\isotm{C}{12}, \isotm{C}{12}\}$ and
$\{B_i\}_{i=1}^{p=2} = \{\isotm{Ne}{20}, \isotm{He}{4}\}$, respectively.
Alternatively, the same set of reaction can be described by the 
two sets of distinct reactants and products together with their multiplicities
-- represented by the tilde notation introduced previously.
In this case, the distinct reactant set is
$\{\widetilde{C}_i\}_{i=1}^{\widetilde{r}=1} = \{\isotm{C}{12}\}$
with the multiplicity $c_{\isotm{C}{12}} = 2$,
while the distinct product set is 
$\{\widetilde{B}_i\}_{i=1}^{\widetilde{p}=2} = \{\isotm{Ne}{20}, \isotm{He}{4}\}$
with a multiplicity $c_{\isotm{Ne}{20}} = c_{\isotm{He}{4}} = 1$.
The generic reaction rate then takes the form
\begin{equation}
    \label{eq:general_forward_rate}
    C_1 + C_2 + \cdots + \ C_{r} \rightarrow B_1 + B_2 + \cdots + B_{p} \enskip .
\end{equation}
The evolution of the number densities of the $k$-th distinct reactant and the $l$-th distinct product
participating in this reaction is governed by
\begin{equation}
    \label{eq:number_density_ode}
    \dfrac{1}{c_{\widetilde{C}_k}}\dfrac{dn_{\widetilde{C}_k}}{dt} = -\dfrac{1}{c_{\widetilde{B}_l}}\dfrac{dn_{\widetilde{B}_l}}{dt} =
    - \dfrac{\lambda_{\{C_i\}}}{N_A^{r-1}}\dfrac{n_{C_1} n_{C_2} \cdots n_{C_r}}{\prod_{i=1}^{\widetilde{r}}c_{\widetilde{C}_i}!} \enskip ,
\end{equation}
where $\lambda_{\{C_i\}}$ is the reactivity of the reaction rate,
e.g.\ $\lambda = N_A \langle\sigma v\rangle$ for a two-body reaction.

Instead of working with number densities, it is often customary to work with molar fractions (e.g.\ \citealt{reaclib}) defined as
\begin{equation}
\label{eq:molar_frac_def1}
Y_i = \frac{n_i}{n} = \frac{n_i}{N_A \rho} \enskip .
\end{equation}
Therefore, Eq.~\ref{eq:number_density_ode} can be written as
\begin{equation}
    \label{eq:molar_fraction_ode}
    \dfrac{1}{c_{\widetilde{C}_k}}\dfrac{dY_{\widetilde{C}_k}}{dt} = -\dfrac{1}{c_{\widetilde{B}_l}}\dfrac{dY_{\widetilde{B}_l}}{dt} =
    - \rho^{r-1}\lambda_{\{C_i\}}
    \dfrac{Y_{C_1} Y_{C_2} \cdots Y_{C_r}}{\prod_{i=1}^{\widetilde{r}} c_{\widetilde{C}_i}!} \enskip .
\end{equation}
Eq.~\ref{eq:molar_fraction_ode} is the general form of the rate equation obeyed by all \rate classes in \pynucastro.
Now consider the inclusion of a custom {\tt stoichiometry}, which is specified as a dictionary that assigns the
custom stoichiometric coefficients to the individual nuclei. In this case, the multiplicity factors,
$c_{\widetilde{C}_k}$ and $c_{\widetilde{B}_l}$, are no longer determined by the number of times a species appears in
the {\tt reactants} or {\tt products} lists. But instead, they are explicitly set by the values provided in {\tt stoichiometry}.
Therefore, the evolution of the molar fraction for the species that have a custom stoichiometry assigned
to them is
\begin{equation}
    \label{eq:molar_fraction_stoichiometry_ode}
    \dfrac{1}{s_{\widetilde{C}_k}}\dfrac{dY_{\widetilde{C}_k}}{dt} = -\dfrac{1}{s_{\widetilde{B}_l}}\dfrac{dY_{\widetilde{B}_l}}{dt} =
    - \rho^{r-1}\lambda_{\{C_i\}}
    \dfrac{Y_{C_1} Y_{C_2} \cdots Y_{C_r}}{\prod_{i=1}^{\widetilde{r}} c_{\widetilde{C}_i}!} \enskip ,
\end{equation}
where $s_{\widetilde{C}_k}$ and $s_{\widetilde{B}_l}$ are the custom 
stoichiometric coefficient for reactant $\widetilde{C}_k$ and product $\widetilde{B}_l$.
It is important to stress that the reaction flux term
on RHS of Eq.~\ref{eq:molar_fraction_ode} and 
Eq.~\ref{eq:molar_fraction_stoichiometry_ode} is exactly the same.
The multiplicity factor, $c_{\widetilde{C}_i}$, that accounts for 
double-counting in the reaction flux term, as well as the density and
molar abundance dependencies are still determined by the set $\{C_i\}_{i=1}^r$,
while {\tt stoichiometry} only affects the multiplicative factor in the 
full rate equation.

To illustrate how custom {\tt stoichiometry} is used, 
we invite the reader to compare Eq.~\ref{eq:molar_fraction_stoichiometry_ode} with the stoichiometry of
Eq.~\ref{eq:modifiedrate_ex} for the CNO-II reaction chain 
discussed in Section~\ref{sec:modified_rates}.
To obtain Eq.~\ref{eq:modifiedrate_ex}, 
we simplified the original reaction chain 
using the rate-limiting process 
${}^{16}\mathrm{O} (p,\gamma){} {}^{17}\mathrm{F}$ 
to motivate the approximation that the production and consumption rates 
in the chain are equal for the intermediate heavy nuclei. 
The RHS of Eq.~\ref{eq:molar_fraction_stoichiometry_ode}, 
by contrast, corresponds to the lone reaction
${}^{16}\mathrm{O} (p,\gamma){} {}^{17}\mathrm{F}$. 
Consequently, we see that we can write 
$dY_{{}^{16}\mathrm{O}}/dt$ and $dY_{p}/dt$
for the reaction chain of Section~\ref{sec:modified_rates} 
by selecting the single ${}^{16}\mathrm{O} (p,\gamma){} {}^{17}\mathrm{F}$
reaction rate in pynucastro, with the modified {\tt stoichiometry} 
$s_{{}^{16}\mathrm{O}} = 1$ and $s_{p} = 2$.

\section{Derived Rate Update}
\label{appendix:nse}

In this section, we address the mistakes we made in \citet{pynucastro2},
causing a slight mismatch between the equilibrium 
mass abundance obtained from integrating the reaction network and what is predicted from Eq.~\ref{eq:nse_eq}.
Following the notation developed in Appendix~\ref{appendix:modifiedrate}, 
now consider both the forward {\it and} reverse reactions between $\{C_i\}_{i=1}^r$ and $\{B_j\}_{j=1}^p$
\begin{equation}
    \label{eq:rate_pair_form}
    C_1 + C_2 + \cdots + \ C_{r} \rightleftharpoons B_1 + B_2 + \cdots + B_{p} \enskip .
\end{equation}

The evolution of the molar fraction of the $k$-th distinct reactant and the $l$-th distinct product
participating in the forward and reverse reactions is governed by
\begin{equation}
    \label{eq:molar_fraction_for_rev_ode}
    \dfrac{1}{c_{\widetilde{C}_k}}\dfrac{dY_{\widetilde{C}_k}}{dt} = -\dfrac{1}{c_{\widetilde{B}_l}}\dfrac{dY_{\widetilde{B}_l}}{dt} =
    - \rho^{r-1}\lambda_{\{C_i\}}
    \dfrac{Y_{C_1} Y_{C_2} \cdots Y_{C_r}}{\prod_{i=1}^{\widetilde{r}} c_{\widetilde{C}_i}!} 
    + \rho^{p-1}\lambda_{\{B_j\}}
    \dfrac{Y_{B_1} Y_{B_2} \cdots Y_{B_p}}{\prod_{j=1}^{\widetilde{p}} c_{\widetilde{B}_j}!} \enskip .
\end{equation}
With the equilibrium condition, the ratio between the reverse and forward rate can be written as 
\begin{equation}
\label{eq:equilibrium_ratio}
    \frac{\lambda_{\{B_j\}}}{\lambda_{\{C_i\}}} = \rho^{r-p} 
    \dfrac{Y_{C_1} Y_{C_2} \cdots Y_{C_r}}{Y_{B_1} Y_{B_2} \cdots Y_{B_p}} 
    \dfrac{\prod_{j=1}^{\widetilde{p}} c_{\widetilde{B}_j}!}{\prod_{i=1}^{\widetilde{r}} c_{\widetilde{C}_i}!} \enskip .
\end{equation}

Until now, we are consistent with Eq.~B4 from \citet{pynucastro2},
except that we have converted number densities to molar fractions explicitly.
To continue, we need to consider the expression for the molar fraction of
the Boltzmann gas. 
By definition, the atomic mass in the units of atomic mass unit
is $\mathcal{A}_i = m_i / m_u$, which is different from the
mass number, $A_i = N_i + Z_i$.
However, Eq.~B5 of \citet{pynucastro2}, assumes the approximation $m_i = \mathcal{A}_i m_u \simeq A_i m_u$,
which introduces an inconsistency in masses used when computing the NSE mass abundance
from Eq.~\ref{eq:nse_eq} and when integrating the reaction network with the inverse
rate defined using Eq.~\ref{eq:equilibrium_ratio}.
Moreover, Eq.~\ref{eq:nse_eq} assumes the relation
\begin{equation}
    \label{eq:molar_frac_def2}
    n_i = \frac{\rho X_i}{m_i} = \frac{\rho Y_i A_i}{m_i} = \rho N_A Y_i \frac{A_i}{\mathcal{A}_i} \enskip .
\end{equation}
Note the extra factor of $A_i/\mathcal{A}_i$ compared to Eq.~\ref{eq:molar_frac_def1}.
This inconsistency can be resolved with a different definition where $Y_i = X_i/\mathcal{A}_i$
or let $X_i = A_i n_i/n$ to represent {\it baryon fraction} as proposed in Appendix A of \citet{mesa:2013}
(see similar discussions on this issue in \citet{winnet}).
Nevertheless, this difference in the relation between molar abundance and number density is {\it still} acceptable
provided that the same molar fraction expression derived under Maxwell-Boltzmann approximation 
(Eq.~\ref{eq:mb_massfrac}) is used when deriving the NSE equation (Eq.~\ref{eq:nse_eq}), 
and when computing the ratio of the forward and reverse rates under equilibrium (Eq.~\ref{eq:equilibrium_ratio}).
Now substituting Eq.~\ref{eq:mb_massfrac} into Eq.~\ref{eq:equilibrium_ratio} and
follow the procedure outlined in \cite{pynucastro2}, we get
\begin{align}
\label{eq:equilibrium_ratio_full}
    \frac{\lambda_{\{B_j\}}}{\lambda_{\{C_i\}}} &= \left(m_u\right)^{r-p}
    \overbrace{\frac{A_{B_1} \cdots A_{B_p}}{A_{C_1} \cdots A_{C_r}} \left(\frac{\mathcal{A}_{C_1} \cdots \mathcal{A}_{C_r}}{\mathcal{A}_{B_1} \cdots \mathcal{A}_{B_p}}\right)^{5/2}}^{\mathrm{Different \ Term}}
    \left(\frac{m_u k_B T}{2\pi\hbar^2}\right)^{\frac{3}{2}(r-p)} 
    \exp{\left(\frac{-Q_{\mathrm{for}}}{k_B T}\right)} \\ \nonumber
    & \times \frac{(2J_{C_1} + 1) \cdots (2J_{C_r} + 1)}{(2J_{B_1} + 1) \cdots (2J_{B_p} + 1)}
    \frac{G_{C_1} \cdots G_{C_r}}{G_{B_1} \cdots G_{B_p}}
    \dfrac{\prod_{j=1}^{\widetilde{p}} c_{\widetilde{B}_j}!}{\prod_{i=1}^{\widetilde{r}} c_{\widetilde{C}_i}!} \enskip ,
\end{align}
where $Q$-value of the forward rate, $Q_{\mathrm{for}}$, is defined as
\begin{equation}
    \label{eq:q-value}
    Q_{\mathrm{for}} \equiv \left(\sum_{i=1}^r m_i - \sum_{j=1}^{p} m_j \right) c^2 \enskip .
\end{equation}

Comparing with Eq.~B9 from \cite{pynucastro2}, the difference appears in the term labeled
as {\it Different Term} in Eq.~\ref{eq:equilibrium_ratio_full}.
This difference arises because the molar fraction definition defined in Eq.~\ref{eq:mb_massfrac}
is $\propto \mathcal{A}_i^{5/2}/A_i$ whereas \citet{pynucastro2} assumes a dependence of $\propto A_i^{3/2}$.
In \pynucastro, \derivedrate is now constructed following Eq.~\ref{eq:equilibrium_ratio_full}.
Additional caveats when constructing \derivedrate include

\begin{enumerate}
    \item It is arbitrary to pick the forward ($Q>0$) or the reverse rate ($Q<0$)
    to use as a source rate to construct the corresponding inverse rate via detailed balance.
    
    \item The binding energy, $B_i$, used in Eq.~\ref{eq:nse_eq} and
    the Q-value of the source rate used to compute the corresponding \derivedrate should
    be computed from the masses to ensure full compatibility between NSE and reaction network.
    This is also discussed in other literature such as \citet{winnet}.
    
    \item In practice, Eq.~\ref{eq:equilibrium_ratio_full} should be evaluated in log-space
    to avoid numerical issues when evaluating the derived-forward rate from the reverse rate at
    low temperatures, $T \sim 10^8 \ K$. 
    Note that this is automatically taken care of when Eq.~\ref{eq:equilibrium_ratio_full} 
    is fitted into the ReacLib format as shown in Eq.~17 from \citet{rauscher:2000}
    and Eq.~B10 from \citet{pynucastro2}.
\end{enumerate}

\section{The {\tt ase} Network}
\label{appendix:ase}

The {\tt ase} network is designed for helium and carbon burning with full NSE compatibility.
It is based on the {\tt aprox13}, but without the carbon and oxygen burning approximation
discussed in Section~\ref{sec:co-burning}.
It also includes an additional rate sequence, 
$\isotm{C}{12}(p, \gamma)\isotm{N}{13}(\alpha, p)\isotm{O}{16}$,
and some {\tt ModifiedRate} that incorporate the effect of neutron release and capture
for carbon and oxygen burning as discussed in Section~\ref{sec:modified_rates}.
The {\tt ase} network is generated as the following

\begin{lstlisting}
reaclib_lib = pyna.ReacLibLibrary()

all_reactants = ["p",
                 "he4", "c12", "o16", "ne20", "mg24", "si28", "s32",
                 "ar36", "ca40", "ti44", "cr48", "fe52", "ni56",
                 "na23", "al27", "p31", "cl35", "k39",
                 "sc43", "v47", "mn51", "co55",
                 "n13"]

lib = reaclib_lib.linking_nuclei(all_reactants)

other_rates = [("c12(c12,n)mg23", "mg24"),
               ("o16(o16,n)s31", "s32"),
               ("o16(c12,n)si27", "si28")]

for r, mp in other_rates:
    _r = reaclib_lib.get_rate_by_name(r)
    forward_rate = pyna.ModifiedRate(_r, new_products=[mp])
    derived_rate = pyna.DerivedRate(forward_rate, use_pf=True)
    lib += pyna.Library(rates=[forward_rate, derived_rate])

rates_to_remove = ["p31(p,c12)ne20",
                   "si28(a,c12)ne20",
                   "ne20(c12,p)p31",
                   "ne20(c12,a)si28",
                   "na23(a,g)al27",
                   "al27(g,a)na23",
                   "al27(a,g)p31",
                   "p31(g,a)al27"]

for r in rates_to_remove:
    _r = net.get_rate_by_name(r)
    lib.remove_rate(_r)

rates_to_derive = lib.backward().get_rates()
for r in rates_to_derive:
    fr = lib.get_rate_by_nuclei(r.products, r.reactants)
    if fr:
        lib.remove_rate(r)
        d = pyna.DerivedRate(fr, use_pf=True)
        lib.add_rate(d)

net = AmrexAstroCxxNetwork(libraries=[lib])
net.make_ap_pg_approx(intermediate_nuclei=["cl35", "k39", "sc43", "v47", "mn51", "co55"])
net.remove_nuclei(["cl35", "k39", "sc43", "v47", "mn51", "co55"])
\end{lstlisting}

\begin{figure}
    \centering
    \plotone{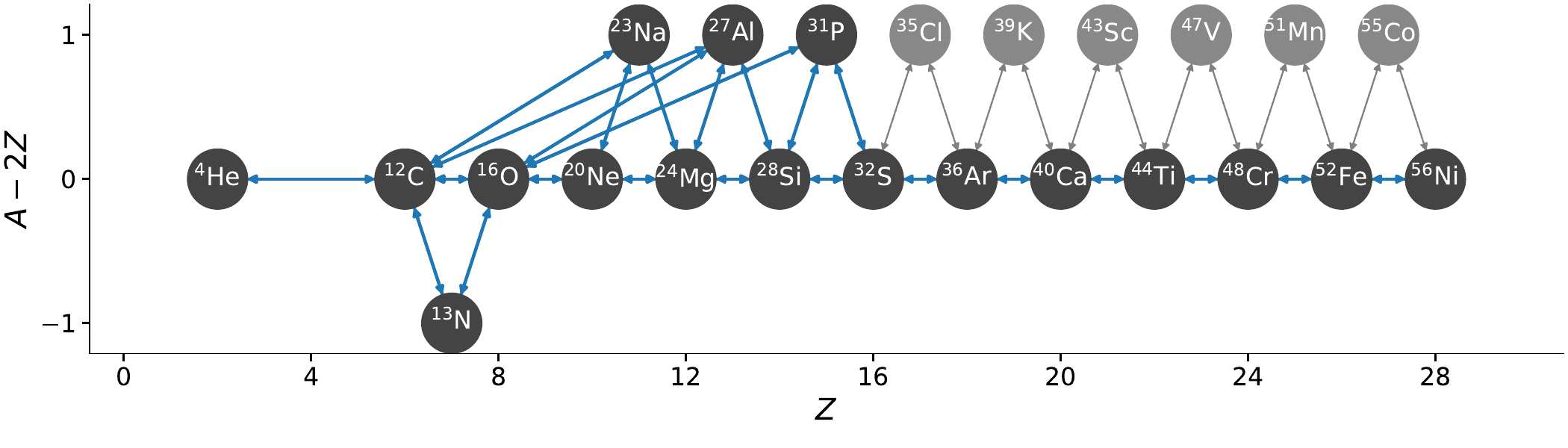}
    \caption{The {\tt ase} network used for testing network integration on both CPUs and GPUs.}
    \label{fig:ase-net}
\end{figure}
A visualization of the {\tt ase} network is shown in Figure~\ref{fig:ase-net}.

\bibliographystyle{aasjournal}


\bibliography{refs}

\end{document}